\documentclass[aps,showpacs,floatfix]{revtex4}
\usepackage[latin9]{inputenc}
\usepackage{amsmath}
\usepackage{amssymb}
\usepackage{graphicx}
\usepackage{esint}
\usepackage{epstopdf}
\usepackage{braket}

\usepackage{hyperref}

\makeatletter

\def\endfmffile{%

	\fmfcmd{\p@rcent\space the end.^^J%
			end.^^J%
			endinput;}%

	\if@fmfio

		\immediate\closeout\@outfmf

	\fi

	\ifnum\pdfshellescape=\@ne

		\immediate\write18{mpost \thefmffile}%

	\fi}

\begin{document}
\unitlength = 1mm

\title{Non-linear stationary solutions in realistic models for analog black-hole lasers}

\author{Juan Ramón Muñoz de Nova}
\affiliation{Department of Physics, Technion-Israel Institute of Technology, Technion City, Haifa 32000, Israel}

\begin{abstract}
From both a theoretical and an experimental point of view, Bose-Einstein condensates are good candidates for studying gravitational analogues of black holes and black-hole lasers. In particular, a recent experiment has shown that a black-hole laser configuration can be created in the laboratory. However, the most considered theoretical models for analog black-hole lasers are quite difficult to implement experimentally. In order to fill this gap, we devote this work to present more realistic models for black-hole lasers. For that purpose, we first prove that, by symmetrically extending every black-hole configuration, one can obtain a black-hole laser configuration with an arbitrarily large supersonic region. Based on this result, we propose the use of an attractive square well and a double delta-barrier, which can be implemented using standard experimental tools, for studying black-hole lasers. We also compute the different stationary states of these setups, identifying the true ground state of the system and discussing the relation between the obtained solutions and the appearance of dynamical instabilities.
\end{abstract}

\pacs{03.75.Kk 04.62.+v 04.70.Dy \volumeyear{2012} \volumenumber{number}
\issuenumber{number} \eid{identifier} \startpage{1}
\endpage{}}

\date{\today}

\maketitle

\section{Introduction}

Hawking radiation is one of the most intriguing results of theoretical Physics; using a semiclassical model in which fields are quantized on top of a classical gravitational background, Hawking predicted the spontaneous emission of radiation by the event horizon of a black hole (BH) \cite{Hawking1974,Hawking1975}. Within a similar scheme, Corley and Jacobson \cite{Corley1999} showed that a bosonic field with superluminal dispersion relation in a metric with two horizons can give rise to a dynamical instability, the so-called {\em black-hole laser} (BHL) effect. The problem is that the observation of such phenomena seems unlikely in the near future due to the small effective temperature of emission, $T_H\simeq 62 M_{\odot}/M~\text{nK}$, with $M_{\odot}$ the mass of the Sun and $M$ the mass of the black-hole. For instance, the microwave background temperature is $2.7$ K, well above the Hawking temperature $T_H$.

An alternative way to study these effects was suggested by Unruh \cite{Unruh1981}, who proved that a subsonic-supersonic interface in a quantum fluid is the acoustic analog of an event horizon in a BH. This pioneering work opened the door to the study of gravitational problems in the laboratory and since then, many analog setups have been proposed in systems such as different as Fermi gases \cite{Giovanazzi2005}, ion rings \cite{Horstmann2010}, polaritons \cite{Nguyen2015} or, in a classical context, surface waves in a water tank \cite{Euve2016}.

Of particular interest are the analogues implemented in Bose-Einstein condensates (BEC), first suggested by Garay {\em et al.} \cite{Garay2000}. The main advantages of this kind of setup are the low temperature, the relative ease of handling, and the deep understanding of the quantum excitations. The analogue of the Hawking radiation in this system is the spontaneous emission of entangled phonons by the acoustic horizon into the subsonic and supersonic regions \cite{Leonhardt2003,Leonhardt2003a,Balbinot2008,Carusotto2008,Recati2009,Macher2009,Coutant2010,deNova2014,Finazzi2013,Busch2014,deNova2014a,deNova2015,Michel2016,Michel2016a}, similar to the particle-antiparticle creation at the event horizon of a BH. In addition, a flowing condensate presenting a finite-size supersonic region (giving rise to a pair of acoustic horizons) provides the analog of a black-hole laser \cite{Barcelo2006,Jain2007,Coutant2010,Finazzi2010,Michel2013,Michel2015,deNova2016}.

Regarding the experimental side, the first acoustic horizon in a BEC was produced by the Technion group \cite{Lahav2010} with the help of a sharp negative potential created by a laser that locally accelerate the atoms. Within this kind of setup, the first observation of the BHL effect was reported \cite{Steinhauer2014}, although there is still some discussion in the community about the interpretation of the experimental results \cite{Tettamanti2016,Wang2016,Steinhauer2017}. Recently, the same group provided the first experimental evidence of the emission of Hawking radiation by measuring the entanglement of the emitted phonons \cite{Steinhauer2016}.

Most of the theoretical works present in the BEC analog literature deal with an extremely idealized model, the so-called {\em  flat-profile} configuration, in which the background condensate is homogeneous and the horizons are created through a very specific spatial dependence of the coupling constant and the external potential. Although this simple model is able to capture the essential features of Hawking radiation, it is quite unrealistic from an experimental point of view. More realistic models study the formation of acoustic BHs considering the flow of a condensate through a localized obstacle, modeled by a delta barrier \cite{Kamchatnov2012}, or an optical lattice \cite{deNova2014a}; the waterfall configuration described in Ref. \cite{Larre2012} is a theoretical model of the actual experimental setups of Refs. \cite{Lahav2010,Steinhauer2016}.

The goal of this Article is to extend the previous results and provide more realistic theoretical models also for analog BHLs. For that purpose, we prove that each BH configuration can be symmetrically extended to provide a BHL configuration. By applying this result to the waterfall and the delta-barrier configurations described above, we obtain two new different black-hole laser configurations that are created by using an attractive square well and a double delta-barrier, respectively. For these configurations, we compute the different families of non-linear stationary states that characterize the stability of the system as well as its long-time behavior. We note that, although stationary transport scenarios in a square well or a double delta-barrier have already been studied in the literature \cite{Leboeuf2001,Zapata2011}, to the best of our knowledge this is the first time that they have been explicitly proposed for modeling black-hole lasers.

Apart from the intrinsic interest of finding new models from a theoretical point of view, these configurations are also expected to be very useful in practice; in particular, the case where the supersonic region is created using an square attractive well can be regarded as a model for studying the experimental BHL of Ref. \cite{Steinhauer2014}.

The scheme of the paper is the following. In Sec. \ref{sec:BHLaserintro} we revisit the basic theory of gravitational analogues in BEC. The general relation between BH and BHL solutions is proven in Sec. \ref{sec:BHLTheorem}. In Secs. \ref{sec:BHLWF}, \ref{sec:BHL2delta} we study the different stationary states for BHL configurations with a square-well and a double delta-barrier, respectively. Conclusions are presented in Sec. \ref{sec:conclu}. Appendix \ref{app:elliptic} is devoted to introduce the different elliptic functions used in this work while Appendices \ref{app:technicalwell}, \ref{app:technicaldelta} are devoted to the technical details of the calculations presented in the main text.

\section{Gravitational analogues in Bose-Einstein condensates}\label{sec:BHLaserintro}

We first provide in this section a general introduction to Bose-Einstein condensates and gravitational analogues. For more details, see for instance Refs. \cite{Macher2009a,Recati2009,Zapata2011,deNova2015Proc}.

\subsection{Effective one-dimensional configurations}\label{subsec:1D}

We begin by reviewing how to reach an effective one-dimensional (1D) configuration, the so-called 1D mean-field regime \cite{Leboeuf2001,Menotti2002}. For that purpose, we consider a 3D gas of $N$ bosons of mass $m$ near $T=0$ (more precisely, $T\ll T_c$, with $T_c$ the critical temperature of the condensate), described by the second-quantization Hamiltonian \cite{Fetter2003,Dickhoff2005}
\begin{equation}\label{eq:2ndHamiltonian}
\hat{H}=\int\mathrm{d}^3\mathbf{x}~\hat{\Psi}^{\dagger}(\mathbf{x})\left[-\frac{\hbar^2}{2m}\nabla^2+V_{\rm{ext}}(\mathbf{x},t)\right]\hat{\Psi}(\mathbf{x})\\
+\frac{g_{\rm{3D}}}{2}\hat{\Psi}^{\dagger}(\mathbf{x})\hat{\Psi}^{\dagger}(\mathbf{x})\hat{\Psi}(\mathbf{x})\hat{\Psi}(\mathbf{x})
\end{equation}
where $V_{\rm{ext}}(\mathbf{x},t)$ is the external potential and the interaction between atoms is taken into account for low momentum by a short-range potential, with $g_{\rm{3D}}=4\pi\hbar^2a_s/m$ the corresponding coupling constant and $a_s$ the $s$-wave scattering length \cite{Pitaevskii2003,Pethick2008}.

In order to obtain an effective 1D configuration along the $x$-axis, we consider a total external potential of the form $V_{\rm{ext}}(\mathbf{x},t)=V(x,t)+V_{\rm{tr}}(y,z)$, where $V(x,t)$ only depends on the $x$ coordinate while $V_{\rm{tr}}(y,z)=\frac{1}{2}m\omega_{\rm{tr}}^2\rho^2$ ($\rho=\sqrt{y^2+z^2}$ being the radial distance to the $x$-axis) represents a transverse harmonic trap, very usual in experimental setups. If the non-linear interacting term is sufficiently small, we can treat it perturbatively and assume that the transverse motion is frozen to the corresponding harmonic oscillator ground state, and hence use the following approximation for the field operator
\begin{equation}\label{eq:1Dansatz}
\hat{\Psi}(\mathbf{x})\simeq\hat{\psi}(x)\frac{e^{-\frac{\rho^2}{2a^2_{\rm{tr}}}}}{\sqrt{\pi}a_{\rm{tr}}},~a_{\rm{tr}}=\sqrt{\frac{\hbar}{m\omega_{\rm{tr}}}}~
\end{equation}
with $a_{\rm{tr}}$ being the transverse harmonic oscillator length. After integrating over the transverse degrees of freedom, we arrive at the following 1D effective Hamiltonian:
\begin{equation}\label{eq:1D2ndHamiltonian}
\hat{H}_{\rm{1D}}=\int\mathrm{d}x~\hat{\psi}^{\dagger}(x)\left[-\frac{\hbar^2}{2m}\partial_x^2+V(x,t)\right]\hat{\psi}(x)\\
+\frac{g_{\rm{1D}}}{2}\hat{\psi}^{\dagger}(x)\hat{\psi}^{\dagger}(x)\hat{\psi}(x)\hat{\psi}(x)
\end{equation}
where we have absorbed the resulting zero-point energy $\hbar\omega_{\rm{tr}}$ of the harmonic oscillator and the effective 1D constant coupling is $g_{\rm{1D}}=2\hbar\omega_{\rm{tr}}a_s$. More specifically, the condition for the approximation of Eq. (\ref{eq:1Dansatz}) to be valid is that the non-linear interacting term is small compared to the transverse confinement energy scale, $g_{1D}n_{1D}(x)\ll \hbar\omega_{\rm{tr}}$, which can be simply put as $n_{1D}(x)a_s\ll 1$, with $n_{1D}(x)$ the 1D-density.

In the same fashion, the 3D canonical commutation rules for the field operator
\begin{equation}\label{eq:canonical}
[\hat{\Psi}(\mathbf{x}),\hat{\Psi}^{\dagger}(\mathbf{x}')]=\delta(\mathbf{x}-\mathbf{x}')~,
\end{equation}
are reduced to the 1D version
\begin{equation}\label{eq:1Dcanonical}
[\hat{\psi}(x),\hat{\psi}^{\dagger}(x')]=\delta(x-x')~,
\end{equation}

As we will only deal with 1D configurations, in the following we omit everywhere the 1D index.

\subsection{Gross-Pitaevskii and Bogoliubov-de Gennes equations}\label{subsec:GPBdG}

Using the Hamiltonian (\ref{eq:1D2ndHamiltonian}) and the corresponding canonical commutation rules (\ref{eq:1Dcanonical}), we write the equation of motion for the field operator $\hat{\psi}(x)$ in the Heisenberg picture:
\begin{equation}\label{eq:Heisenberg}
i\hbar\partial_t\hat{\psi}(x,t)=[\hat{\psi}(x,t),\hat{H}]=\left[-\frac{\hbar^2}{2m}\partial_x^2+V(x,t)\right]\hat{\psi}(x,t)+
g\hat{\psi}^{\dagger}(x,t)\hat{\psi}(x,t)\hat{\psi}(x,t)
\end{equation}
Since there is a condensate, we can perform a mean-field approximation
\begin{equation}\label{eq:meanfieldGP}
\hat{\psi}(x,t)=\psi(x,t)+\hat{\varphi}(x,t),
\end{equation}
with $\psi(x,t)$ the Gross-Pitaevskii (GP) wave function \cite{Pitaevskii2003} describing the condensate and $\hat{\varphi}(x,t)$ representing the quantum fluctuations of the field operator.
The time evolution of the GP wave function is described by the {\em time-dependent} GP equation, a non-linear Schrödinger equation of the form:
\begin{equation}\label{eq:TDGP}
i\hbar\partial_t\psi(x,t)=\left[-\frac{\hbar^2}{2m}\partial_x^2+V(x,t)+g|\psi(x,t)|^2\right]\psi(x,t)
\end{equation}
Assuming that the depletion cloud (i.e., the cloud formed by the atoms outside the condensate) is negligible, $\psi(x,t)$ is normalized to the total number of particles:
\begin{equation}\label{eq:normalization}
N=\int\mathrm{d}x~|\psi(x,t)|^{2}
\end{equation}
The conservation of the norm of the GP wave function is guaranteed by the same relation as in the usual linear Schrodinger equation:
\begin{equation}\label{eq:GPnormconservation}
\partial_t|\psi(x,t)|^2+\partial_x J(x,t)=0,~J(x,t)=-\frac{i\hbar}{2m}\left[\psi^*(x,t)\partial_x \psi(x,t)-\psi(x,t)\partial_x \psi^*(x,t)\right]
\end{equation}
with $J(x,t)$ the current.

It is quite instructive to rewrite these equations in terms of the amplitude and phase of the wave function, $\psi(x,t)=A(x,t)e^{i\phi(x,t)}$,
\begin{eqnarray}\label{eq:PhaseAmplitude}
\partial_tn(x,t)+\partial_x[n(x,t)v(x,t)]&=&0\\
\nonumber -\hbar \partial_t\phi(x,t) &=&-\frac{\hbar^2}{2mA(x,t)}\partial_x^2A(x,t)+\frac{1}{2}mv^2(x,t)+V(x,t)+gn(x,t)
\end{eqnarray}
where $J(x,t)=n(x,t)v(x,t)$ is the current and
\begin{equation}\label{eq:CV}
n(x,t)= A^2(x,t),~v(x,t)=\frac{\hbar\partial_x\phi(x,t)}{m}~,
\end{equation}
are the mean-field density and flow velocity, respectively. Interestingly, the first line of Eq. (\ref{eq:PhaseAmplitude}) is the equivalent of the continuity equation for a hydrodynamical fluid. On the other hand, taking spatial derivative in the second line gives:
\begin{equation}\label{eq:Hydrodynamics}
m\partial_t v(x,t)=-\partial_x\left[\frac{1}{2}mv^2(x,t)+V(x,t)\right]-\frac{1}{n(x,t)}\partial_x P(x,t)+\partial_x\left[\frac{\hbar^2}{2m\sqrt{n(x,t)}}\partial_x^2\sqrt{n(x,t)}\right],~P(x,t)=\frac{gn^2(x,t)}{2}
\end{equation}
which can be regarded as the analog of the Euler equation for the velocity of a potential flow since the pressure of a uniform condensate at equilibrium is $P=\frac{gn^2}{2}$. The only difference is the rightmost term, which is a genuine quantum feature as it contains $\hbar$ and it is often called the {\em quantum pressure} term. However, in the {\em hydrodynamic} regime, where the density of the condensate varies on a large scale compared to the other terms, one can neglect the contribution of the quantum pressure and recover the same equations as for an ideal potential fluid flow; this is the key point of the gravitational analogy since the original analogy was precisely established for ideal potential fluid flows \cite{Unruh1981} (see also discussion in the next subsection).

On the other hand, to lowest order in the quantum fluctuations of the field operator, $\hat{\varphi}(x,t)$, one finds from Eq. (\ref{eq:Heisenberg}) that:
\begin{eqnarray}\label{eq:BdGfieldequation}
\nonumber i\hbar\partial_t\hat{\Phi}(x,t)&=&M(x,t)\hat{\Phi}(x,t), \\
M(x,t)&=&\left[\begin{array}{cc}G(x,t) & L(x,t)\\
-L^{*}(x,t)&-G(x,t)\end{array}\right],~\hat{\Phi}(x,t)=\left[\begin{array}{c}\hat{\varphi}(x,t)\\ \hat{\varphi}^{\dagger}(x,t)\end{array}\right]\\
\nonumber G(x,t)&=&-\frac{\hbar^2}{2m}\partial_x^2+V(x,t)+2g|\psi(x,t)|^2,~L(x,t)=g\psi^2(x,t)
\end{eqnarray}
which are known as the Bogoliubov-de Gennes (BdG) equations.

For time-independent potentials, $V(x,t)=V(x)$, we can look for particular solutions of the form
\begin{equation}
\hat{\psi}(x,t)=[\psi_0(x)+\hat{\varphi}(x,t)]e^{-i\frac{\mu}{\hbar}t}
\end{equation}
that are of special interest as they describe stationary configurations. In particular, the stationary wave function $\psi_0(x)$ obeys the {\em time-independent} GP equation:
\begin{equation}\label{eq:TIGP}
\mu\psi_0(x)=\left[-\frac{\hbar^2}{2m}\partial_x^2+V(x)+g|\psi_0(x)|^2\right]\psi_0(x)
\end{equation}
Note that the above equation is a {\em non-linear} eigenvalue problem. The presented amplitude-phase decomposition greatly simplifies the stationary problem as the continuity equation is reduced to
\begin{equation}
\partial_xJ(x)=0
\end{equation}
so the current $J(x)=n(x)v(x)=J$ is constant. Using this fact, we can rewrite the equation for the amplitude as a purely real second-order differential equation:
\begin{equation}\label{eq:GPNewton}
\mu A(x) =-\frac{\hbar^2}{2m}A''(x)+\frac{mJ^2}{2A^3(x)}+V(x)A(x)+gA^3(x)
\end{equation}
with $'$ the spatial derivative. The phase is simply obtained from the relation:
\begin{equation}\label{eq:stationaryphase}
\phi(x)=\int\mathrm{d}x~\frac{mv(x)}{\hbar}=\int\mathrm{d}x~\frac{mJ}{\hbar A^2(x)}
\end{equation}
Note that, if $J\neq 0$, neither the amplitude or the flow velocity vanish.

For a fixed value of the number of particles, there can be several different solutions for the GP equation (\ref{eq:TIGP}). The true ground state of the system is that minimizing the grand-canonical energy, $K=E-\mu N$, with $E$ the energy of the state (i.e., the expectation value of the Hamiltonian evaluated for the GP wave function) and $N$ the total number of particles. Indeed, by rewriting the expression for $K$ as a functional for the GP wave function:
\begin{equation}\label{eq:grandcanonicalenergy}
K[\psi]=\int\mathrm{d}x~\psi^*(x)\left[-\frac{\hbar^2}{2m}\partial^2_x+V(x)-\mu\right]\psi(x)+\frac{g}{2}|\psi(x)|^4,
\end{equation}
it can be seen that Eq. (\ref{eq:TIGP}) is precisely the condition for $\psi_0(x)$ to be an extreme of $K$. In that case, $K$ takes the simple form:
\begin{equation}\label{eq:grandcanonicalenergypsi0}
K[\psi_0]=-\int\mathrm{d}x~\frac{g}{2}|\psi_0(x)|^4.
\end{equation}
Solutions of Eq. (\ref{eq:TIGP}) which are not a local minimum of $K$ are energetically unstable as any perturbation would induce the system to decay to a lower energy state. In physical terms, we can understood the minimization of $K$ as the minimization of the expectation value of Hamiltonian $H$ with the constraint of fixed total number of particles $N$, with the chemical potential $\mu$ playing the role of Lagrange multiplier.

With respect to the quantum fluctuations, $\hat{\varphi}(x,t)$, Eq. (\ref{eq:BdGfieldequation}) is now a stationary problem of the form:
\begin{eqnarray}\label{eq:BdGfieldequationTI}
\nonumber i\hbar\partial_t\hat{\Phi}(x,t)&=&M(x)\hat{\Phi}(x,t), \\
M(x)&=&\left[\begin{array}{cc}G(x) & L(x)\\
-L^{*}(x)&-G(x)\end{array}\right],\\
\nonumber G(x)&=&-\frac{\hbar^2}{2m}\partial_x^2+V(x)+2g|\psi_0(x)|^2-\mu,~L(x)=g\psi_0^2(x)
\end{eqnarray}
As it is a linear equation, we can expand the field operator in terms of eigenmodes:
\begin{eqnarray}\label{eq:BdGmodesexpansion}
\nonumber\hat{\Phi}(x,t)&=&\sum_n \hat{\gamma}_nz_n(x)e^{-i\omega_nt}+\hat{\gamma}^{\dagger}_n\bar{z}_n(x)e^{i\omega_nt}\\
z_n(x)&=&\left[\begin{array}{c}u_n(x)\\ v_n(x)\end{array}\right],\bar{z}_n(x)=\left[\begin{array}{c}v^*_n(x)\\ u^*_n(x)\end{array}\right]
\end{eqnarray}
where the spinors $z_n$ satisfy the {\it time-independent} BdG equations:
\begin{equation}\label{eq:BdGeigenvalueequation}
M(x)z_n(x)=\epsilon_nz_n(x), \, \epsilon_n=\hbar\omega_n
\end{equation}
Due to the structure of the equations, the conjugate $\bar{z}_n$ is also a mode with energy $-\epsilon^*_n$.

An interesting property of the eigenvalue problem (\ref{eq:BdGeigenvalueequation}) is that it is non-Hermitian and thus it can yield complex eigenvalues. In particular, eigenvalues with positive imaginary part correspond to {\it dynamical instabilities}, i.e., exponentially growing modes: the presence of such dynamical instabilities in a finite region of a condensate flow are the origin of the black-hole laser effect, discussed in the next section.

Moreover, there is a Klein-Gordon type scalar product associated to the BdG eigenvalue problem, given by:
\begin{equation}\label{eq:KGProduct}
(z_1|z_2)\equiv\braket{z_1|\sigma_z|z_2}=\int\mathrm{d}^3\mathbf{x}~z_1^{\dagger}(\mathbf{x})\sigma_zz_2(\mathbf{x})=\int\mathrm{d}^3\mathbf{x}~u^*_1(\mathbf{x})u_2(\mathbf{x})-v^*_1(\mathbf{x})v_2(\mathbf{x}) \, ,
\end{equation}
with $\sigma_z=\rm{diag}(1,-1)$ the corresponding Pauli matrix. Note that this scalar product is not positive definite so the norm of a given solution $z_n$, defined as $(z_n|z_n)$, can be positive, negative or zero. In fact, the norm of the conjugate $\bar{z}_n$ has the opposite sign to that of $z_n$, $(\bar{z}_n|\bar{z}_n)=-(z_n|z_n)$.

The utility of this scalar product is that, as usual, two modes $z_n,z_m$ with different eigenvalues $\epsilon_n,\epsilon_m$ are orthogonal, as seen from the relation
\begin{equation}\label{eq:orthogonal}
(\epsilon_n-\epsilon^*_m)(z_m|z_n)=0,
\end{equation}
from which it also follows that modes with complex frequency have zero norm.

\subsection{Analog configurations}\label{subsec:BHBEC}

Gravitational analogues in BEC appear when considering stationary condensate flows. We note that, although for illustrative purposes we restrict here to 1D configurations, the following discussion can be straightforwardly adapted for general 3D stationary flows. First, we analyze 1D homogeneous stationary flows, characterized by GP plane waves of the form $\psi_0(x)=\sqrt{n}e^{iqx+\phi_0}$, with $n$ the density of the condensate, $q$ its momentum and $\phi_0$ some phase. After removing the phase of the condensate from the field operator, $\hat{\varphi}(x,t)\rightarrow e^{iqx+\phi_0}\hat{\varphi}(x,t) $, it is straightforward to show that the eigenmodes of the BdG equations (\ref{eq:BdGeigenvalueequation}) are plane waves with wave vector $k$ and frequency $\omega$, giving rise to the following dispersion relation:
\begin{equation}\label{eq:dispersionrelation}
\left[\omega-vk\right]^{2}=\Omega^2(k)=c^{2}k^{2}+\frac{\hbar^2k^{4}}{4m^2}=c^{2}k^{2}\left[1+\frac{(k\xi)^2)}{4}\right]
\end{equation}
with $c=\sqrt{gn/m}$ the sound velocity, $v=\hbar q/m$ the constant flow velocity, $\xi\equiv \hbar/mc$ the so-called healing length and $\Omega$ the comoving frequency. The above dispersion relation gives four different wavevectors for a given value of the frequency. In fact, Eq. (\ref{eq:dispersionrelation}) is just the usual Bogoliubov dispersion relation for phonons in a condensate at rest, $\Omega(k)$, shifted by the Doppler effect due to the fluid velocity $v$. For convention, we take the flow velocity and comoving frequency as positive ($v,\Omega>0$) throughout this work. In this way, the flow is {\it supersonic} when $v>c$ and {\it subsonic} when $v<c$.

\begin{figure*}[htb]
\begin{tabular}{@{}lr@{}}
    \includegraphics[width=0.45\columnwidth]{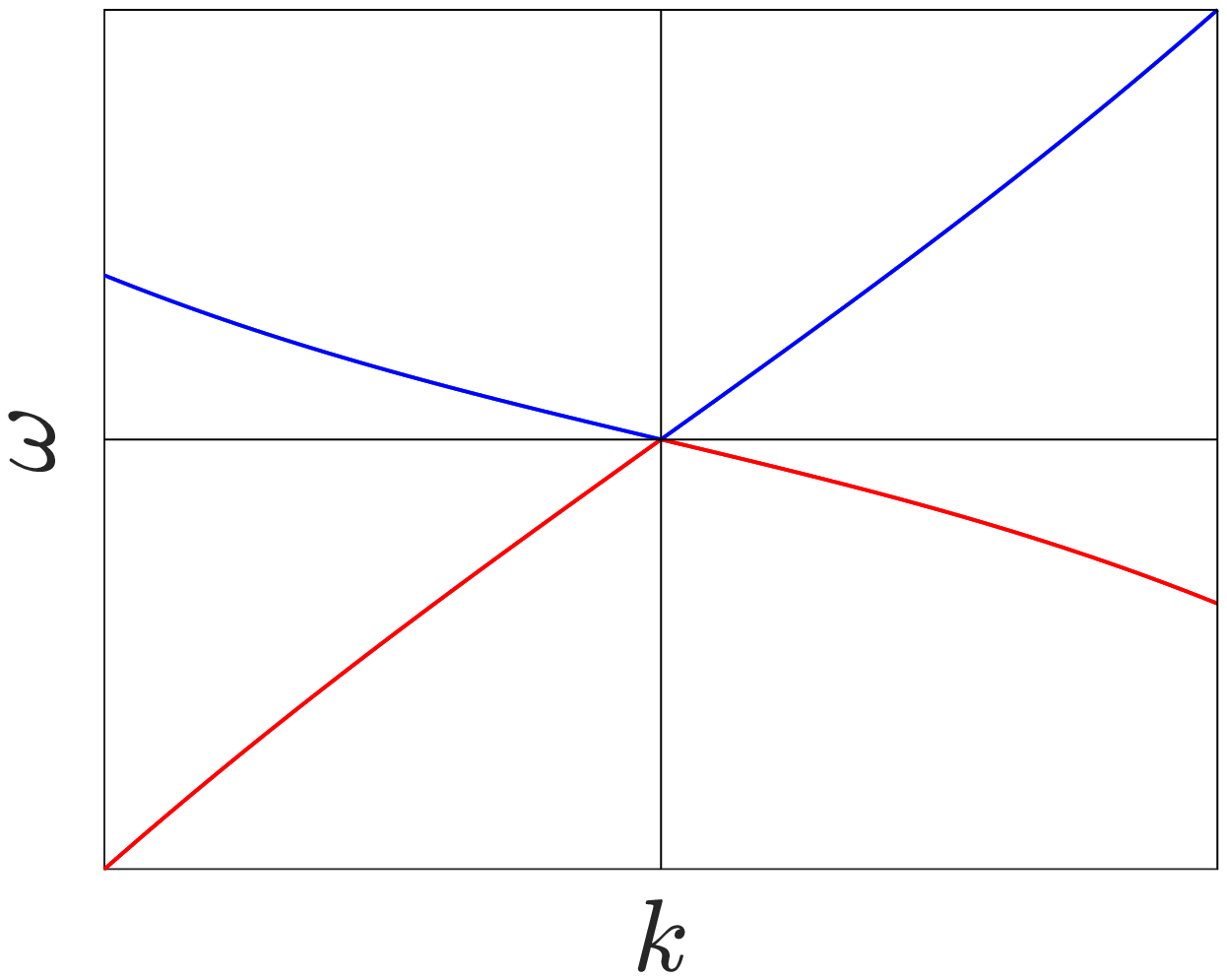} &
    \includegraphics[width=0.45\columnwidth]{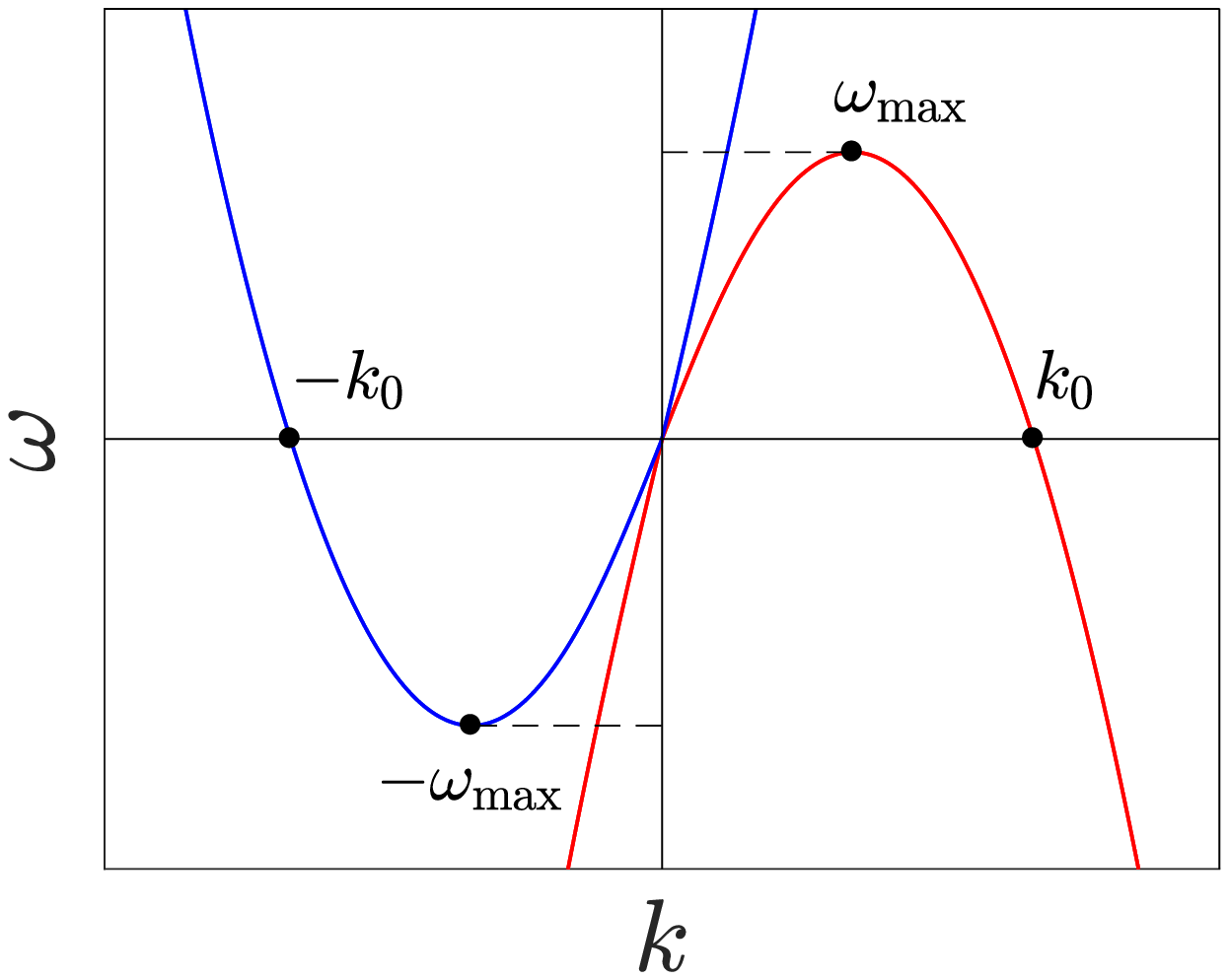} \\
\end{tabular}
\caption{Schematic plot of the BdG dispersion relation for a subsonic (left) and a supersonic (right) flow. }
\label{fig:Dispersion}
\end{figure*}

The dispersion relation for subsonic (supersonic) flows is schematically represented in left (right) panel of  Fig. \ref{fig:Dispersion}, where the blue (red) curves represents the sign $+(-)$ branches of the dispersion relation, $\omega_{\pm}(k)=vk\pm \Omega(k)$, and also positive (negative) normalization according to the scalar product of Eq. (\ref{eq:KGProduct}). Indeed, the $-$ branch is just the dispersion relation of the conjugate modes of the $+$ branch, $\omega_{-}(k)=-\omega_{+}(-k)$ [see Eq. (\ref{eq:BdGeigenvalueequation}) and related discussion]. For subsonic flows, for a given real frequency, there are only two propagating modes (i.e., modes with purely real wavevector) and the other two solutions have complex wave vector. On the other hand, for supersonic flows, in the window $-\omega_{\rm max}<\omega<\omega_{\rm max}$ all the four modes are propagating, where the threshold frequency $\omega_{\rm max}$ is $\omega_{\rm max}=\max_k \omega_{-}(k)$ and is marked by a horizontal dashed line in right Fig. \ref{fig:Dispersion}. Outside this window, we recover essentially the same structure of subsonic flows and only two modes are propagating. The presence of negative energy modes for $-k_0<k<0$ in the $+$ branch of the supersonic dispersion relation, with $\hbar k_0=2m\sqrt{v^2-c^2}$, arises due the energetic instability of supersonic flows, as first argued by Landau. As a result, the introduction of a time-independent perturbation in a supersonic flow gives rise to the emission of Bogoliubov-\v{C}erenkov radiation \cite{Carusotto2006}, characterized by the wave vector $k_0$.

The previous magnitudes can be extended to non-homogeneous configurations by taking $c(x)\equiv\sqrt{gn(x)/m}$ and $v(x)$ as defined in Eq. (\ref{eq:CV}). In an similar way, we say that the flow is subsonic where $v(x)<c(x)$ and supersonic where $v(x)>c(x)$. It is precisely in this context where the gravitational analogy with astrophysical black holes arises. For that purpose, we rewrite Eq. (\ref{eq:BdGfieldequationTI}) in terms of the relative quantum fluctuations, $\hat{\varphi}(x,t)\equiv\psi_0(x)\hat{\chi}(x,t)$,
\begin{eqnarray}\label{eq:BdGfieldequationTIHydrodynamic}
i\hbar D_t\hat{\chi}(x,t)&=&\left[T(x)+mc^2(x)\right]\hat{\chi}(x,t)+mc^2(x)\hat{\chi}^{\dagger}(x,t) \\
\nonumber T(x)&=&-\frac{\hbar^2}{2mn(x)}\partial_xn(x)\partial_x
\end{eqnarray}
with $D_t=\partial_t+v(x)\partial_x$ the {\em comoving} derivative. Gathering this equation with its complex conjugate and defining the hermitic fields
\begin{eqnarray}\label{eq:BdGfieldequationTIHydrodynamic2}
\hat{\chi}_{+}(x,t)&\equiv&\frac{1}{2}\left[\hat{\chi}(x,t)+\hat{\chi}^{\dagger}(x,t)\right]\\
\nonumber \hat{\chi}_{-}(x,t)&\equiv&-\frac{i}{2}\left[\hat{\chi}(x,t)-\hat{\chi}^{\dagger}(x,t)\right]
\end{eqnarray}
gives a pair of equations of the form:
\begin{eqnarray}\label{eq:BdGfieldequationTIHydrodynamicPhase}
\hbar D_t\hat{\chi}_{+}(x,t)&=&T(x)\hat{\chi}_{-}(x,t)\\
\nonumber \hbar D_t\hat{\chi}_{-}(x,t)&=&-\left[T(x)+2mc^2(x)\right]\hat{\chi}_{+}(x,t)
\end{eqnarray}
The above fields are related to physical magnitudes as $\hat{\rho}(x,t)=2n(x)\hat{\chi}_{+}(x,t)$ and $\hat{\phi}(x,t)=\hat{\chi}_{-}(x,t)$, with $\hat{\rho},\hat{\phi}$ the density and phase fluctuations, respectively. Note that the first line of the above equation results from linearizing the continuity equation while the second line results from linearizing the equation for the phase [see Eq. (\ref{eq:PhaseAmplitude})].

Now, if we assume that the background condensate varies on a sufficiently large scale, in the long-wavelength limit we can neglect the contribution of $T(x)$ at the r.h.s. of the second line of Eq. (\ref{eq:BdGfieldequationTIHydrodynamicPhase}), which precisely amounts to work in the hydrodynamic regime where all the contributions arising from the quantum pressure are neglected. In this approximation, we can write the equation for the phase fluctuations as:
\begin{equation}\label{eq:RelativisticBdG}
\left[\frac{1}{n(x)}\partial_x n(x)\partial_x-D_t\frac{1}{c^2(x)}D_t\right]\hat{\phi}(x,t)=0
\end{equation}
which can be rewritten as the relativistic Klein-Gordon equation for a massless scalar field $\hat{\phi}$ on a metric $g_{\mu\nu}$,
\begin{equation}\label{eq:RelativisticBdGKleinGordon}
\square\hat{\phi}\equiv\frac{1}{\sqrt{-g}}\partial_{\mu}(\sqrt{-g}g^{\mu\nu}\partial_{\nu}\hat{\phi})=0
\end{equation}
with the effective stationary metric $g_{\mu\nu}$ given by
\begin{equation}\label{eq:RelativisticMetric}
g_{\mu\nu}(x)=\frac{n(x)}{c(x)}\left[\begin{array}{cccc}-[c^2(x)-v^2(x)]& -v(x) & 0 & 0\\
-v(x)& 1 & 0 & 0\\
0& 0 & 1 & 0\\
0& 0 & 0 & 1\\
\end{array}\right]
\end{equation}
Thus, the points where $c(x)=v(x)$ are the horizons of the acoustic metric $g_{\mu\nu}(x)$, analog to astrophysical event horizons. Using a simple physical picture, for acoustic phonons (long-wavelength excitations) the dispersion relation (\ref{eq:dispersionrelation}) has the form $\omega_{\pm}(k)\simeq (v\pm c)k$, so they are dragged away by a supersonic flow and hence trapped in the supersonic side of an acoustic horizon in the same way as light is trapped inside the event horizon of a black hole. Nevertheless, although the above derivation was done for phonons, the analog of the Hawking effect still holds when considering the complete superluminal dispersion relation in a black hole (BH) configuration \cite{Macher2009a,Recati2009}, where modes with a sufficiently large wave vector in the supersonic region can travel upstream and escape unlike in gravitational black holes, where nothing escapes. A BH configuration is defined as that with two asymptotic homogeneous regions, one subsonic and one supersonic, with flow traveling from subsonic to supersonic, while if the flow goes from supersonic to subsonic, we have a white-hole (WH) configuration, the time reversal of a BH (which just amounts to take the complex conjugate of the GP wave function). Invoking the continuity of the wave function, a BH configuration implies that, at least, one acoustic horizon is formed, that is, a point where $v(x)=c(x)$.

In the same fashion, a configuration with two asymptotic homogeneous subsonic regions and displaying a pair of acoustic horizons (corresponding to a black and a white hole) is the analog of a black-hole laser (BHL). Specifically, the BHL effect in this setup is characterized by the appearance of dynamical instabilities in the BdG spectrum. As supersonic flows are energetically unstable, one can expect this instability to occur for sufficiently large supersonic regions between the two horizons. A more physical insight of the process can be given using a semiclassical picture \cite{Finazzi2010}: negative energy radiation emitted at the BH impacts at the WH and, due to the superluminal dispersion relation, some of the reflected modes are able to travel upstream and hit the BH again, stimulating further emission. This process originates a self-amplifying emission that gives rise to a dynamical instability in the flow.

\subsection{Solutions of the homogeneous Gross-Pitaevskii equation}\label{subsec:GP}

We finally end this section by reviewing the different stationary solutions of the homogeneous GP equation as they are the building blocks of most of the theoretical analog models due to their analytical tractability, serving also as a basis for the calculations presented in this work. In homogeneous problems, the external potential $V(x)$ is constant and it can be reabsorbed into the definition of the chemical potential. In that case, the resulting equation for the amplitude, Eq. (\ref{eq:GPNewton}), is analog to the Newtonian equation of motion of a classical particle in a potential with the role of position and time played here by the amplitude of the wave function and the spatial coordinate $x$, respectively. Then, it can be integrated to obtain
\begin{eqnarray} \label{eq:GPpotential}
\frac{1}{2}A'^{2}+W(A) &=& E_A \\
\nonumber W(A)&=&\frac{m}{\hbar^{2}}\left(\frac{mJ^2}{2A^2}+\mu A^{2}-\frac{g}{2}A^{4}\right)
\end{eqnarray}
The quantity $E_A$ is the ``energy'' of the classical particle and we refer to it as the amplitude energy while $W(A)$ is the corresponding amplitude potential.

As a first step, we study the equilibrium points of $W(A)$ as they give the homogeneous plane wave solutions, $A(x)=A$, which can be obtained from the zeros of Eq. (\ref{eq:GPNewton}):
\begin{equation}\label{eq:1Dhomogeneous}
gn^3-\mu n^2+\frac{mJ^2}{2}=0,~n=A^2
\end{equation}
This polynomial equation for the density only has (two) real positive roots whenever:
\begin{equation}\label{eq:currentcondition}
J\leq \sqrt{\frac{8\mu^3}{27mg^2}}
\end{equation}
In the rest of the work, we will assume that condition (\ref{eq:currentcondition}) is fulfilled and denote the largest root as $n=n_0$; the associated flow velocity is constant and equal to $v_0=J/n_0$. In order to simplify the calculations, we rescale the wave function as $\psi_0(x)\rightarrow \sqrt{n_0}\psi_0(x)$ so it become dimensionless and take units such that $\hbar=m=c_0=1$, with $c_0=\sqrt{gn_0/m}$ the sound velocity associated to the density $n_0$. Length, time and energy are measured in units of $\xi_0=\hbar/mc_0$, $t_0=\xi_0/c_0$ and $E_0=mc^2_0$, respectively. Also, we will refer to $v_0$ as simply $v$ for simplicity.

In this system of units, the amplitude of the homogeneous solution with density $n_0$ is just $A=A_0=1$ and the associated current simply reads $J=v$, while the chemical potential is $\mu=1+v^2/2$. Indeed, $v$ also represents now the value of the Mach number of the flow [that is, the dimensionless ratio between the flow velocity and the speed of sound, $v(x)/c(x)$].

With the help of these considerations, we rewrite Eq. (\ref{eq:1Dhomogeneous}) as
\begin{eqnarray}\label{eq:1Dhomogeneousrescaled}
0&=&n^3-\mu n^2+\frac{J^2}{2}=(n-1)\left(n^2-\frac{v^2}{2}n-\frac{v^2}{2}\right)
\end{eqnarray}
from which we immediately obtain the density of the other homogeneous solution:
\begin{equation}\label{eq:1Dhomogeneousroots}
n_p=A^2_p=\frac{v^2+\sqrt{v^4+8v^2}}{4}
\end{equation}
By construction, $n_p<n_0=1$, which implies that $v<1$ and hence the homogeneous solution $n=n_0=1$ is necessarily subsonic (note that the limit value $v=1$ corresponds to the degenerate case $A_p=A_0=1$). The flow velocity of the solution $n=n_p$, $v_p$, is obtained from the conserved current $n_pv_p=J=v$. As in these units the sound speed is just the square root of the density, $c(x)=\sqrt{n(x)}=A(x)$, the Mach number of the solution $A=A_p$ is $v_p/c_p=v/n_p^{3/2}$. By defining $z\equiv 8/v^2$ and observing that the function $f(z)=z^{2/3}-1-\sqrt{1+z}$ increases monotonically for $z>0$, we conclude that the solution $A=A_p$ corresponds to a supersonic flow since $f(z)$ only has one zero at $z=8$ ($v=1$).

\begin{figure}[!htb]
\begin{tabular}{@{}lr@{}}
    \includegraphics[width=0.5\columnwidth]{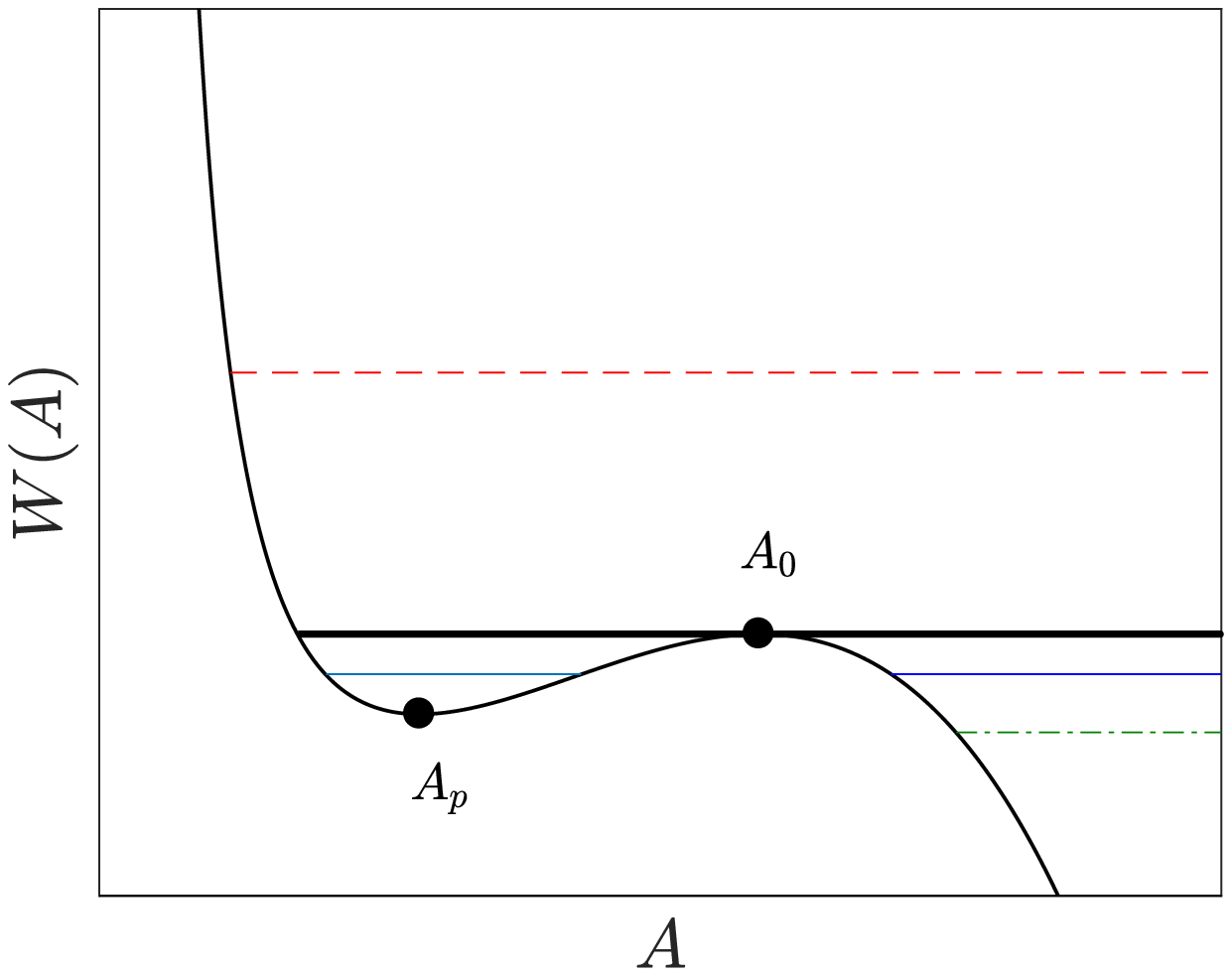} & \includegraphics[width=0.5\columnwidth]{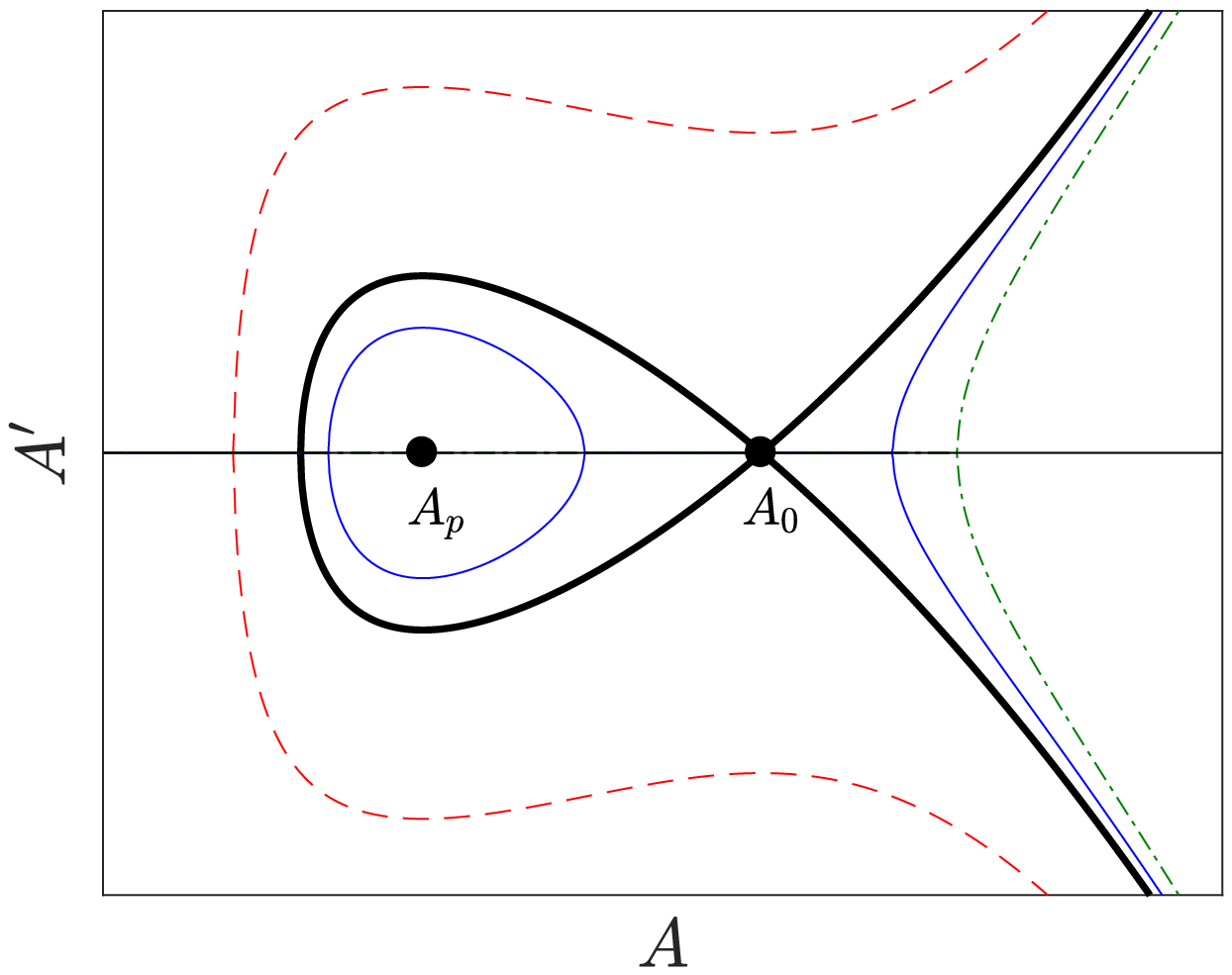}
\end{tabular}
\caption{Left plot: representation of the amplitude potential $W(A)$ (black solid line). We depict some values of the amplitude energy $E_A$ corresponding to qualitatively different solutions; $E_A=W(A_0)$ is plotted in black thick line, $W(A_p) < E_A < W(A_0)$ is plotted in solid blue line, $E_A>W(A_0)$ is plotted in red dashed line and $E_A<W(A_p)$ is plotted in green dashed-dotted line. The homogeneous solutions $A=A_0$ and $A=A_p$ are marked with a black dot. Right panel: orbits in the space $(A,A')$ associated to the values of the amplitude energy shown in the left panel, which can be obtained from the relation $A'^2=2[E_A-W(A)]$. The horizontal black line marks $A'=0$.}
\label{fig:AmplitudePotential}
\end{figure}

We also study the non-homogeneous solutions of Eq. (\ref{eq:GPpotential}). For that purpose, we represent the amplitude potential $W(A)$ in Fig. \ref{fig:AmplitudePotential}. The local minimum corresponds to the homogeneous supersonic solution $A_p=\sqrt{n_p}$ (which means that it is an stable fixed point) and the local maximum to the homogeneous subsonic solution $A_0=\sqrt{n_0}=1$ (which means that it is an unstable fixed point).

Rewriting Eq. (\ref{eq:GPpotential}) in terms of the density gives the simplified equation:
\begin{equation}\label{eq:ellipticdensityenergy}
n'^2=4(n-n_1)(n-n_2)(n-n_3),
\end{equation}
with the densities $n_i,~i=1,2,3$ computed from the zeros of the equation $W(A)=E_A$, equivalent to obtaining the roots of the following polynomial equation in terms of the density
\begin{equation}\label{eq:roots}
n^3-2\mu n^2+2E_An-v^2=(n-n_1)(n-n_2)(n-n_3)=0
\end{equation}

Several cases can be distinguished depending on the value of the amplitude energy $E_A$. First, for $W(A_p) < E_A < W(1)$, the three roots of Eq. (\ref{eq:roots}) are real and we order them such that $0<n_1 < n_2 < n_3$. The case $n_1<n(x)<n_2$ corresponds to the oscillating solution represented by the the closed blue line of right Fig. \ref{fig:AmplitudePotential}. By integrating Eq. (\ref{eq:ellipticdensityenergy}), we find that:
\begin{equation}\label{eq:ellipticintegration}
\int~\frac{\mathrm{d}n}{2\sqrt{(n-n_1)(n_2-n)(n_3-n)}}=\pm(x-x_0)
\end{equation}
with $x_0$ some integration constant arising from the translational invariance of the problem. The solution of the previous indefinite integral is given in term of elliptic functions. The resulting phase of the wave function is computed from Eq. (\ref{eq:stationaryphase}), obtaining:
\begin{equation}
\psi_0(x)=\Lambda(x,n_1,n_2,n_3,-\sqrt{n_3-n_1}x_0)
\end{equation}
where:
\begin{eqnarray}\label{eq:cnoidalwave}
\Lambda(x,n_1,n_2,n_3,\alpha)&\equiv&\sqrt{n(x,n_1,n_2,n_3,\alpha)}e^{i\phi(x,n_1,n_2,n_3,\alpha)}\\
\nonumber n(x,n_1,n_2,n_3,\alpha)&\equiv&n_1+(n_2-n_1)\text{sn}^2(\sqrt{n_3-n_1}x+\alpha,\nu),~\nu\equiv\frac{n_2-n_1}{n_3-n_1}\\
\nonumber \phi(x,n_1,n_2,n_3,\alpha)&\equiv&\phi(0)+\frac{v}{n_1\sqrt{n_3-n_1}}\Theta\left(\sqrt{n_3-n_1}x+\alpha,\alpha,n_1,n_2,n_3,\nu\right)\\
\nonumber \Theta\left(u_2,u_1,n_1,n_2,n_3,\nu\right)&\equiv&\Pi\left[\text{am}(u_2,\nu),1-\frac{n_2}{n_1},\nu\right]-\Pi\left(\text{am}(u_1,\nu),1-\frac{n_2}{n_1},\nu\right)
\end{eqnarray}
with $\phi(0)$ some global phase. We refer the reader to Appendix \ref{app:elliptic} for the precise definition of the different elliptic functions used along this work.

The case $n(x)>n_3$ for the same value of $E_A$ corresponds to a solution that grows indefinitely, see blue curve for $A>1$ in right plot of Fig. \ref{fig:AmplitudePotential}). Moreover, for high values of $n$, $n'\propto n^{3/2}$ and then the solution blows up at some finite value $x_{bu}$ as $n(x)\sim (x-x_{bu})^{-2}$. The same reasoning holds for $E_A<W(A_p)$ (green dashed-dotted line of Fig. \ref{fig:AmplitudePotential}) or $E_A>W(1)$ (red dashed line of Fig. \ref{fig:AmplitudePotential}). These exploding solutions are not relevant for the present work so we ignore them in the following.

Finally, we consider the degenerate cases $E_A=W(A_p)$ and $E_A=W(A_0)$. For $E_A=W(A_p)$, $n_1=n_2=n_p$. One possible solution corresponds to the stable fixed point of the homogeneous supersonic solution $n(x)=n_p$, described by the plane wave:
\begin{equation}\label{eq:supersonicplanewave}
\psi_0(x)=\sqrt{n_p}e^{iv_px}
\end{equation}
The other possible solution corresponds to $n(x)>n_3$, which blows up in a similar way to the exploding solutions described above.

The other degenerate case is $E_A=W(1)$, where the roots satisfy $n_2=n_3=1$ and $n_1=v^2$. For $n(x)=1$, the solution is the subsonic plane wave
\begin{equation}\label{eq:subsonicplanewave}
\psi_0(x)=e^{ivx}
\end{equation}
For $n(x)\neq 1$, we get from Eq. (\ref{eq:ellipticintegration})
\begin{equation}\label{eq:solitonintegration}
\int~\frac{\mathrm{d}n}{2\sqrt{(n-1)^2(n-v^2)}}=\pm(x-x_0)
\end{equation}
For $n(x)<1$, we obtain:
\begin{equation}\label{eq:solitondensity}
n(x)=v^2+(1-v^2)\tanh^2\left[\sqrt{1-v^2}(x-x_0)\right]
\end{equation}
which is of the same form of Eq. (\ref{eq:cnoidalwave}) after taking into account that $\text{sn}(u,1)=\tanh(u)$. The phase of the wave function can be obtained analytically in a simple form and we can write the total wave function as:
\begin{equation}\label{eq:solitonwavefunction}
\psi_0(x)=e^{ivx}e^{-i\phi_0}\left(v+i\sqrt{1-v^2}\tanh\left[\sqrt{1-v^2}(x-x_0)\right]\right)
\end{equation}
being $\phi_0$ some constant phase. This solution represents a soliton with zero velocity \cite{Pethick2008}.

On the other hand, taking $n>1$ in Eq. (\ref{eq:solitonintegration}) gives the so-called shadow soliton solution \cite{Michel2013}:
\begin{equation}\label{eq:shadowsolitonwavefunction}
\psi_0(x)=e^{ivx}e^{-i\phi_0}\left(v+i\sqrt{1-v^2}\text{cotanh}\left[\sqrt{1-v^2}(x-x_0)\right]\right)
\end{equation}
Although this solution also blows up at a finite value of $x$, it is quite relevant for the computation of stationary states in BHL configurations; see Secs. \ref{sec:BHLWF}, \ref{sec:BHL2delta}.

\section{General relation between black holes and black-hole lasers}\label{sec:BHLTheorem}

After introducing the basic concepts and techniques of gravitational analogues in BEC, we proceed to prove one of the central results of this work, which states that every {\em compact} BH solution can be used to produce a BHL configuration with an arbitrary large homogeneous supersonic region, explicitly showing the mechanism to construct such BHL configuration. We define a {\em compact} BH solution as that in which a homogeneous supersonic flow is reached at a finite point, $x=x_H$. Indeed, this is the situation of all the BH configurations usually appearing in the literature \cite{Larre2012}.

The proof is straightforward. Consider a compact BH configuration, which satisfies a time-independent GP equation of the form:
\begin{equation}\label{eq:TIGPCompact}
\mu\psi_0(x)=\left[-\frac{\hbar^2}{2m}\partial_x^2+V_C(x)+g|\psi_0(x)|^2\right]\psi_0(x)
\end{equation}
For simplicity, we consider that the BH is produced only with the help of an external potential but the generalization to situations in which the coupling constant (like the flat-profile configuration) or the mass are space-dependent is trivial. By definition of compact BH configuration, $V_C(x>x_H)=V_{sp}$ is homogeneous and the GP wave function is of the form:
\begin{equation}\label{eq:CompactBH}
\psi_{0}(x)= \left\{ \begin{array}{cc}
\psi_C(x) & x< x_H\\
A_{sp}e^{iq_{sp}x}, & x\geq x_H
\end{array}\right.
\end{equation}
with $A_{sp},q_{sp}$ the supersonic amplitude and momentum and $\Psi_C(x)$ the part of the wave function that describes the subsonic-supersonic transition. Without loss of generality, we choose the origin of coordinates such that $x_H=-X/2$, with $X>0$.

The idea for obtaining a BHL configuration is to replicate the same structure of the potential and the GP wave function for $x>0$. This can be done by taking the spatial and time reverse of the wave function and the potential. Explicitly, we consider the GP wave function
\begin{equation}\label{eq:CompactBHL}
\psi_{0}(x)= \left\{ \begin{array}{cc}
\psi_C(x) & x< -\frac{X}{2}\\
A_{sp}e^{iq_{sp}x}, & -\frac{X}{2}\leq x \leq \frac{X}{2}\\
\psi_C^*(-x) & x>\frac{X}{2}
\end{array}\right.
\end{equation}
which satisfies the following GP equation
\begin{equation}\label{eq:TIGPCompactBHL}
\mu\psi_0(x)=\left[-\frac{\hbar^2}{2m}\partial_x^2+V_{BHL}(x)+g|\psi_0(x)|^2\right]\psi_0(x)
\end{equation}
where the potential $V_{BHL}(x)$ is given by:
\begin{equation}\label{eq:CompactBHLPotential}
V_{BHL}(x)= \left\{ \begin{array}{cc}
V_C(x) & x<0\\
V_C(-x) & x>0
\end{array}\right.
\end{equation}
The wave function of Eq. (\ref{eq:CompactBHL}) describes a BHL configuration with a homogeneous supersonic flow in a region of size $X$. Indeed, since $X$ is not fixed, we can construct a supersonic region of arbitrary length with this solution. Therefore, we conclude that every BH solution can be extended to produce a BHL configuration.

\begin{figure}[!htb]
\begin{tabular}{@{}crc@{}}
    \includegraphics[width=0.3\columnwidth]{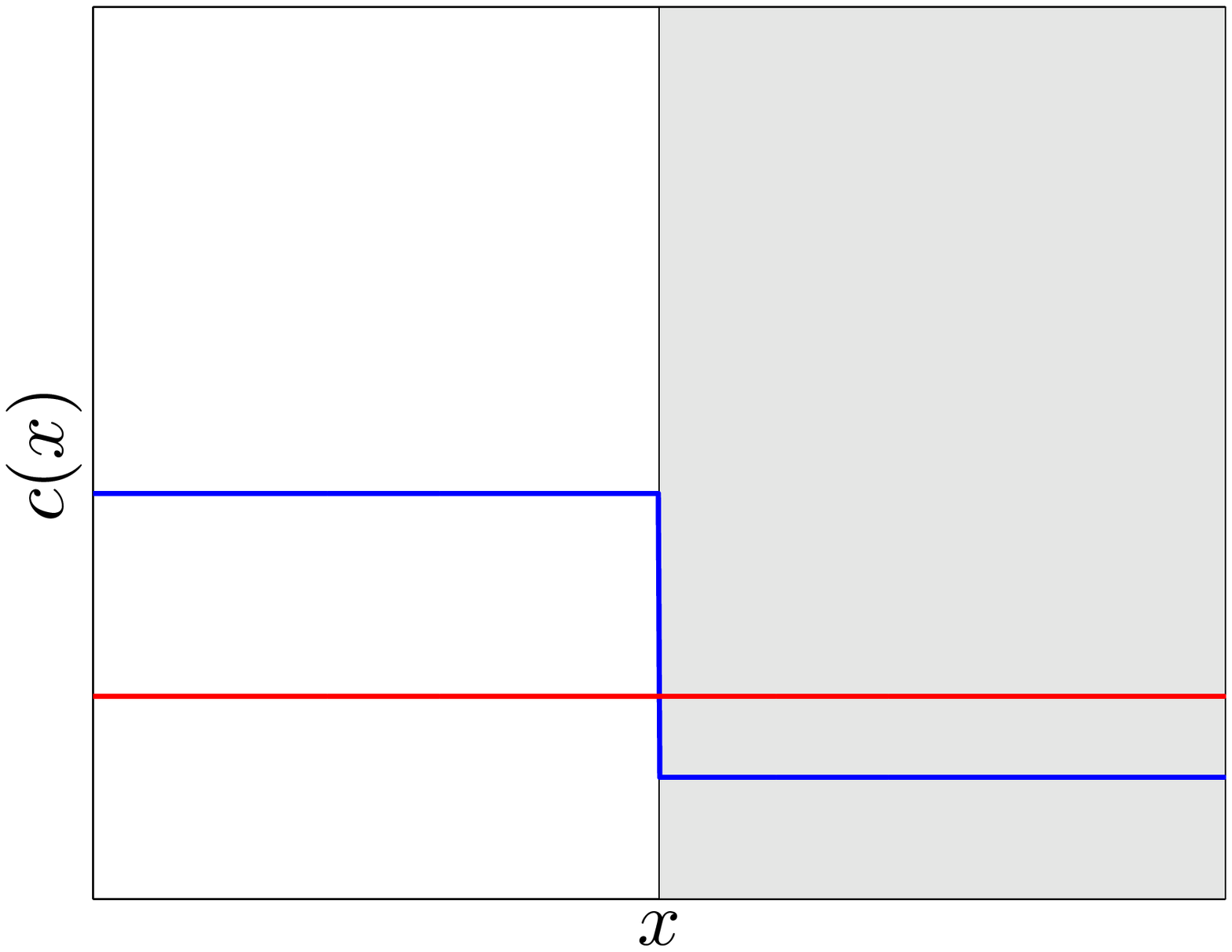} & \includegraphics[width=0.3\columnwidth]{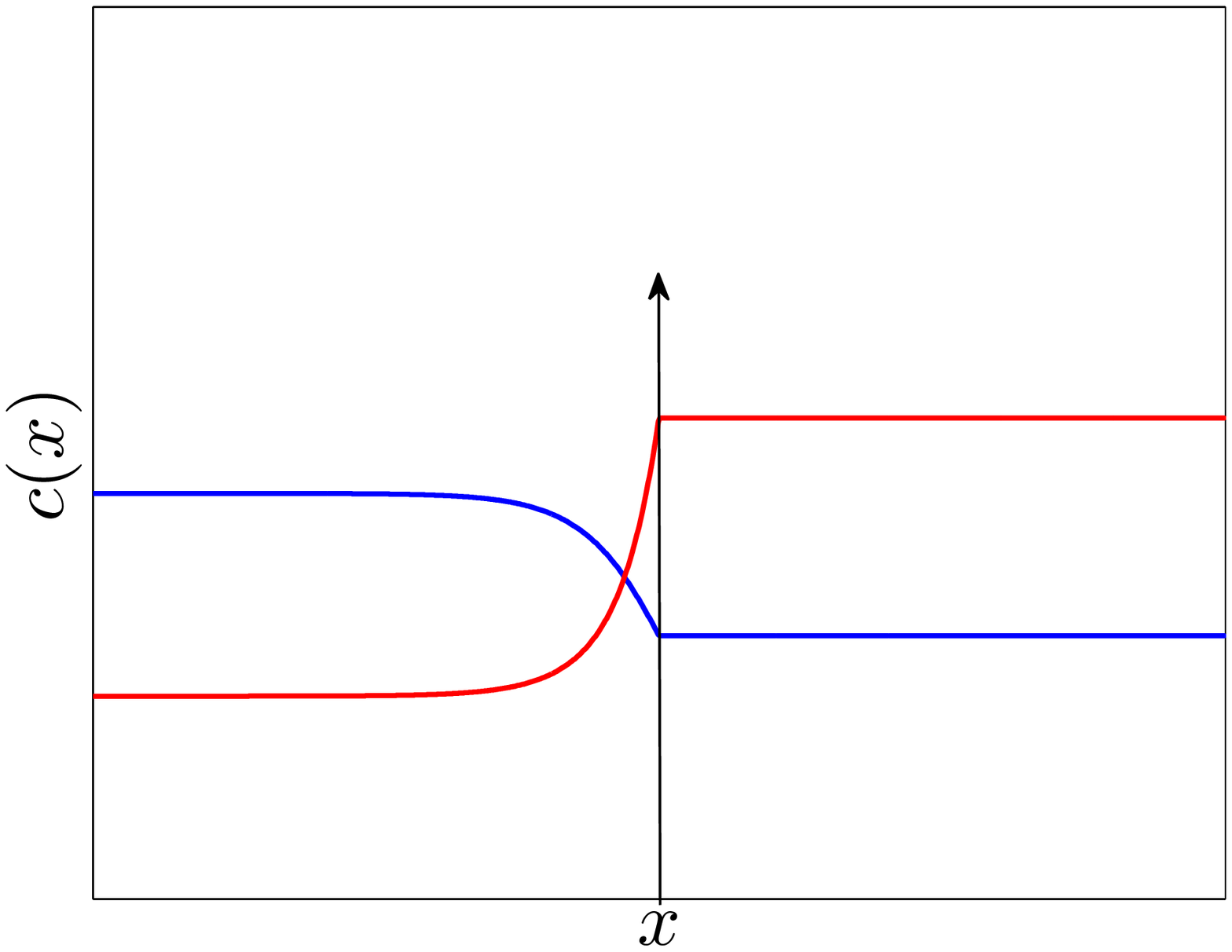} & \includegraphics[width=0.3\columnwidth]{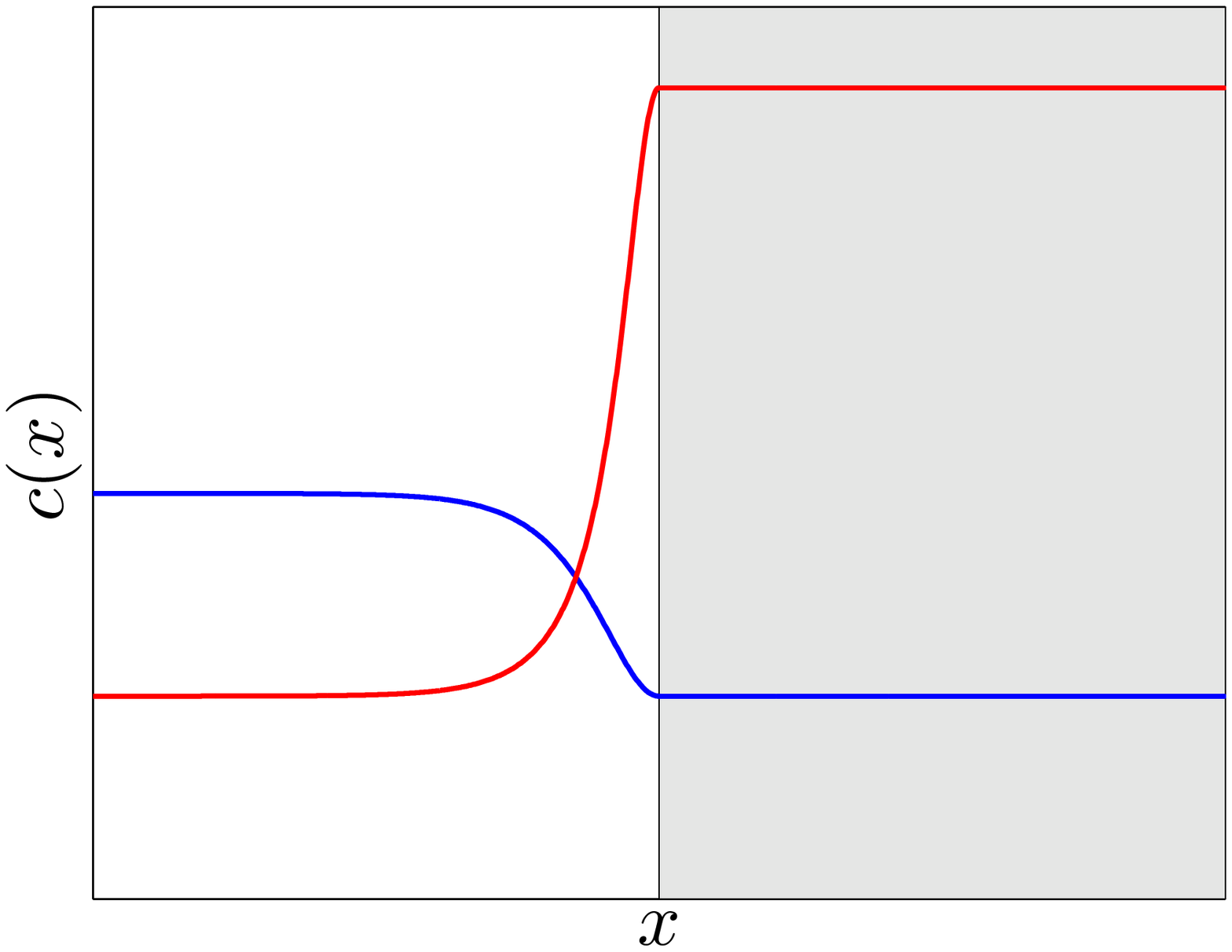}\\
    \includegraphics[width=0.3\columnwidth]{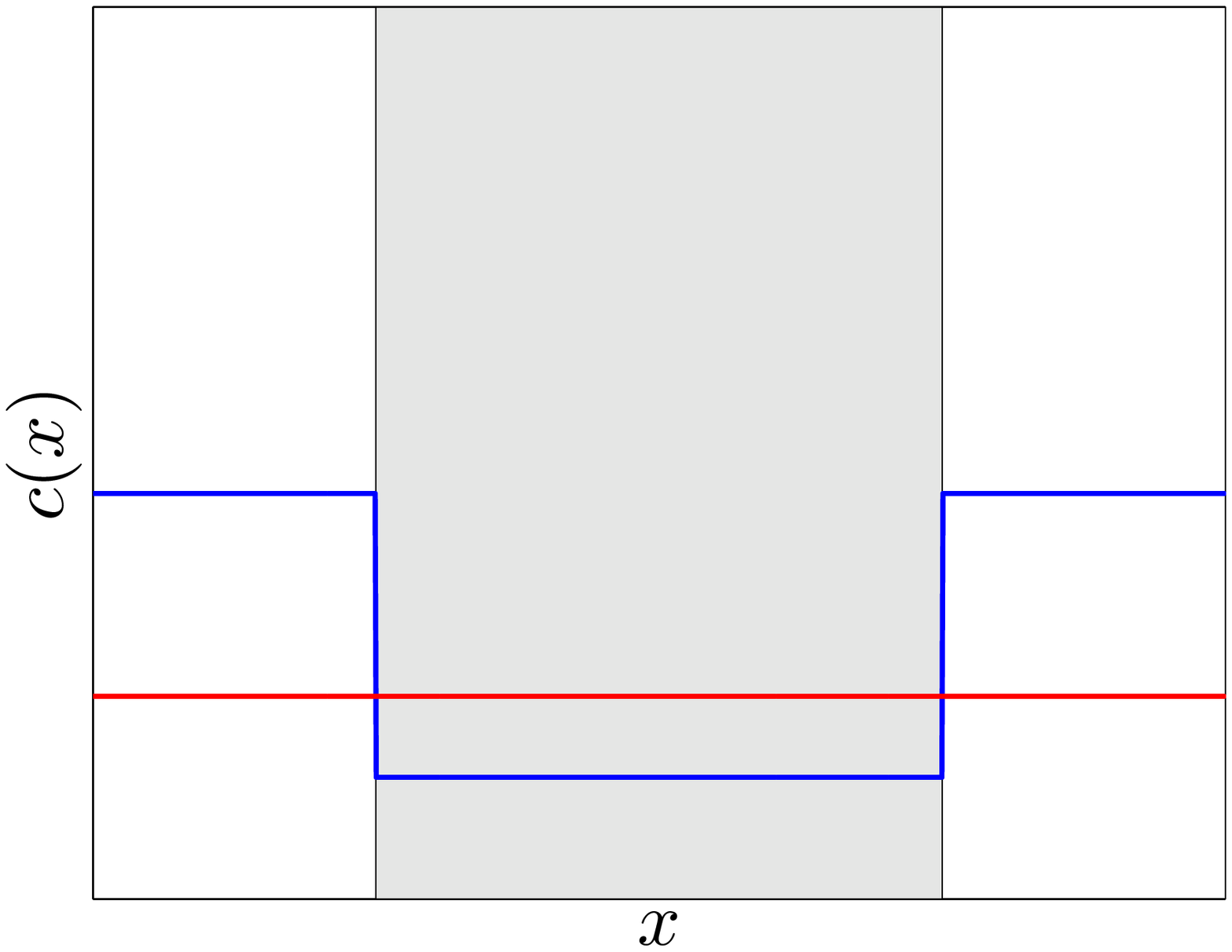} & \includegraphics[width=0.308\columnwidth]{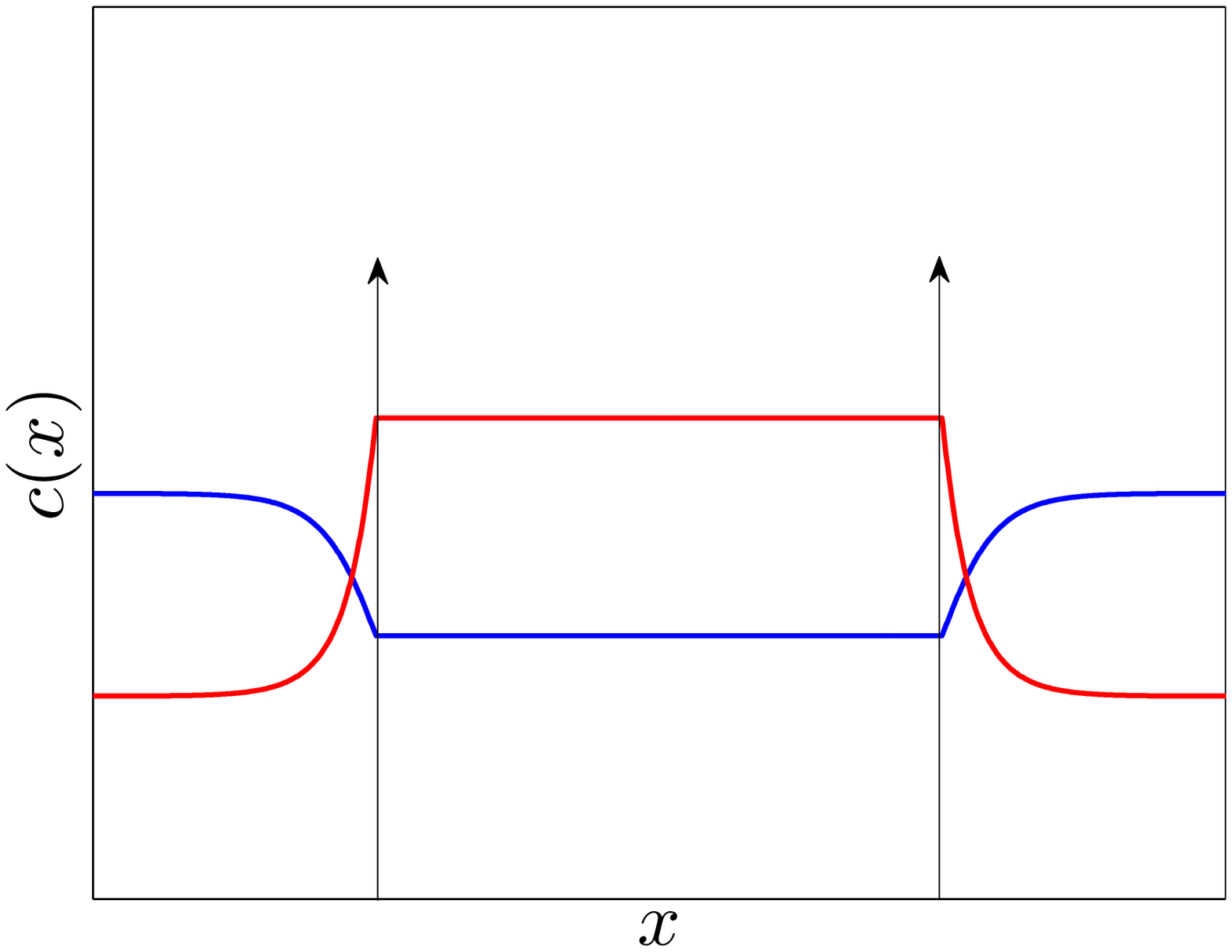} & \includegraphics[width=0.3\columnwidth]{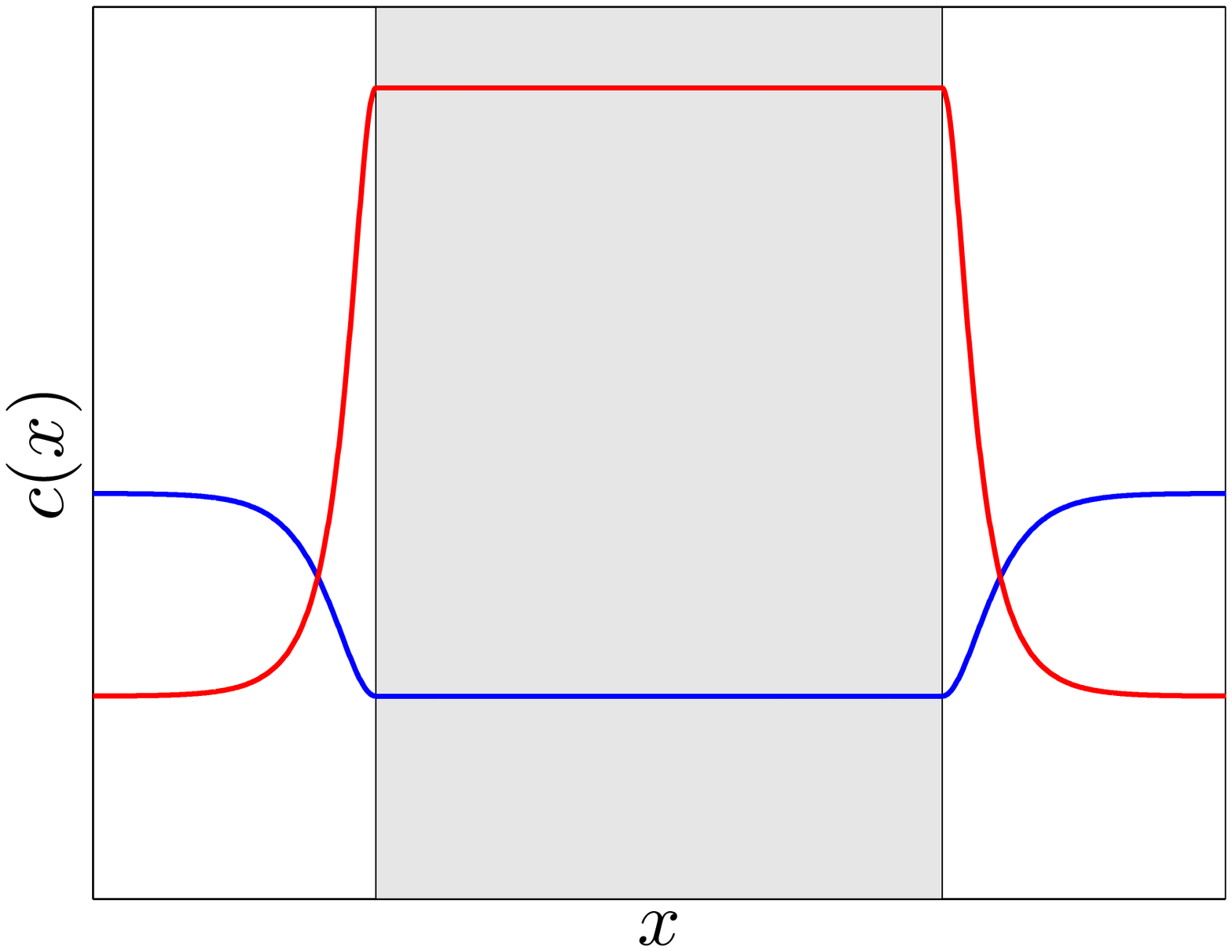}
\end{tabular}
\caption{Schematic plot of the spatial profile of sound (blue) and flow (red) velocities for typical BH (upper row) and the corresponding BHL (lower row) configurations. Left column: flat-profile configuration, where the shaded area represents the region with different value of the coupling constant, $g(x)=g_2$. Central column: delta-barrier configuration, where the arrows represent the position of the delta potential. Right column: waterfall configuration, where the shaded area represents the region in which the negative step potential $V(x)=-V_0$ is present.}
\label{fig:BHConfigurations}
\end{figure}

Now, with the help of the previous result, we study the BHL configurations arising from some well-known BH configurations, shown in the lower and upper row of Fig. \ref{fig:BHConfigurations}, respectively. One of the most considered BH configurations is the flat-profile configuration (upper left of Fig. \ref{fig:BHConfigurations}), in which the GP wave function is a global plane wave so the condensate density $n$ and flow velocity $v$ are homogeneous in all of the space. In order to fulfill the homogeneity condition, the 1D coupling strength $g\left(x\right)$ and the external potential $V\left(x\right)$ must satisfy that $g(x)n+V(x)$ is constant. In particular, in order to construct a BH solution, $g(x)$ is chosen to be a step function with a downstream value $g(x)=g_2$ such that the resulting flow is supersonic. Although the experimental implementation of this configuration is extremely challenging due to the required high precision in the control of both the external potential and the local coupling constant, it is considered in many theoretical works \cite{Balbinot2008,Carusotto2008,Recati2009} because of its analytical simplicity.

More realistic configurations are displayed in central and right panels of the upper row of Fig. \ref{fig:BHConfigurations}, corresponding to the delta-barrier and waterfall configurations, respectively; the point is that these configurations only require simple external potentials that are achievable with the use of standard experimental tools as blue-detuned (for repulsive potentials) or red-detuned (for attractive potentials) lasers. For instance, the delta-barrier configuration models the BH arising from the flow of a condensate through a localized obstacle \cite{Kamchatnov2012}, represented by a repulsive delta potential; by Galilean invariance, this configuration is similar to launching the obstacle against the condensate. On the other hand, the waterfall configuration uses an attractive step potential to accelerate the flow and create a supersonic current. In fact, this model provides a realistic description of the actual setups of Refs. \cite{Lahav2010,Steinhauer2016}, in which a negative step potential created with the help of a laser is swept along a trapped condensate, finding a good agreement with the experimental data \cite{Steinhauer2016}.

Regarding the BHL side, the associated flat-profile BHL of lower left Fig. \ref{fig:BHConfigurations} has been already considered in the literature in both analytical and numerical studies \cite{Michel2013,Michel2015,deNova2016}. Specifically, Ref. \cite{Michel2013} provided a detailed study not only of the associated dynamical instabilities of the flow but also of the different non-linear stationary solutions existing for fixed chemical potential and current as they describe the potential quasi-stationary states of the system for long times, once the dynamical instability has already grown up. In fact, it was shown the appearance of each dynamical instability is associated to the appearance of a stationary solution with lower grand-canonical energy $K$ than the initial homogeneous plane-wave solution.

On the other hand, Refs. \cite{Michel2015,deNova2016} extended the previous analytical work with numerical simulations of the time evolution of the initially unstable homogeneous solution in order to study the non-linear saturation of the instability. In particular, Ref. \cite{deNova2016} found that the system only presents two kind of asymptotic behaviors: it either reaches the GP ground state or a regime of continuous emission of solitons (CES) in which the system emits trains of solitons with perfect periodicity, providing in this way the closest analog of an actual optical laser.

Following these results, in order to provide more realistic models for black-hole lasing, we study in the rest of the work the BHL configurations associated to the delta-barrier and the waterfall configurations, depicted in the lower center and right panel of Fig. \ref{fig:BHConfigurations}, respectively. Specifically, due to energy and particle number conservation \cite{Michel2013}, we only aim at the GP stationary solutions asymptotically matching at $\pm \infty $ the corresponding subsonic plane wave solution. Several reasons motivate this choice: first, solutions with lower grand-canonical energy and continuously connected to the initial BHL solution are expected to also characterize the appearance of dynamically unstable modes \cite{Michel2013}. In addition, there should not be substantial differences in the linear regime with respect to the usual flat-profile BHL. Moreover, thinking in a realistic implementation, the computed stationary solutions should still govern the late time dynamics once the system enters into the non-linear regime \cite{Michel2013,deNova2016} regardless the specific mechanism giving rise to the initial growth of the instability at short times \cite{Tettamanti2016,Steinhauer2017,Wang2016}.

Although the explained protocol to create BHL configurations applies for any arbitrary compact BH configuration, for illustrative purposes we focus on these two particular cases as they provide simple analytical models of realistic experimental scenarios and extend well-known models in the literature. Of particular interest is the BHL resulting from the waterfall configuration as it corresponds to an attractive potential well and hence it is expected to capture the essential features of the actual BHL configuration of the experiment of Ref. \cite{Steinhauer2014}, in which the laser cavity is created by sweeping along the condensate the effective potential well arising from the combination of the background trap and the negative step potential.

In order to simplify the notation and match the results of Sec. \ref{subsec:GP}, we set units in the rest of this work such that $\hbar=m=c_0=1$, where $c_0$ is the asymptotic subsonic speed of sound. Once in these units, it is easy to check that for the BHL configurations considered in this work, the problem is completely determined by only two parameters: the asymptotic subsonic flow velocity $v$ (which is also the subsonic Mach number in these units) and the size of the supersonic region $X$, since the amplitudes of the different potentials are functions of $v$. This contrasts to the case of the flat-profile BHL configuration, where there are three degrees of freedom, $v,X,c_2$, with $c_2$ the supersonic sound speed \cite{deNova2016}.

For both configurations, we first present the general structure of the problem and then describe the main features and the conditions of existence for the different families of stationary solutions; technical details of the computations are given in Appendices \ref{app:technicalwell}, \ref{app:technicaldelta}, to which the interested reader is referred.

\section{Attractive square well}\label{sec:BHLWF}

\subsection{General structure}

As a first step, we present the BH solution corresponding to the waterfall configuration, where the external potential is given by:
\begin{equation}\label{eq:waterfall}
V(x)=-V_0\theta(x-x_H),~V_0=\frac{1}{2}\left(v^2+\frac{1}{v^2}\right)-1
\end{equation}
where $\theta(x)$ is the step function and $x_H$ the point where the step is placed. The corresponding GP wave function is given by:
\begin{equation}\label{eq:CompactBHWF}
\psi_0(x)=\psi^{BH}(x)=\left\{ \begin{array}{cc}
e^{ivx}e^{-i\phi_0}\left(v+i\sqrt{1-v^2}\tanh\left[\sqrt{1-v^2}\left(x-x_H\right)\right]\right) & x< x_H\\
ve^{i\frac{x}{v}}, &  x \geq x_H
\end{array}\right.
\end{equation}
with $\phi_0$ some phase to make the wave function continuous at $x=x_H$. Following the procedure described in Sec. \ref{sec:BHLTheorem}, the associated BHL configuration is created by using an attractive square well potential of size $X$,
\begin{equation}\label{eq:square}
V(x)=-V_0\theta\left(x+\frac{X}{2}\right)\theta\left(\frac{X}{2}-x\right).
\end{equation}
We note that the different stationary solutions of a condensate flowing through an attractive square well were first addressed in Ref. \cite{Leboeuf2001}. Here we restrict to the specific case where $V_0$ is given by Eq. (\ref{eq:waterfall}), so a BHL solution as that of lower right Fig. \ref{fig:BHConfigurations} exists, described by the GP wave function:
\begin{equation}\label{eq:CompactBHLWF}
\psi_0(x)=\psi^{BHL}(x)=\left\{ \begin{array}{cc}
e^{ivx}e^{-i\phi_0}\left(v+i\sqrt{1-v^2}\tanh\left[\sqrt{1-v^2}\left(x+\frac{X}{2}\right)\right]\right) & x< -\frac{X}{2}\\
ve^{i\frac{x}{v}}, & -\frac{X}{2}\leq x \leq \frac{X}{2}\\
e^{ivx}e^{i\phi_0}\left(v+i\sqrt{1-v^2}\tanh\left[\sqrt{1-v^2}\left(x-\frac{X}{2}\right)\right]\right) & x>\frac{X}{2}
\end{array}\right.
\end{equation}

In order to find the remaining stationary solutions that asymptotically match the subsonic plane-wave solution $e^{ivx}$ on both sides, we use the phase-amplitude decomposition and consider Eq. (\ref{eq:GPpotential}). The asymptotic boundary conditions fix the current to $J=v$ and the amplitude energy outside the well to $E_A=E_1=\frac{1}{2}+v^2$. Accordingly, two different regions can be distinguished: region 1 corresponds to the exterior of the square well, $|x|>X/2$, while region 2 corresponds to its interior, $|x|<X/2$. The equation for the amplitude reads in each region as:
\begin{equation}\label{eq:MechanicalEnergyConservationWF}
\frac{A'^2}{2}+W_i(A)=E_i,~W_i(A)=\frac{v^2}{2A^2}-\frac{A^4}{2}+\mu_i A^2
\end{equation}
where $W_i(A),E_i$ are the amplitude potential and the conserved amplitude energy for the $i=1,2$ regions, with $\mu_1=\mu=1+v^2/2$ and  $\mu_2=\mu_1+V_0=v^2+\frac{1}{2v^2}$. Invoking the continuity of the wave function and its derivative, we find the matching condition at both edges
\begin{equation}\label{eq:matchingWF}
E_2-E_1=V_0n_W~,n_W\equiv n\left(\pm\frac{X}{2}\right)
\end{equation}
As $E_1$ is fixed, we only need to find the possible values of $E_2$ which make that the GP wave function satisfy both Eq. (\ref{eq:MechanicalEnergyConservationWF}) and Eq. (\ref{eq:matchingWF}).

The situation is schematically depicted in left Fig. \ref{fig:BHLHomogenous}: outside the well, the orbits follow the dashed black line as they must asymptotically match the subsonic plane wave on both sides, while the other curves represent possible solutions inside the well.

Thus, there are only three possible solutions outside the well: shadow solitons, with amplitude $A(x)>1$; regular solitons, with $A(x)<1$; and the homogeneous subsonic plane wave, $A(x)=1$. On the other hand, as the amplitude at $x=\pm\frac{X}{2}$ must be the same, the only possible solution inside the well is a cnoidal wave as that of Eq. (\ref{eq:cnoidalwave}), where $n_i,~i=1,2,3$, are now the roots of the equation
\begin{equation}\label{eq:rootsWF}
n^3-2\mu_2 n^2+2E_2n-v^2=0
\end{equation}
For convention, we choose the wave function such that it is real at $x=0$, $\phi(0)=0$.

The matching of the cnoidal wave (\ref{eq:cnoidalwave}) at $x=\pm\frac{X}{2}$ gives two equations:
\begin{equation}\label{eq:matchingWFelliptic}
n\left(\pm\frac{X}{2},n_1,n_2,n_3,\alpha\right)=n_W
\end{equation}
Since $n_1,n_2,n_3,n_W$ are functions of $E_2$, the above system gives two conditions for two variables, $\alpha$ and $E_2$. Due to the periodicity of the elliptic functions, the possible solutions for a given length $X$ are discretized by an index $m=0,1,2\ldots$ representing the number of complete periods inside the well. As a result, we only need to compute the corresponding values of $E_2,\alpha$, labeled as $E^m_2,\alpha_m$, and the associated parameters $n^m_W,n^m_i,\nu_m$ [see Eqs. (\ref{eq:cnoidalwave}), (\ref{eq:matchingWF})] to determine the wave function; the details of this calculation are given in Appendix \ref{app:technicalwell}.

We now discuss the different families of stationary solutions depending on the three possible cases for the wave function outside the well.

\subsection{Homogeneous plane wave}

\begin{figure}[!htb]
\begin{tabular}{@{}ccccc@{}}
    \includegraphics[width=0.2\columnwidth]{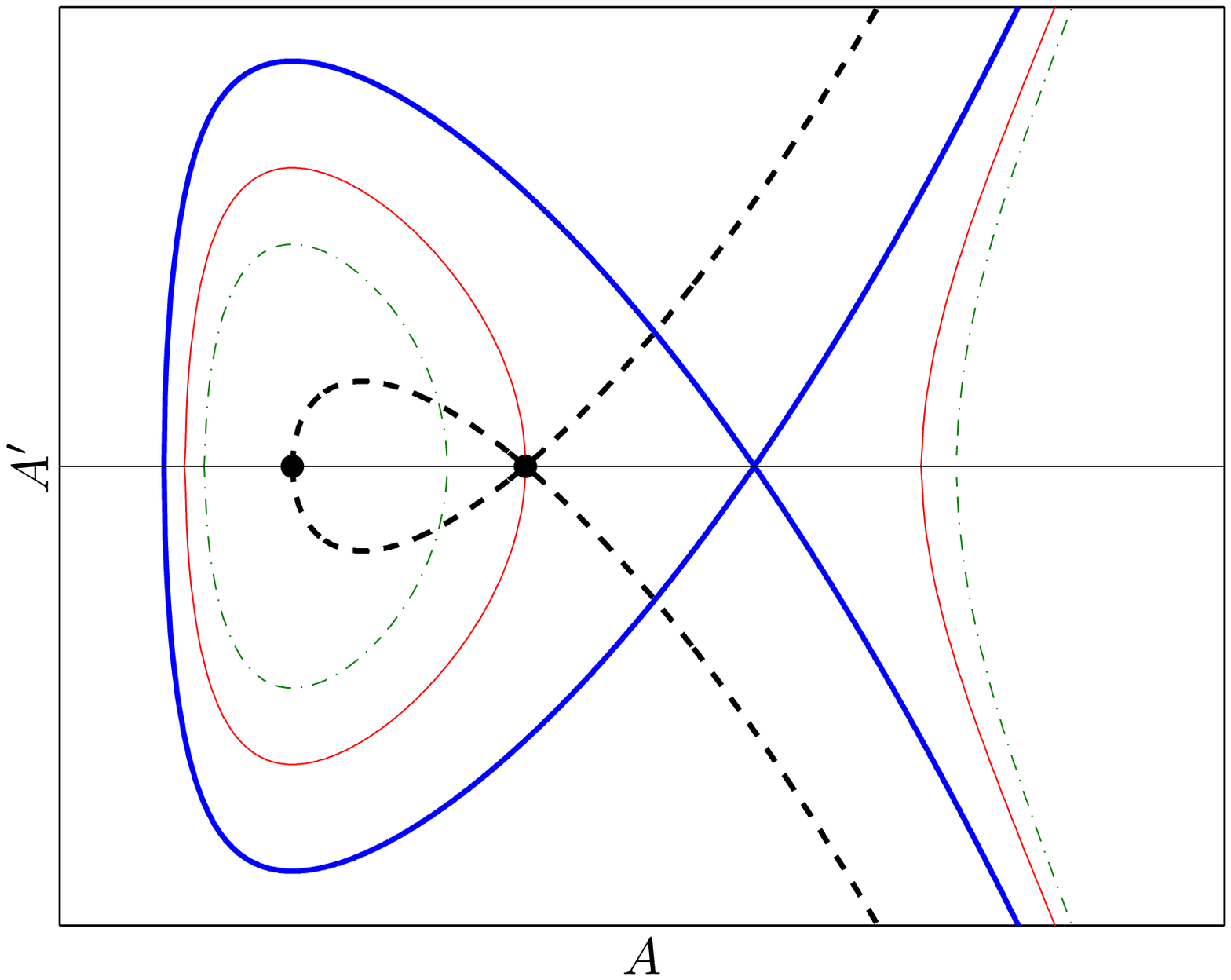} & \includegraphics[width=0.2\columnwidth]{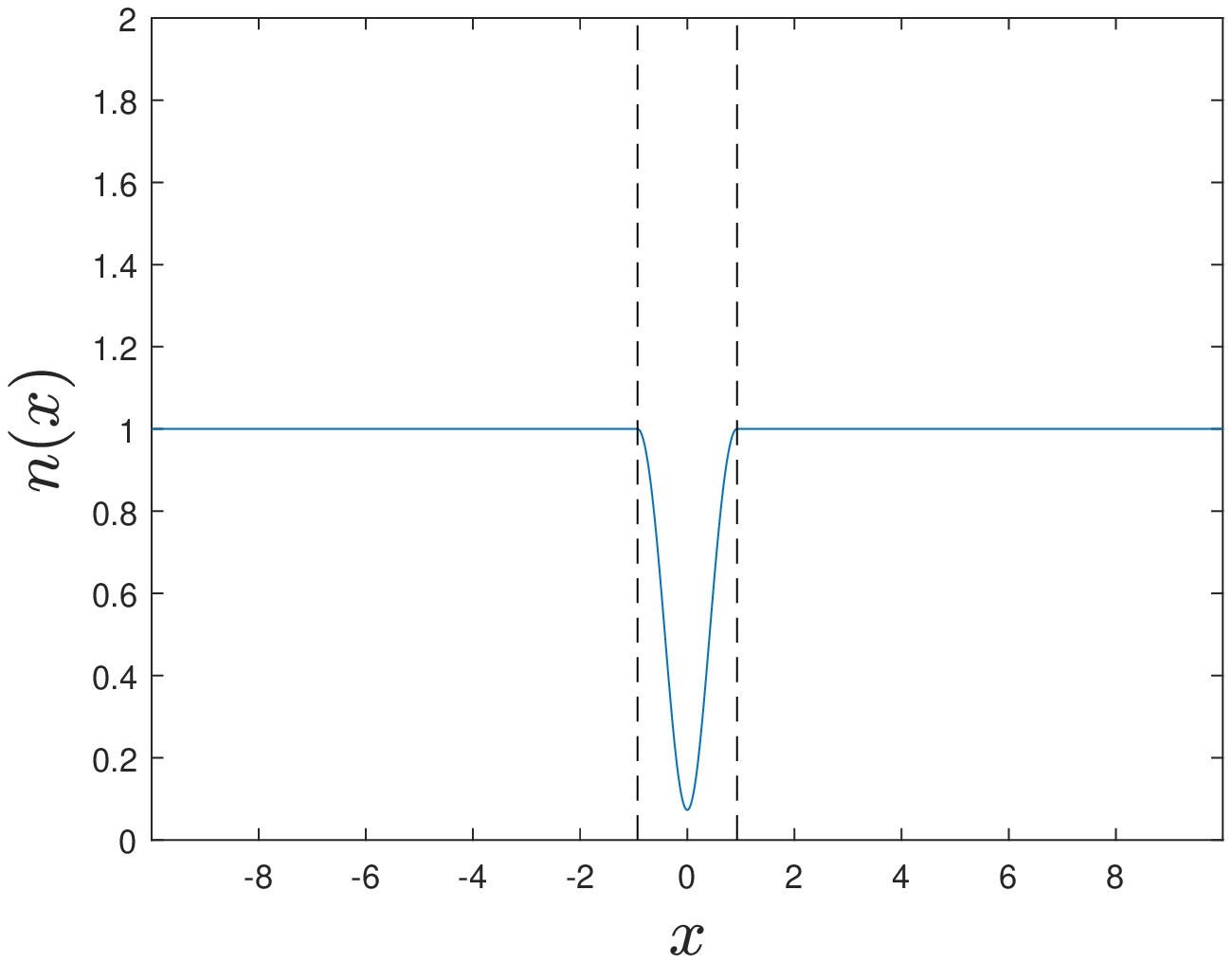} & \includegraphics[width=0.2\columnwidth]{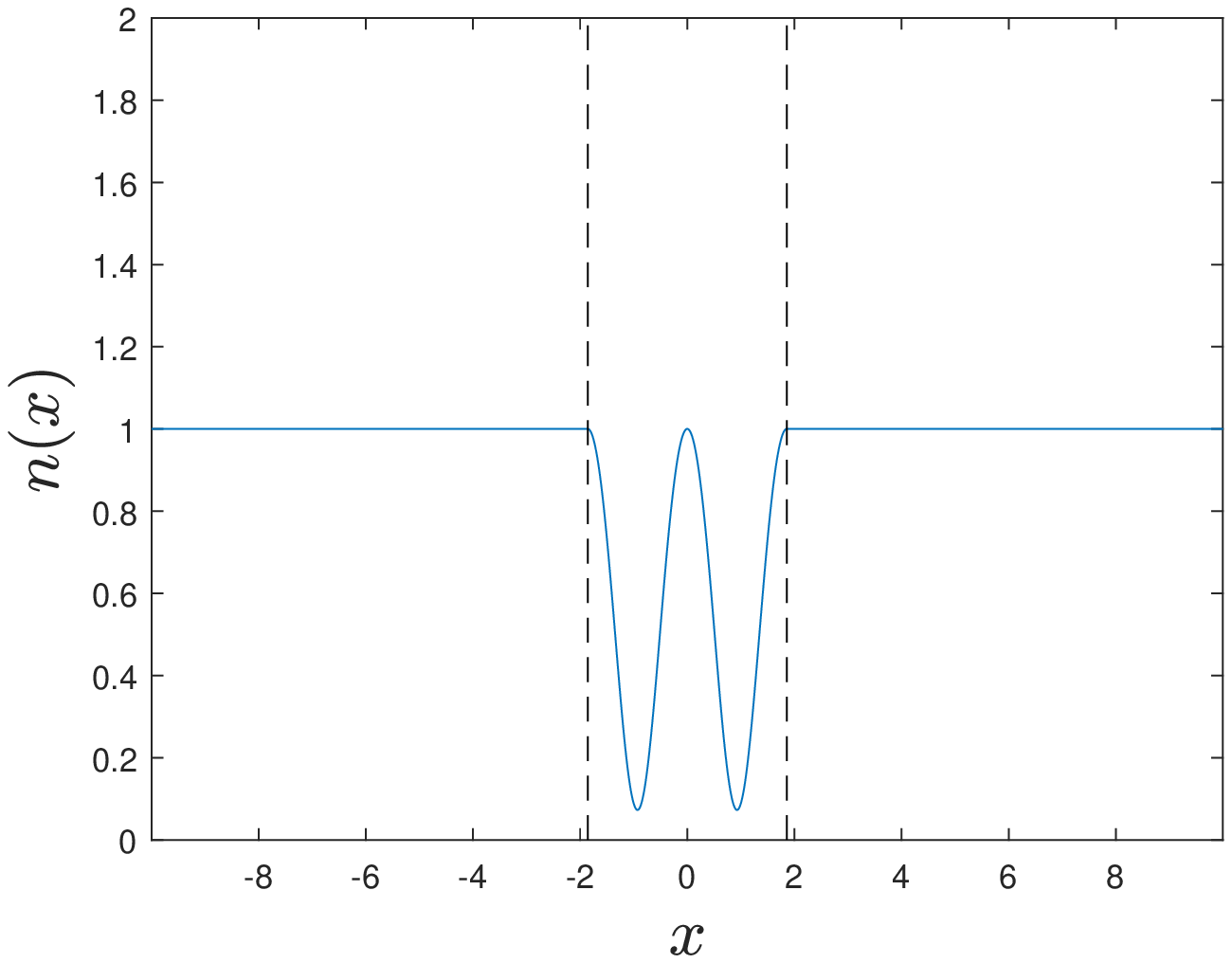} & \includegraphics[width=0.2\columnwidth]{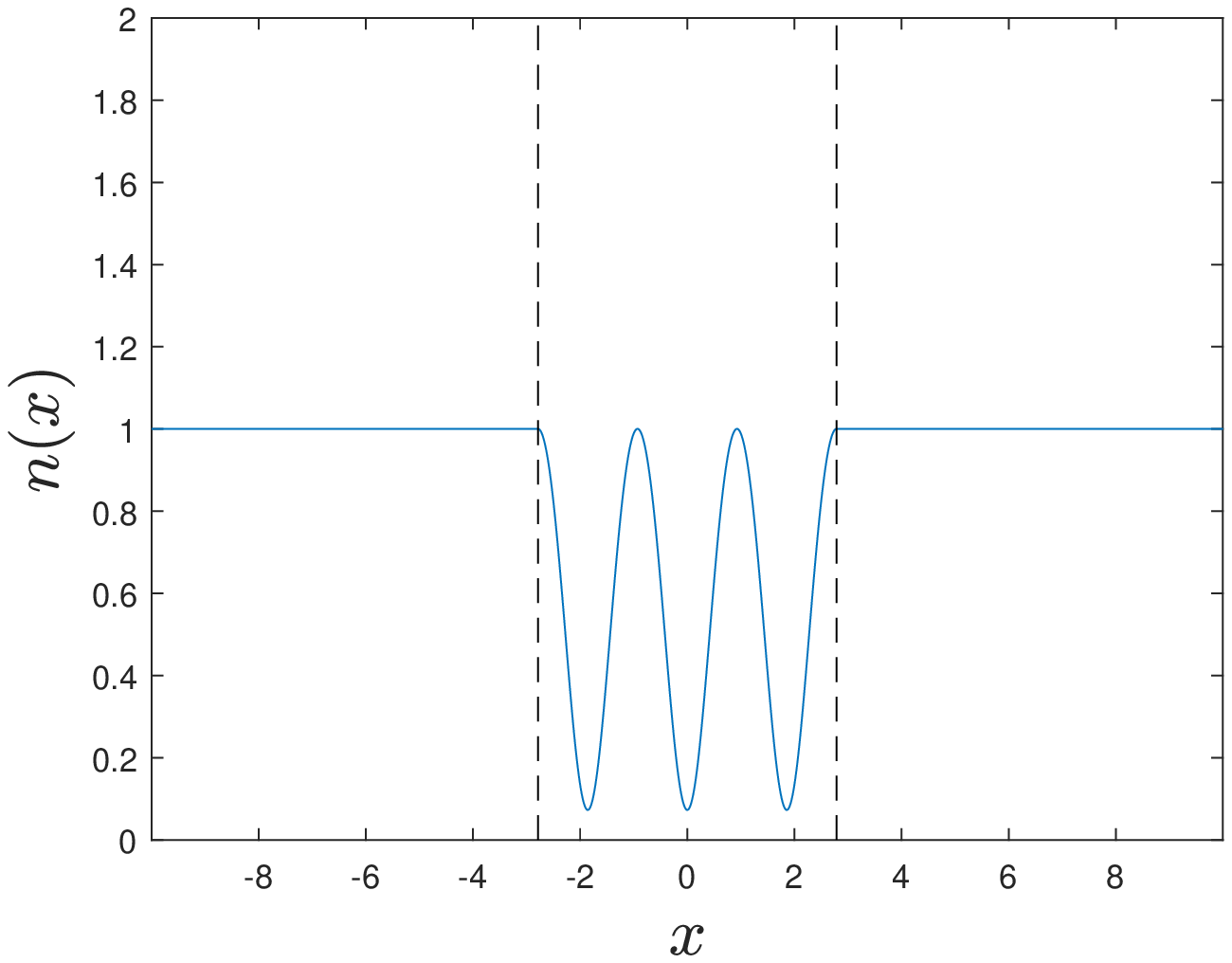} & \includegraphics[width=0.2\columnwidth]{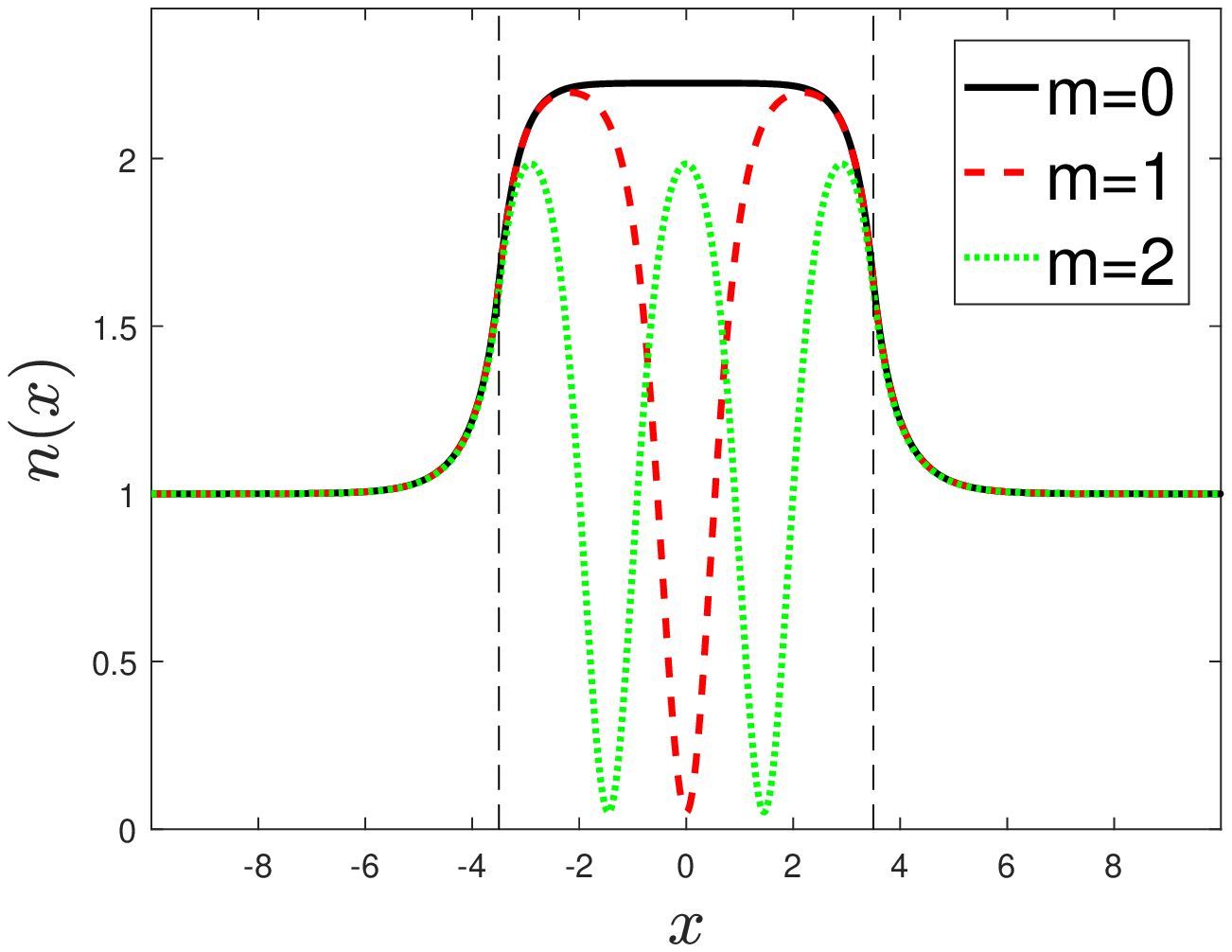}
\end{tabular}
\caption{Leftmost panel: orbits in the space $(A,A')$ corresponding to Eq. (\ref{eq:MechanicalEnergyConservationWF}). The thick dashed black line gives the possible solutions outside the well with amplitude energy $E_1$. The dashed-dotted green line describes arbitrary oscillatory solution inside the well while the solid red line corresponds to the limit solution within the well that matches the homogeneous subsonic plane wave outside of it. The solid thick blue line marks the limit orbit associated to the subsonic fixed point, analog to that of the dashed curve, inside the well. Central panels: density profile of $\psi^{H}_m(x)$, $m=1,2,3$ for $v=0.5$, corresponding to well lengths $X^H_1=1.8568$, $X^H_2=3.7137$ and $X^H_3=5.5705$. The vertical dashed lines mark the limits of the square well. Rightmost panel: density profile of $\psi^{SH}_m(x)$, $m=0,1,2$ for $v=0.5$ and $X=7$.}
\label{fig:BHLHomogenous}
\end{figure}

In this case, the wave function outside the well is the homogeneous subsonic plane wave and it is the most simple one as it represents a limit of the other two families of solutions. This solution only appears for some critical values of the length $X=X^H_m$, $m=0,1,2\ldots$, and the corresponding GP wave function is:
\begin{equation}\label{eq:HomoWFSolution}
\psi_0(x)=\psi^{H}_m(x)=\left\{ \begin{array}{cc}
e^{ivx}e^{-i\phi_0} & x< -\frac{X^H_m}{2}\\
\Lambda(x,n^H_1,n^H_2,n^H_3,0) & -\frac{X^H_m}{2}\leq x \leq \frac{X^H_m}{2}\\
e^{ivx}e^{i\phi_0} & x>\frac{X^H_m}{2}
\end{array}\right.
\end{equation}
where the critical values $n^H_{1,2,3},X^H_m$ are given in Appendix \ref{appsub:homogeneous}. Note that $m=0$ corresponds to $X=X^H_0=0$ and the trivial solution of a homogeneous subsonic plane wave in whole space as there is no attractive well.

The orbit in phase space inside the well associated to this family of solutions is depicted as a red solid line in left Fig. \ref{fig:BHLHomogenous} while the density profile for $\psi^{H}_m(x)$, $m=1,2,3$, is shown in the central panels of the same figure.

\subsection{Asymptotic shadow solitons}\label{subsec:subsonicsh}

The resulting GP wave function for this family of solutions reads:
\begin{equation}\label{eq:SHWF}
\psi_0(x)=\psi^{SH}_m(x)\equiv\left\{\begin{array}{cc}
e^{ivx}e^{-i\phi_0}\left(v+i\sqrt{1-v^2}\text{cotanh}\left[\sqrt{1-v^2}(x+x_0)\right]\right)& x< -\frac{X}{2}\\
\Lambda(x,n^{m}_1,n^{m}_2,n^{m}_3,\alpha_m) & -\frac{X}{2}\leq x \leq \frac{X}{2}\\
e^{ivx}e^{i\phi_0}\left(v+i\sqrt{1-v^2}\text{cotanh}\left[\sqrt{1-v^2}(x-x_0)\right]\right) & x>\frac{X}{2}
\end{array}\right.
\end{equation}
with $\phi_0,x_0$ chosen such that the wave function and its derivative are continuous, $x_0$ satisfying \linebreak $x_0-X/2<0$.

%[[Wave function. Order for Shadow Solitons. Keywords. Conclusions. AD???]]

As the solutions outside the well are shadow solitons, $1\leq n^m_W < n^{SH}$, where $n^{SH}$ is the density obtained from the intersection between the dashed black line and the solid blue line in left Figure \ref{fig:BHLHomogenous},
\begin{equation}
n^{SH}=\frac{E^{SH}_2-E_1}{V_0}, E^{SH}_2=W_2(A_2), A_2=\frac{\sqrt{1+\sqrt{1+8v^4}}}{2v}
\end{equation}

The limit case $n^m_W=1$ corresponds to $E^m_2=E^H_2$ and $\Psi^{SH}_m(x)=\Psi^{H}_m(x)$. In particular, the function $\Psi^{SH}_0(x)$ is continuously connected to the homogeneous subsonic plane wave solution, and since it has the larger amplitude inside the well, it represents the true ground state of the system, as can be inferred from Equation (\ref{eq:grandcanonicalenergypsi0}). Hence, the BHL solution $\Psi^{BHL}(x)$ of Equation (\ref{eq:CompactBHLWF}) should be dynamically unstable for any length $X>0$ of the well. However, since $\Psi^{BHL}(x)$ for $X=0$ corresponds to a perfect soliton, and it is not continuously connected to the ground state, the expected instability should be just the acceleration of the soliton \cite{Busch2000}, rather than the growth of a lasing mode, at least for small cavity lengths $0<X<X^C_0$ in which there is no room for other solutions (see discussion in the next section). A detailed computation of the BdG spectrum should be carried out in order to confirm the previous hypothesis.

On the other hand, the upper limit $n^m_W=n^{SH}$ is a strict inequality, as it corresponds to $E^m_2=E^{SH}_2$, so the wave function is a soliton inside the well, giving rise to an infinite size $X$. Therefore, $E^H_2\leq E^m_2 < E^{SH}_2$, and the $m$-th solution only exists for lengths:
\begin{equation}\label{eq:lengthlimitshWF}
X^{H}_m\leq X< \infty
\end{equation}

The density profile of $\psi^{SH}_m(x)$ for $m=0,1,2$ is represented in the rightmost panel of Fig. \ref{fig:BHLHomogenous}.

\subsection{Asymptotic solitons}\label{subsec:subsonicsolitons}

When the solutions outside the well corresponds to solitons, things are more intriguing than in the previous cases. We mainly distinguish between symmetric solutions, where the density has {\em even} parity and the GP wave function satisfies $\psi_0(x)=\psi^*_0(-x)$, and asymmetric solutions, with no spatial symmetry.

\begin{figure}[!htb]
\begin{tabular}{@{}ccc@{}}
    \includegraphics[width=0.3\columnwidth]{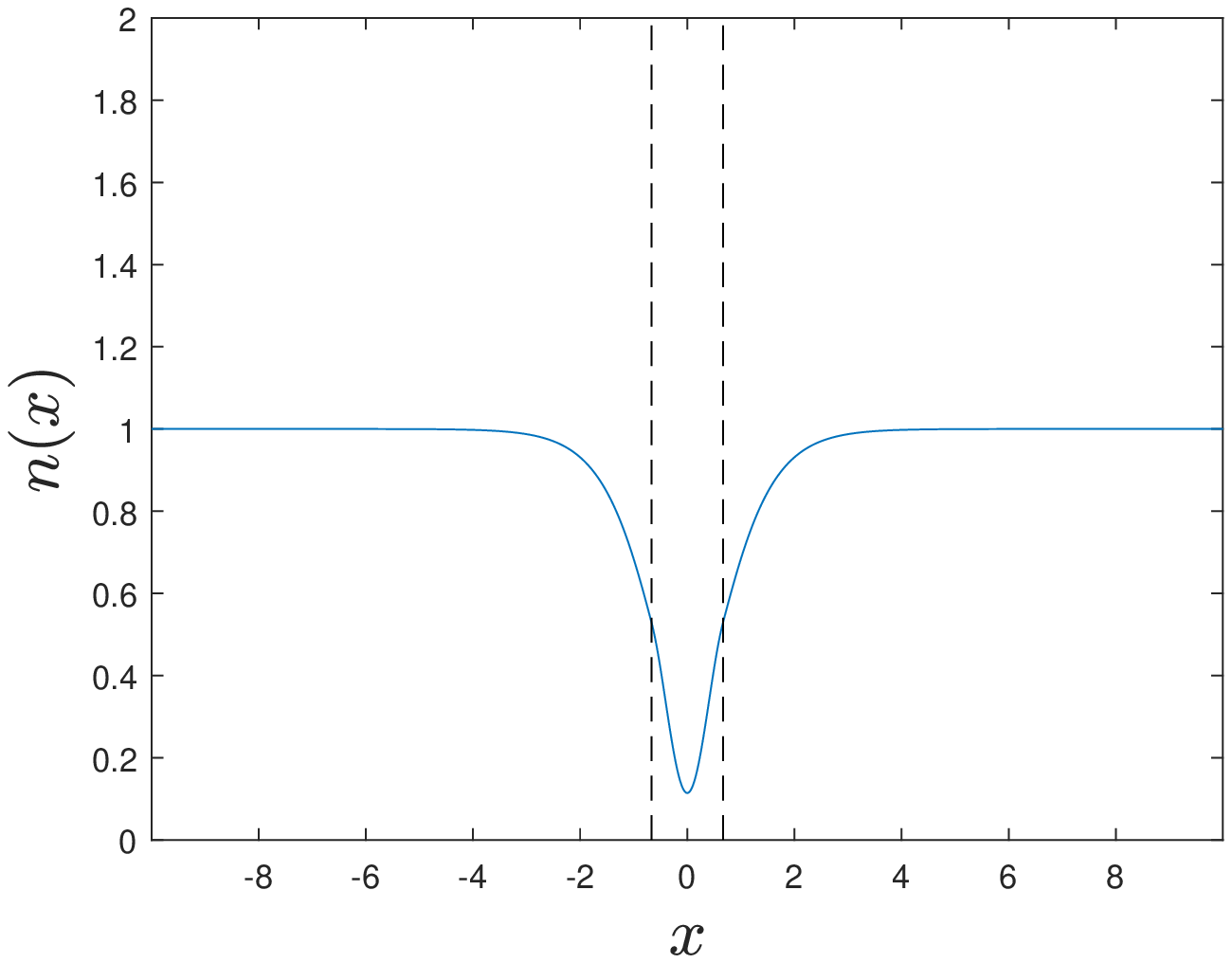} & \includegraphics[width=0.3\columnwidth]{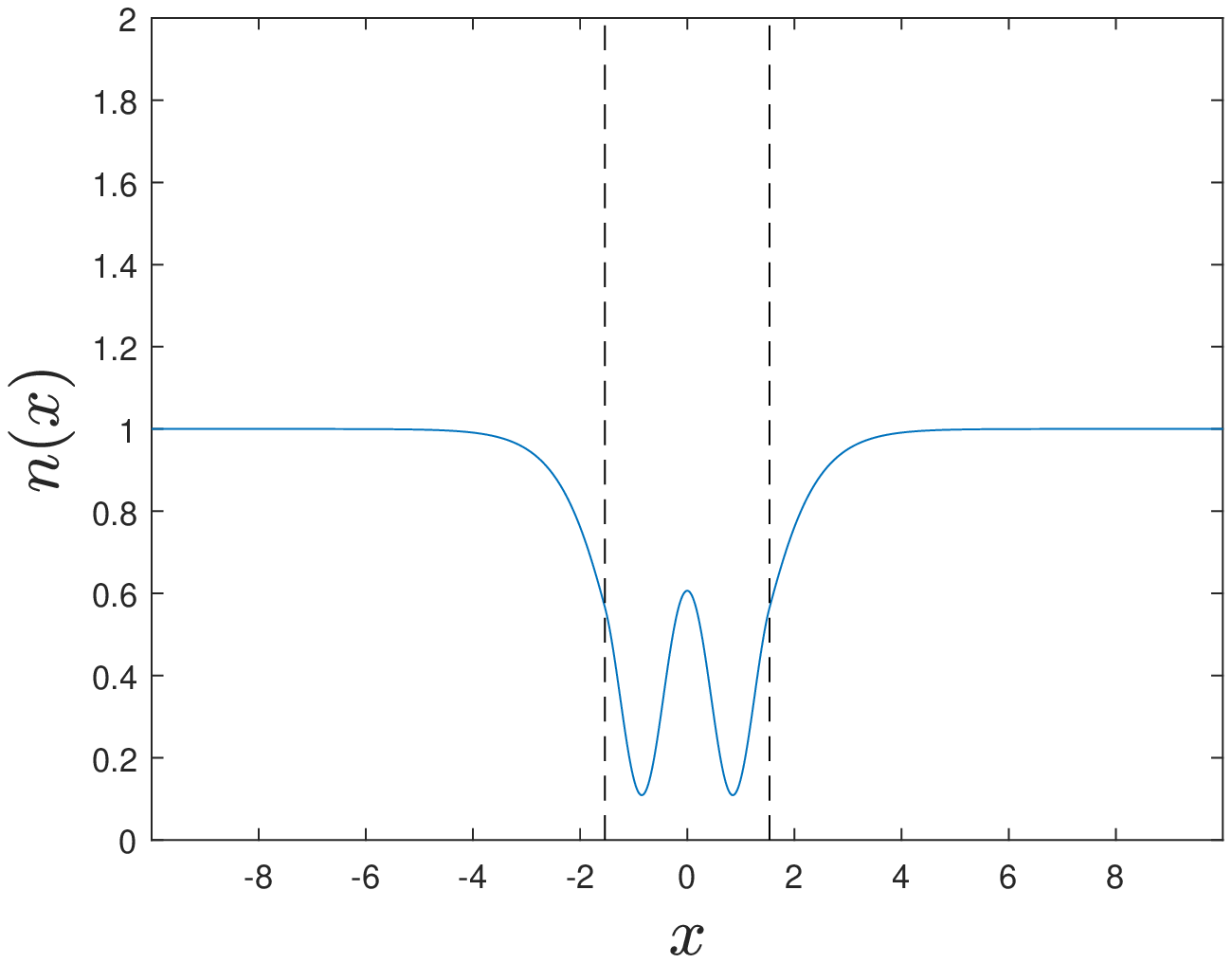} & \includegraphics[width=0.3\columnwidth]{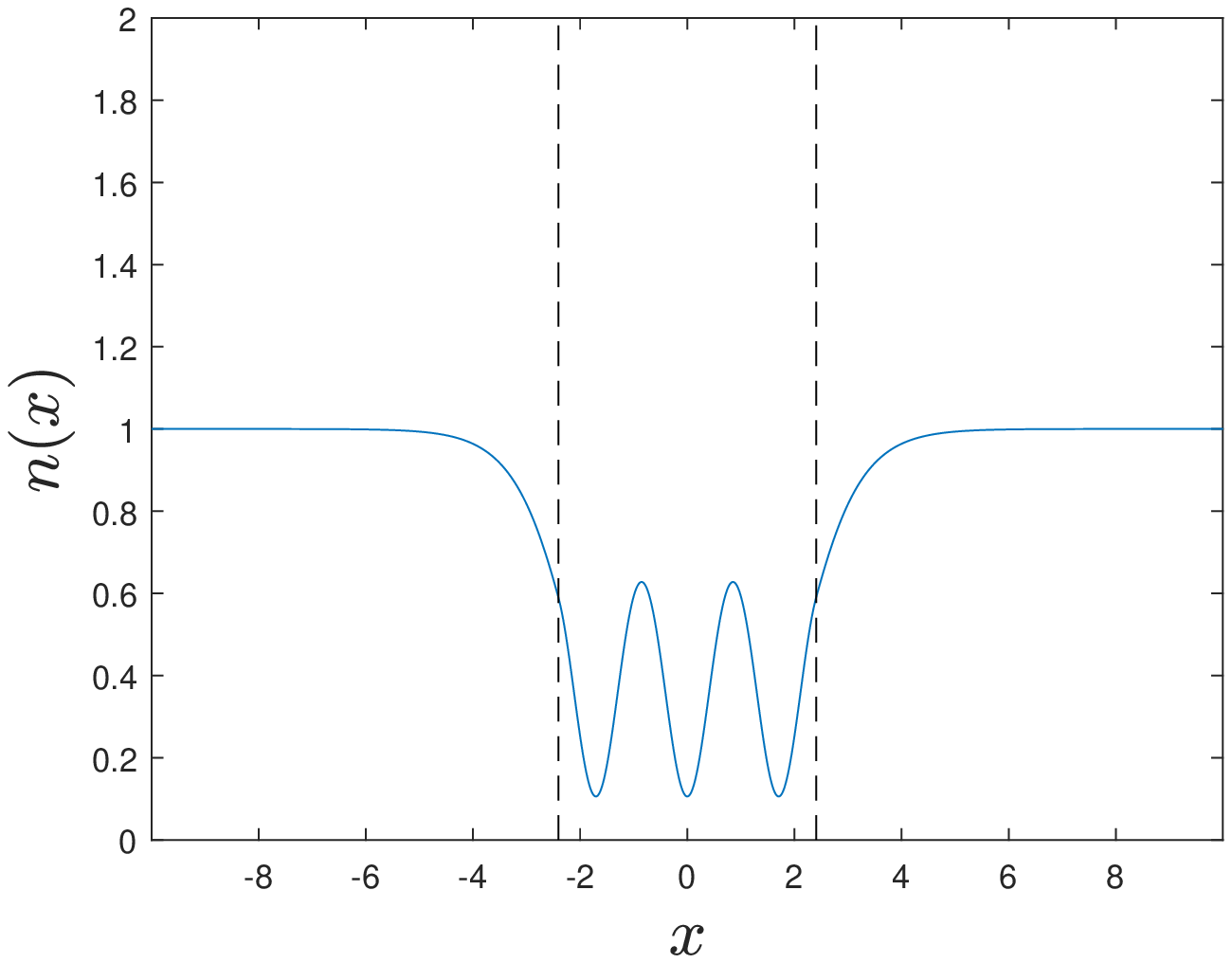} \\
    \includegraphics[width=0.3\columnwidth]{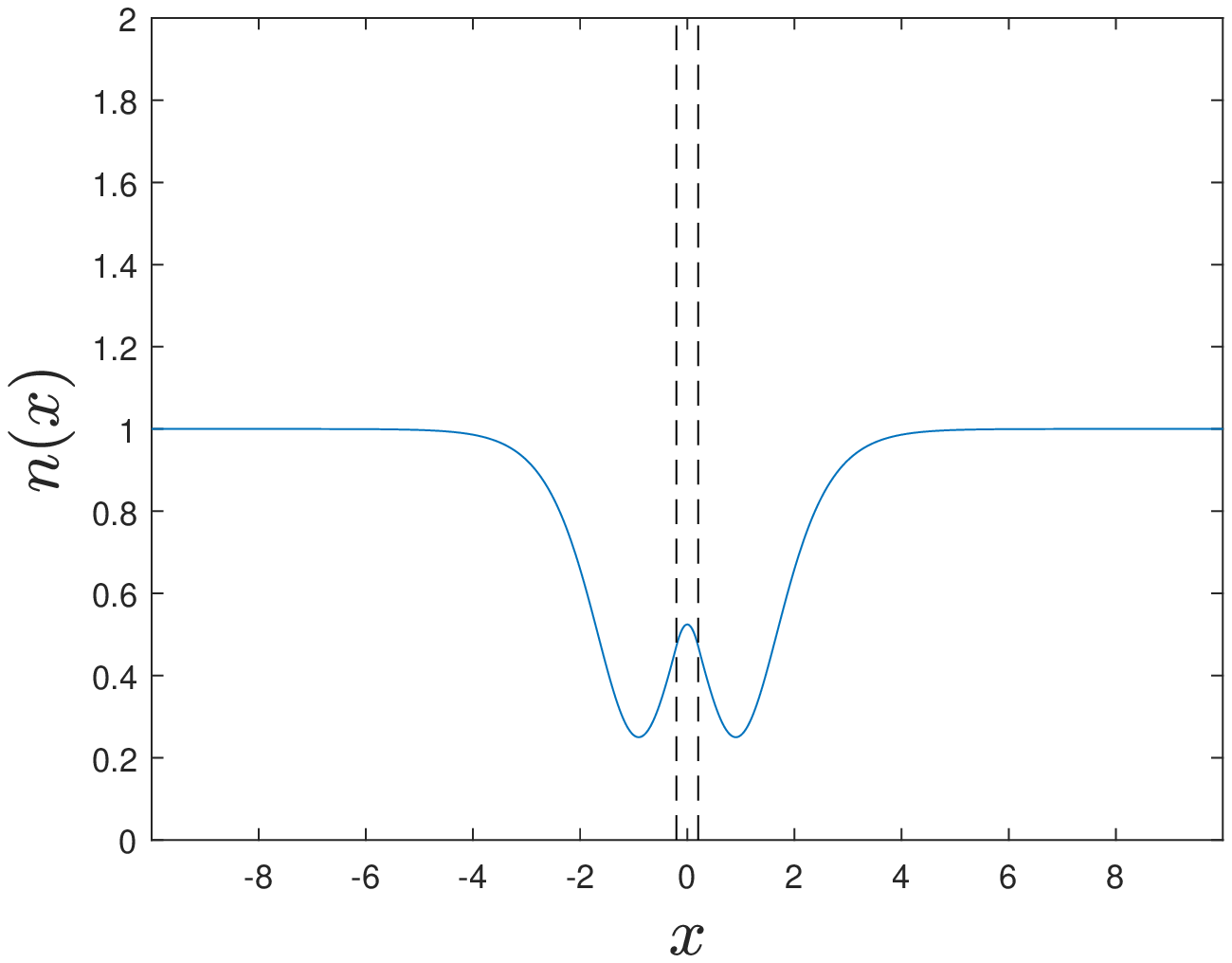} & \includegraphics[width=0.3\columnwidth]{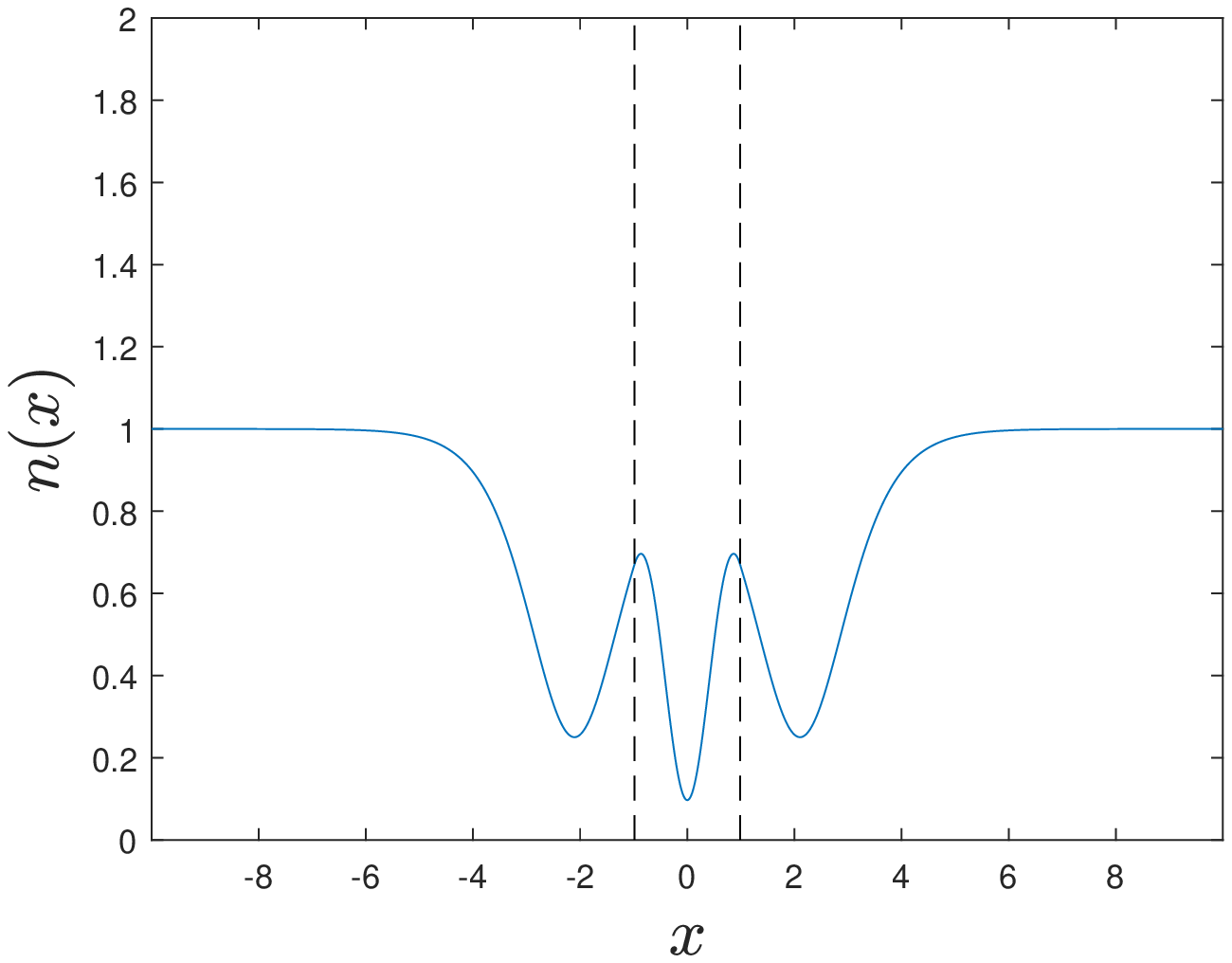} & \includegraphics[width=0.3\columnwidth]{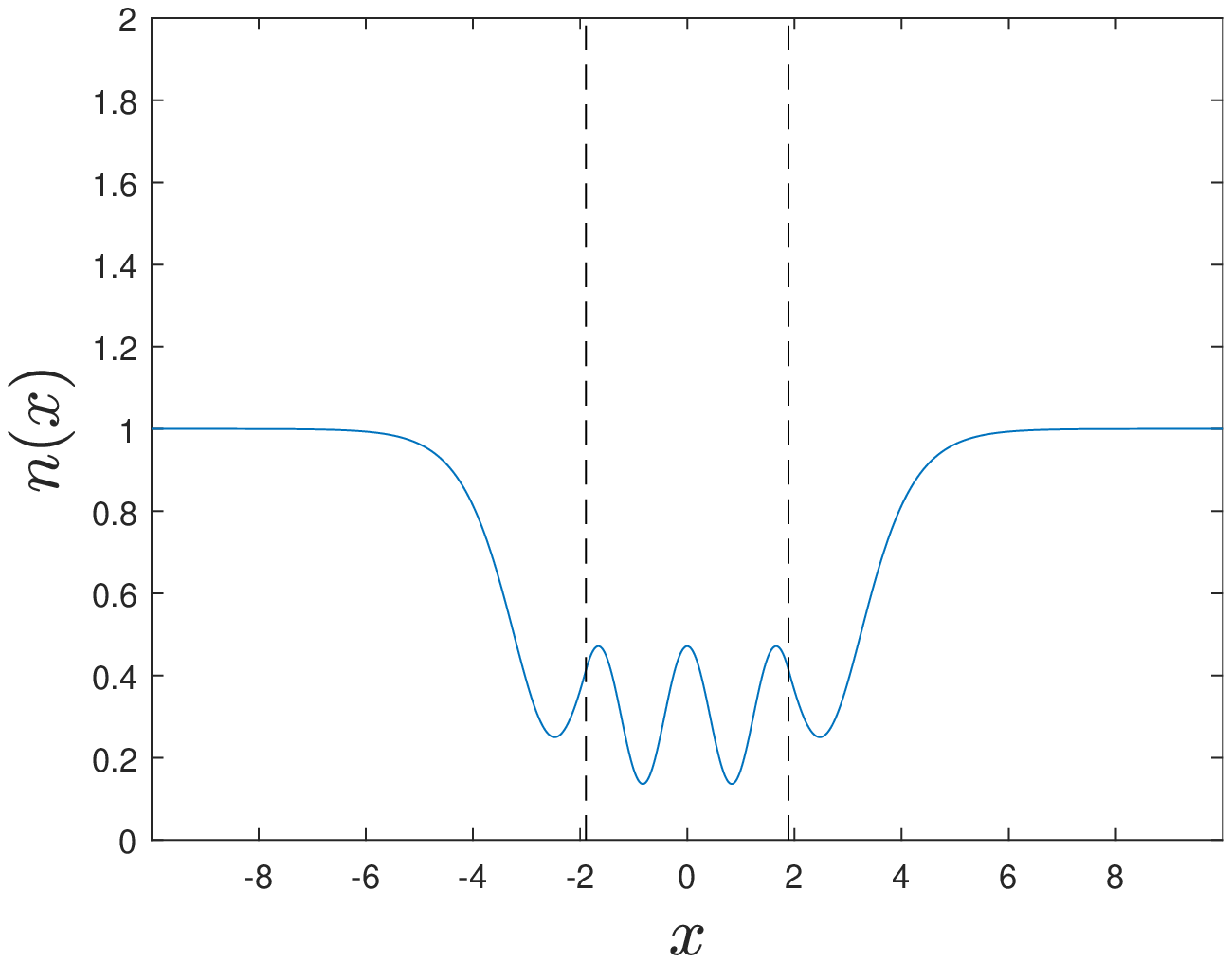} \\
    \includegraphics[width=0.3\columnwidth]{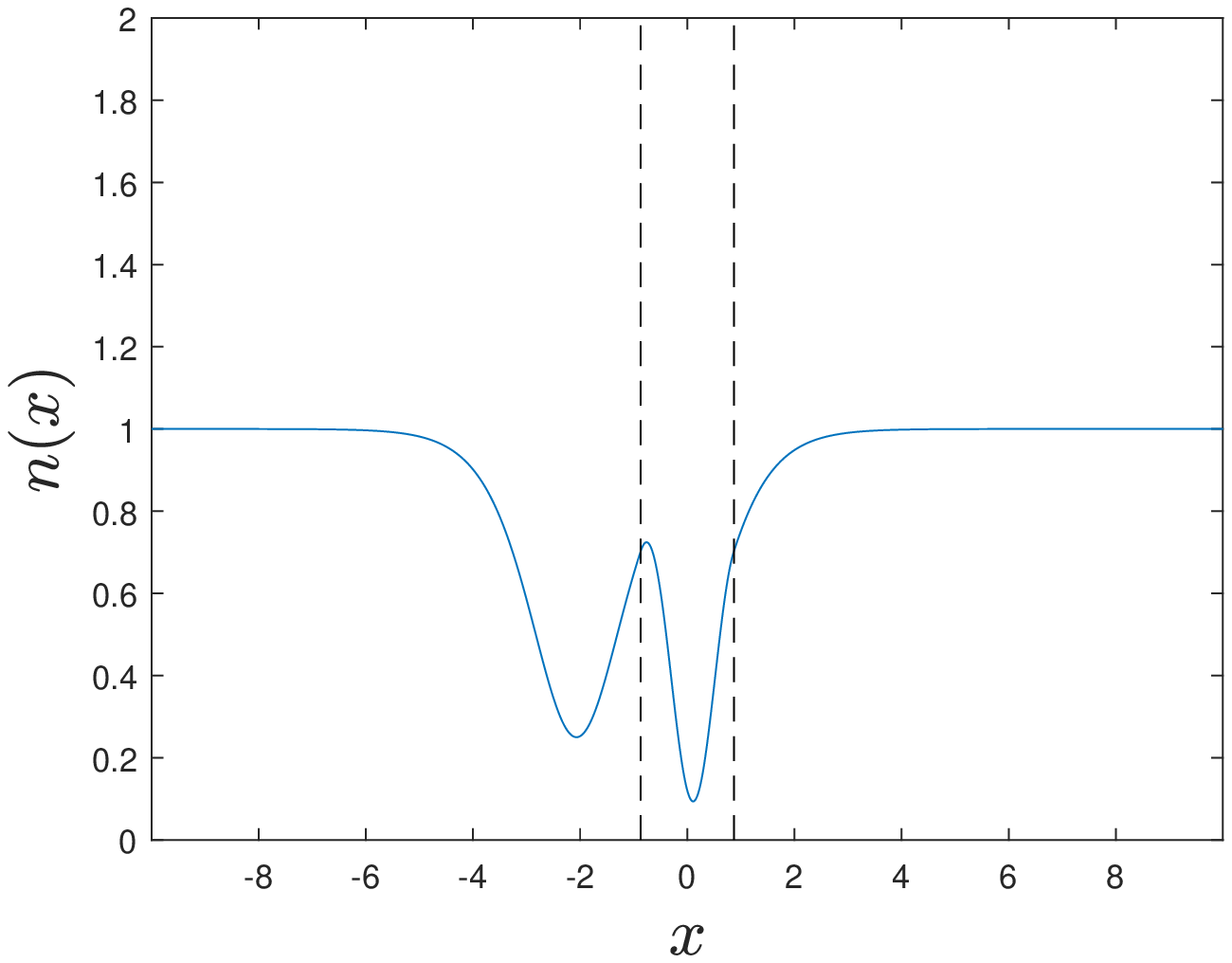} & \includegraphics[width=0.3\columnwidth]{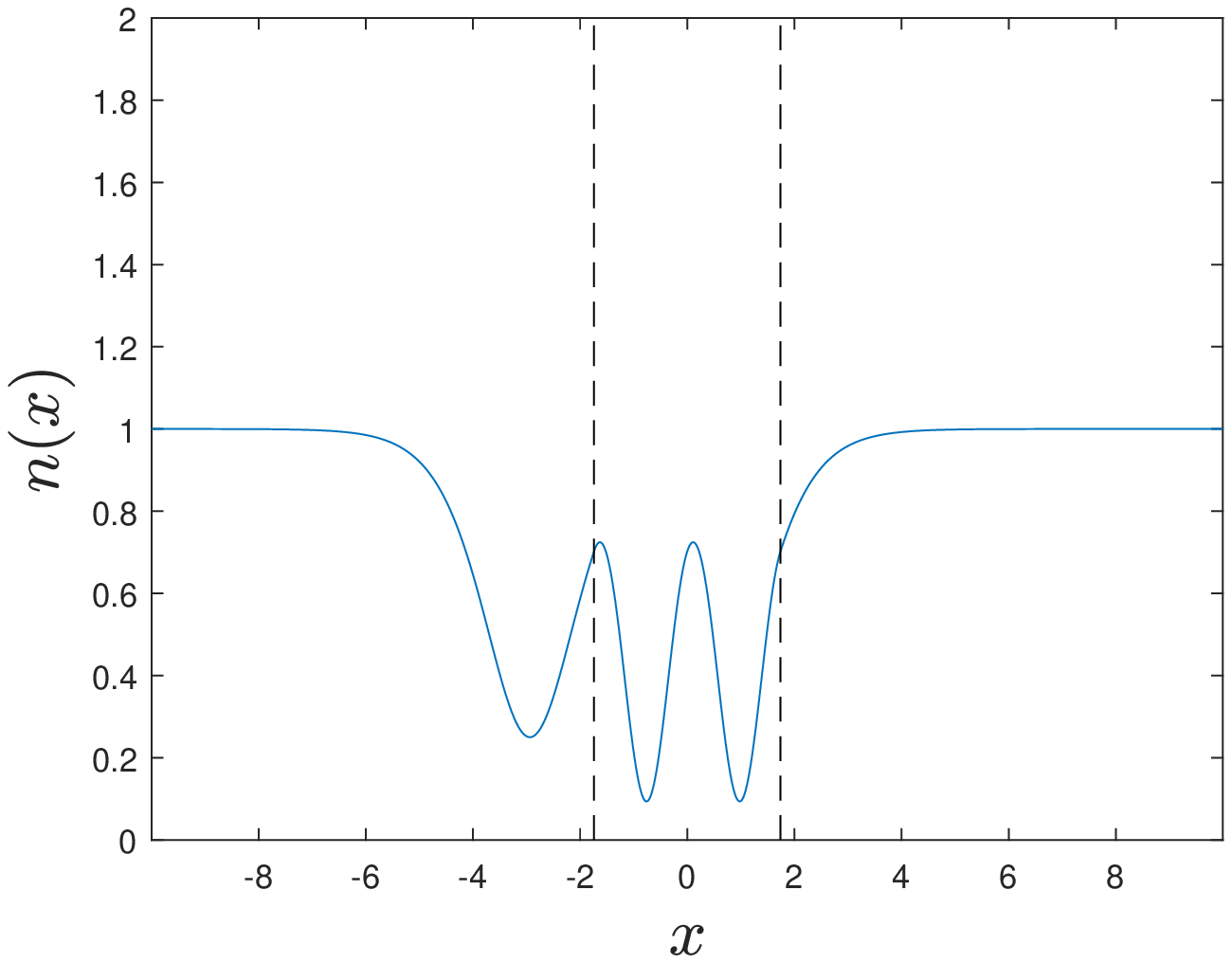} & \includegraphics[width=0.3\columnwidth]{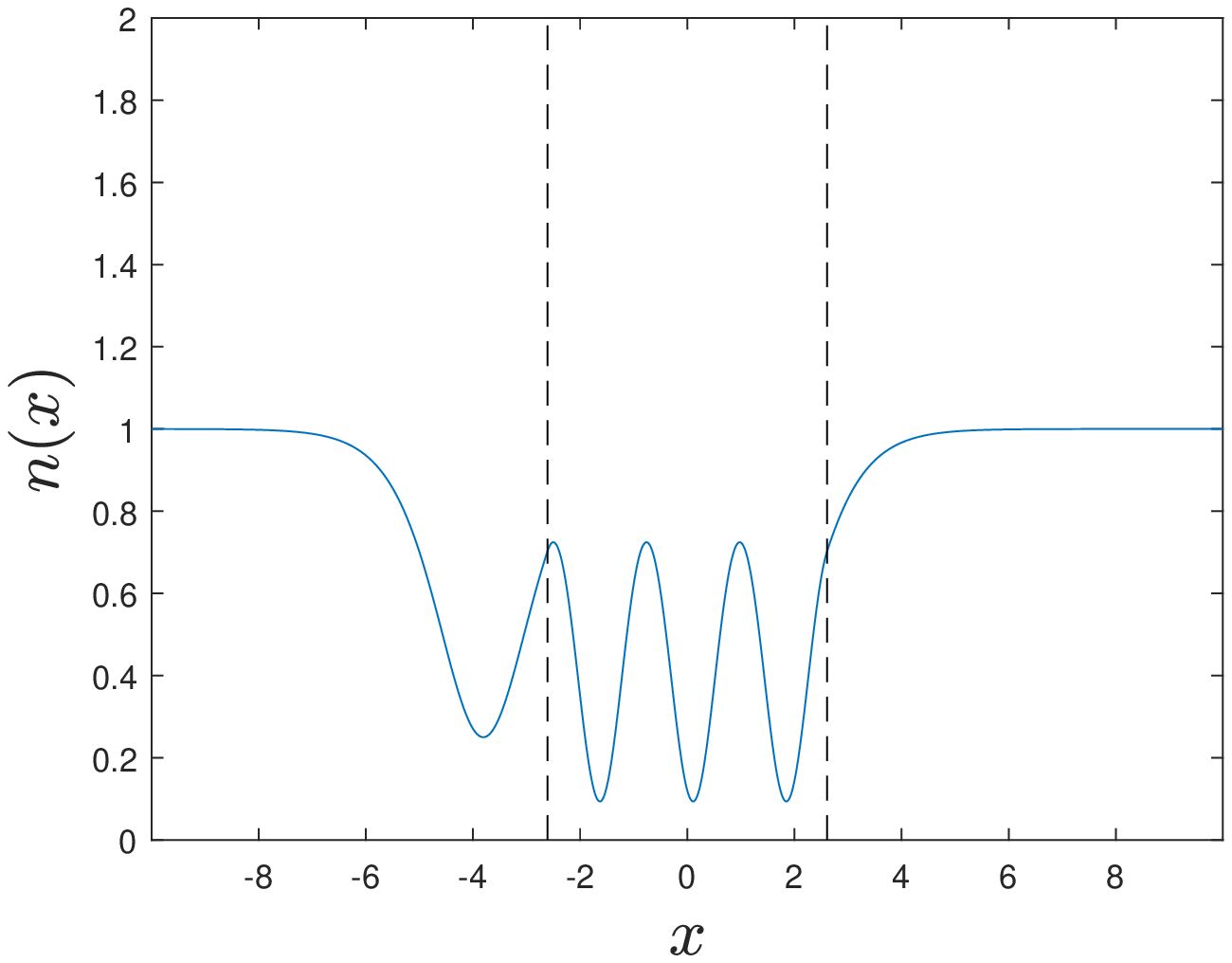}
\end{tabular}
\caption{Density profile of the solutions containing solitons outside the square well. The vertical dashed lines mark the limits of the cavity. Upper row: $m=0,1,2$ incomplete-soliton solutions for $v=0.5$ and well lengths $X=1.33$, $X=3.07$ and $X=4.81$, respectively.
Central row: $m=0,1,2$ complete-soliton solutions for $v=0.5$ and well lengths $X=0.46$, $X=1.97$ and $X=3.78$, respectively.
Lower row: $m=1,2,3$ asymmetric solutions for $v=0.5$ and well lengths $X=1.74$, $X=3.48$ and $X=5.22$, respectively. }
\label{fig:BHLSolitons}
\end{figure}

\subsubsection{Symmetric solutions}

For solutions with symmetric character, the solitons out of the well either both contain a minimum in the density ({\em complete}-soliton solutions) or not ({\em incomplete}-soliton solutions).

\paragraph{Incomplete-soliton solutions.}

In this case, the GP wave functions reads:
\begin{equation}\label{eq:SolitonWF}
\psi_0(x)=\psi^{S}_m(x)\equiv\left\{\begin{array}{cc}
e^{ivx}e^{-i\phi_0}\left(v+i\sqrt{1-v^2}\text{tanh}\left[\sqrt{1-v^2}(x+x_0)\right]\right)& x< -\frac{X}{2}\\
\Lambda(x,n^{m}_1,n^{m}_2,n^{m}_3,\alpha_m) & -\frac{X}{2}\leq x \leq \frac{X}{2}\\
e^{ivx}e^{i\phi_0}\left(v+i\sqrt{1-v^2}\text{tanh}\left[\sqrt{1-v^2}(x-x_0)\right]\right) & x>\frac{X}{2}
\end{array}\right.
\end{equation}
For incomplete-soliton solutions, $x_0$ satisfies $X/2-x_0>0$.

Since the solutions outside the well are solitons, $v^2\leq n^m_W \leq 1$. The lower boundary $n_W=v^2$ gives the same solution of Eq. (\ref{eq:CompactBHLWF}), in which $n^m_1=n^m_2=n^m_W=v^2$ and $E^m_2=E^C_2\equiv E_1+v^2V_0=1+\frac{v^4}{2}$, existing for arbitrary length $X$. Hence, this family of solutions is continuously connected to the BHL configuration described by $\psi^{BHL}(x)$. A more detailed analysis of the limit $n^m_W\rightarrow v^2$ (see Appendix \ref{appsub:subsonicsolitons}) shows that these solutions appear at the critical lengths $X^C_m$, given by Eq. (\ref{eq:criticallengths}), and can be understood as small amplitude perturbations on top of the GP wave function of Eq. (\ref{eq:CompactBHLWF}), described by the BdG equations. Precisely, in this limit, all the elliptic functions are reduced to trigonometric functions and the cnoidal wave to a regular sinusoidal wave with wavevector $k_0$, resulting from the corresponding BdG plane-wave solutions with zero frequency for the supersonic flow inside the well (see right Fig. \ref{fig:Dispersion} and related discussion). Indeed, as the dynamically unstable modes arising from the BHL effect are expected to first show as zero-frequency BdG modes \cite{Michel2013}, the critical lengths $X^C_m$ should also signal the appearance of a new dynamical instability.

The upper limit, $n^m_W=1$ gives $\psi^{SOL}_m(x)=\psi^{H}_{m+1}(x)$, merging with the $\psi^{SH}_{m+1}(x)$ solutions. Then, the $m$-th solution of Eq. (\ref{eq:SolitonWF}) only exists for
\begin{equation}\label{eq:incompletelengths}
X\in[X^C_m,X^H_{m+1}]
\end{equation}

The density profile for the incomplete-soliton solutions $m=0,1,2$ is represented in the upper row of Fig. \ref{fig:BHLSolitons}.

\paragraph{Complete-soliton solutions.}

The wave function is of the same form as that of Eq. (\ref{eq:SolitonWF}) but with $x_0$ satisfying $X/2-x_0<0$. This family of solutions is also continuously connected to $\psi^{BHL}(x)$ and the limit $n^m_W=v^2$ gives the same length as for incomplete-soliton solutions, $X=X^C_m$. The upper limit $n^m_W=1$ gives $X=X^H_m$ and $\psi^{SOL}_m(x)=\psi^{H}_{m}(x)$, merging with the $\psi^{SH}_{m}(x)$ solutions. For $m=0$, $X^C_0>X^H_0=0$; however, for $m$ sufficiently large, $X^C_m<X^H_m$. Thus, the $m=0$ complete-soliton solution is limited to the range
\begin{equation}\label{eq:completelengths0}
X\in[0,X^C_0]
\end{equation}
while for $m\geq 1$ we can only generally say that
\begin{equation}\label{eq:completelengths}
X\in[\tilde{X}_m,\max (X^C_m,X^H_m)]
\end{equation}
with $\tilde{X}_m\leq\min (X^C_m,X^H_m)$; check Appendix \ref{appsub:subsonicsolitons} for the details.

The density profile for the complete-soliton solutions $m=0,1,2$ is represented in the central row of Fig. \ref{fig:BHLSolitons}.

\subsubsection{Asymmetric solutions}

These solutions are not symmetric with respect to the well and are characterized by one complete (incomplete) soliton at the left and one incomplete (complete) soliton at the right, one case corresponding to the spatial reverse of the density profile of the other. n particular, they contain an exact integer number of periods $m=1,2,3$ inside the well (see Eq. (\ref{eq:matchingasWF})), with $m=0$ giving the trivial solution of no well, $X=0$. The corresponding GP wave function reads:
\begin{equation}\label{eq:SolitonWF}
\psi_0(x)=\psi^{A}_m(x)\equiv\left\{\begin{array}{cc}
e^{ivx}e^{-i\phi_{L}}\left(v+i\sqrt{1-v^2}\text{tanh}\left[\sqrt{1-v^2}\left(x+\frac{X}{2}\pm\delta x\right)\right]\right)& x< -\frac{X}{2}\\
\Lambda(x,n^{m}_1,n^{m}_2,n^{m}_3,\alpha_m) & -\frac{X}{2}\leq x \leq \frac{X}{2}\\
e^{ivx}e^{i\phi_{R}}\left(v+i\sqrt{1-v^2}\text{tanh}\left[\sqrt{1-v^2}\left(x-\frac{X}{2}\pm\delta x\right)\right]\right) & x>\frac{X}{2}
\end{array}\right.
\end{equation}
where $\delta x>0,\phi_{L},\phi_{R}$ are chosen such that the wave function and its derivative are continuous and $\pm$ corresponds to the case of complete-incomplete (incomplete-complete) solitons.

In the same fashion of the symmetric families, in the limit $n^m_W\rightarrow v^2$ the asymmetric solutions are continuously connected to $\psi^{BHL}(x)$, appearing at the critical lengths $X^A_m$ given by Eq. (\ref{eq:criticallengthsAs}), while in the upper limit $n^m_W\rightarrow 1$ they converge to the homogeneous $\psi^{H}_{m}(x)$ solutions.

Hence, the $m$-th asymmetric solution is restricted for lengths
\begin{equation}\label{eq:lengthsXas}
X\in[X^A_m,X^H_m],~
\end{equation}
The density profile for the asymmetric solutions $m=1,2,3$ is represented in the lower row of Fig. \ref{fig:BHLSolitons}.

\section{Double delta-barrier}\label{sec:BHL2delta}

\subsection{General structure}

The potential corresponding to a {\em single} delta-barrier configuration is
\begin{equation}\label{eq:1Delta}
V(x)=Z\delta(x-x_H),~Z=\sqrt{\frac{W(1)-W(A_p)}{2A^2_p}}
\end{equation}
with $W(A)$ given by Eq. (\ref{eq:GPpotential}), $A_p$ the supersonic amplitude of Eq. (\ref{eq:1Dhomogeneousroots}) and $x_H$ the point where the barrier is placed. Such a delta potential originates a discontinuity in the derivative of $\psi_0(x)$ of the form $\psi'_0(x_H^{+})-\psi'_0(x_H^{-})=2Z\psi_0(x_H)$. The resulting GP wave function is:
\begin{equation}\label{eq:CompactBHDelta}
\psi_0(x)=\psi^{BH}(x)=\left\{ \begin{array}{cc}
e^{ivx}e^{-i\phi_0}\left(v+i\sqrt{1-v^2}\tanh\left[\sqrt{1-v^2}(x+x_0)\right]\right) & x< x_H\\
A_pe^{iv_px}, &  x \geq x_H
\end{array}\right.
\end{equation}
The constants $x_0,\phi_0$ are fixed by imposing the continuity of $\psi_0(x)$ and its derivative at $x=x_H$.

The associated BHL configuration is described by a cavity of length $X$ placed between two delta barriers,
\begin{equation}\label{eq:2Delta}
V(x)=Z\left[\delta\left(x+\frac{X}{2}\right)+\delta\left(x-\frac{X}{2}\right)\right]
\end{equation}
This configuration was studied in Ref. \cite{Zapata2011} in order to look for resonant BH configurations, which enhance the spontaneous Hawking signal \cite{deNova2014}. Here, we focus only on looking for stationary BHL solutions, as in the previous section. In particular, by construction, a BHL solution as that of lower central Fig. \ref{fig:BHConfigurations} exists, described by the GP wave function:
\begin{equation}\label{eq:CompactBHLDelta}
\psi_0(x)=\psi^{BHL}(x)=\left\{ \begin{array}{cc}
e^{ivx}e^{-i\phi_0}\left(v+i\sqrt{1-v^2}\tanh\left[\sqrt{1-v^2}\left(x+x_0\right)\right]\right) & x< -\frac{X}{2}\\
A_{p}e^{iv_px} & -\frac{X}{2}< x < \frac{X}{2}\\
e^{ivx}e^{i\phi_0}\left(v+i\sqrt{1-v^2}\tanh\left[\sqrt{1-v^2}\left(x-x_0\right)\right]\right) & x>\frac{X}{2}
\end{array}\right.
\end{equation}
with $x_0,\phi_0$ chosen such that the wave function and its derivative are continuous.

As in the computation for the square well, we distinguish two different regions: region 1 corresponds to the exterior of the cavity, $|x|>X/2$, while region 2 corresponds to its interior, $|x|<X/2$. In each region, we have that
\begin{equation}\label{eq:MechanicalEnergyConservationDelta}
\frac{A'^2}{2}+W(A)=E_i,
\end{equation}
where $E_i$ is the conserved amplitude energy for the $i=1,2$ regions, $E_1=\frac{1}{2}+v^2$ fixed by the asymptotic subsonic behavior.

The wave function is continuous everywhere and the only effect of the two delta barriers is to introduce a discontinuity in the derivative of the wave function given by
\begin{equation}\label{eq:matchingDelta}
\psi'_0\left(\pm\frac{X^+}{2}\right)-\psi'_0\left(\pm\frac{X^-}{2}\right)=2Z\psi_0\left(\pm\frac{X}{2}\right)
\end{equation}
which, in terms of the amplitude, reads:
\begin{equation}\label{eq:matchingDeltadensity}
A'\left(\pm\frac{X^+}{2}\right)-A'\left(\pm\frac{X^-}{2}\right)=2ZA\left(\pm\frac{X}{2}\right)\equiv 2ZA_{\pm}
\end{equation}
Thus, we can understood the effect of the delta barriers as ``instantaneously accelerating'' the classical particle described by Eq. (\ref{eq:MechanicalEnergyConservationDelta}).

As a result of the above considerations, the only possible choice for the wave function outside the cavity is the soliton solution as the other solutions would monotonically increase. The same reasoning restrict even more the possibilities and the solitons must satisfy
\begin{equation}\label{eq:matchingsigndelta}
A'\left(\pm\frac{X^{\pm}}{2}\right)\gtrless0
\end{equation}
Inside the cavity, the solution corresponds to a cnoidal wave, with the phase chosen such that $\phi(0)=0$ for convention.

By joining Eqs. (\ref{eq:MechanicalEnergyConservationDelta}), (\ref{eq:matchingDeltadensity}) and (\ref{eq:matchingsigndelta}), we find that the energy inside the cavity is related to the amplitude at the edges through:
\begin{equation}\label{eq:energyDelta}
E_2=E_1-2ZA_{\pm}\left(\sqrt{2\left[E_1-W(A_{\pm})\right]}-ZA_{\pm}\right)
\end{equation}
Hence, the value of $E_2$ is determined by the cnoidal waves arising from Eq. (\ref{eq:MechanicalEnergyConservationDelta}) that are compatible with Eq. (\ref{eq:energyDelta}). In particular, as $E_p\leq E_2<E_1$, with $E_p\equiv W(A_p)$, the amplitudes at the edges of the cavity must be in the range
\begin{equation}\label{eq:rangeDelta}
A_{\inf}<A_{\pm}<A_{\sup}
\end{equation}
where $A_{\inf}<A_{\sup}<1$ are obtained from the roots of the equation:
\begin{equation}\label{eq:infsupDelta}
2\left[E_1-W(A)\right]-Z^2A^2=0
\end{equation}

As for the square well, the possible solutions are labeled by a discrete index $m=0,1,2\ldots$ representing the number of complete periods inside the cavity and the wave function is determined once $E^m_2,\alpha_m$ are obtained; the details of this computation are provided in Appendix \ref{app:technicaldelta}.

In order to classify the different families of solutions, we first distinguish between symmetric and asymmetric solutions, in analogy to the discussion presented in Sec. \ref{subsec:subsonicsolitons}.

\subsection{Symmetric solutions}\label{subsec:DeltaSymmetric}

Symmetric solutions satisfy:
\begin{eqnarray}\label{eq:deltasymmetricdensity}
A_{+}&=&A_{-}\equiv A_W\\
\nonumber A'\left(\frac{X^{\pm}}{2}\right)&=&-A'\left(-\frac{X^{\pm}}{2}\right)
\end{eqnarray}
and then the matching equations at the edges of the cavity read:
\begin{eqnarray}\label{eq:deltasymmetricmatching}
E_2&=&E_1-2ZA_W\left(\sqrt{2\left[E_1-W(A_W)\right]}-ZA_W\right)\\
\nonumber n_W&=&n\left(\pm\frac{X}{2},n_1,n_2,n_3,\alpha\right)
\end{eqnarray}
with $n_W=A^2_W$. Since the first line gives the density at the edges $n_W$ as an implicit function of $E_2$, the second line is a similar matching condition as that of Eq. (\ref{eq:matchingWFelliptic}); the only difference is that now $n_W$ is a much more complicated function of $E_2$. In fact, there are two different solutions for $A_{W}<1$ for a given value of the amplitude energy $E_2$ in the range $E_p<E_2<E_1$, one satisfying $A_p<A_{W}<A_{\sup}$ and the other one $A_{\inf}<A_{W}<A_{p}$.

As the signs of the derivatives at the edges outside the cavity are fixed by Eq. (\ref{eq:matchingsigndelta}), we classify the solutions according to the sign od the derivatives at the internal side of the cavity edges, $x=\pm X^{\mp}/2$. The limit values between the different solutions are obtained from
\begin{equation}
0=A'\left(\pm\frac{X^{\mp}}{2}\right)=2ZA\left(\pm\frac{X}{2}\right)\mp A'\left(\pm\frac{X^{\pm}}{2}\right)=2ZA_{\pm}-\sqrt{2\left[E_1-W(A_{\pm})\right]}
\end{equation}
which yields a similar equation to Eq. (\ref{eq:infsupDelta}):
\begin{equation}\label{eq:criticalvaluesDelta}
\left[E_1-W(A)\right]-2Z^2A^2=0
\end{equation}
This equation has two solutions for $A<1$, $A=A_q$ and, by construction, $A=A_p$, with $A_q<A_p$. The energies associated to these solutions are $E_2=E_q\equiv W(A_q)$ and $E_2=E_p<E_q$.

Following the above considerations, we distinguish three families of solutions: $S+$, for $A_{p}<A_W<A_{\sup}$; $S-$, for $A_{q}< A_W<A_{p}$; and $SD$, for $A_{\inf}< A_W<A_{q}$.

\subsubsection{$S+$ solutions}\label{subsubsec:S+}

\begin{figure}[!htb]
\begin{tabular}{@{}ccccc@{}}
    \includegraphics[width=0.2\columnwidth]{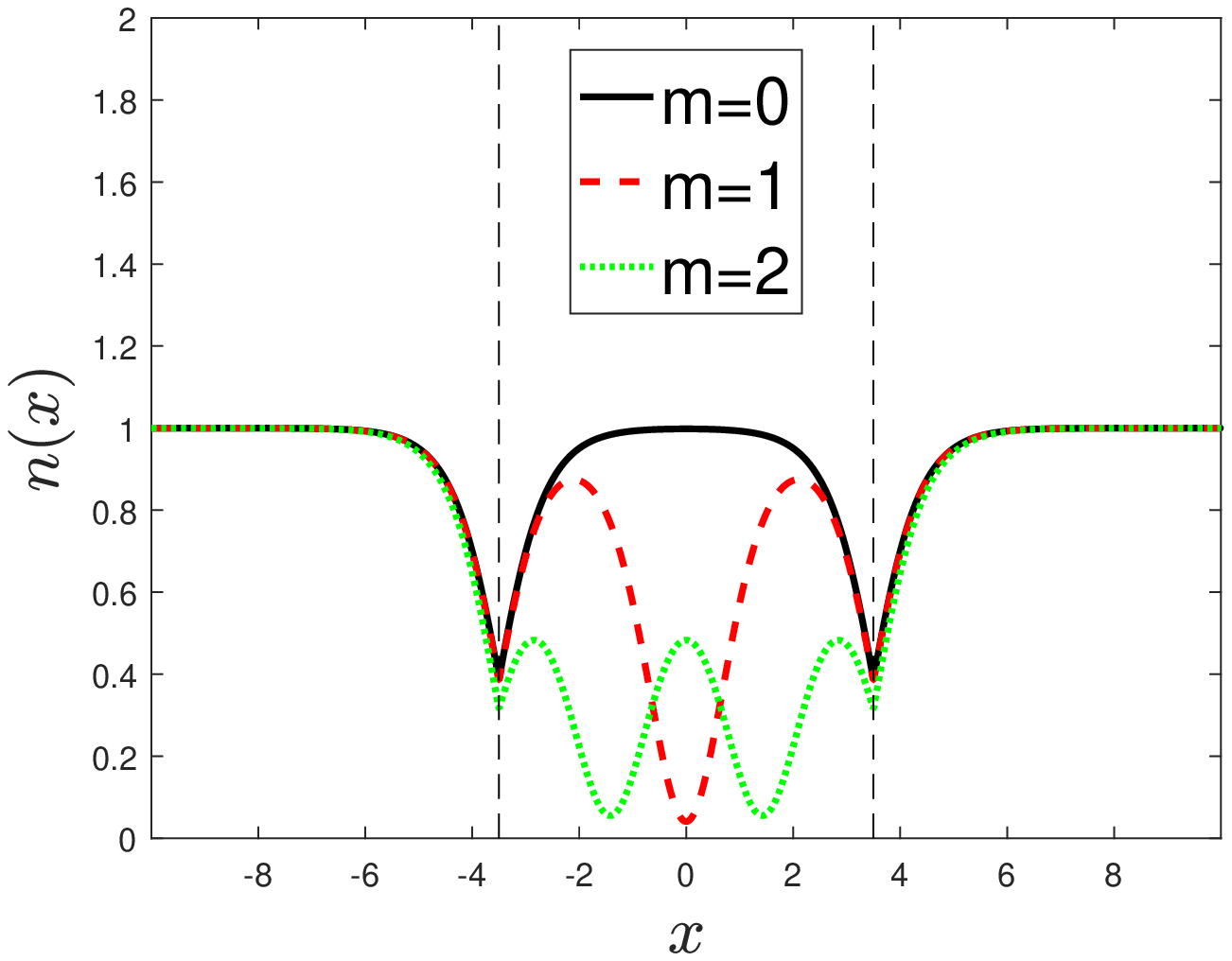} &   \includegraphics[width=0.2\columnwidth]{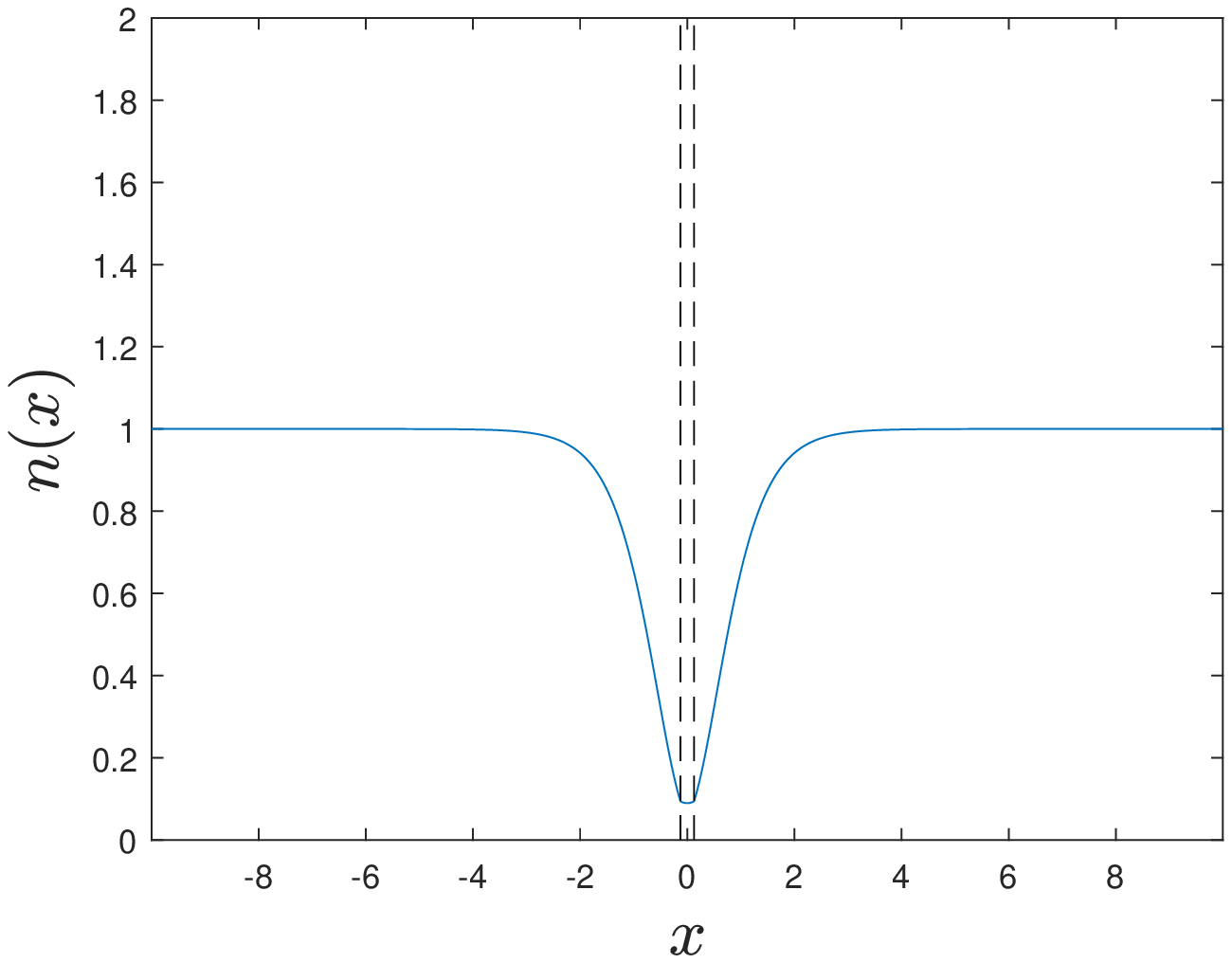} & \includegraphics[width=0.2\columnwidth]{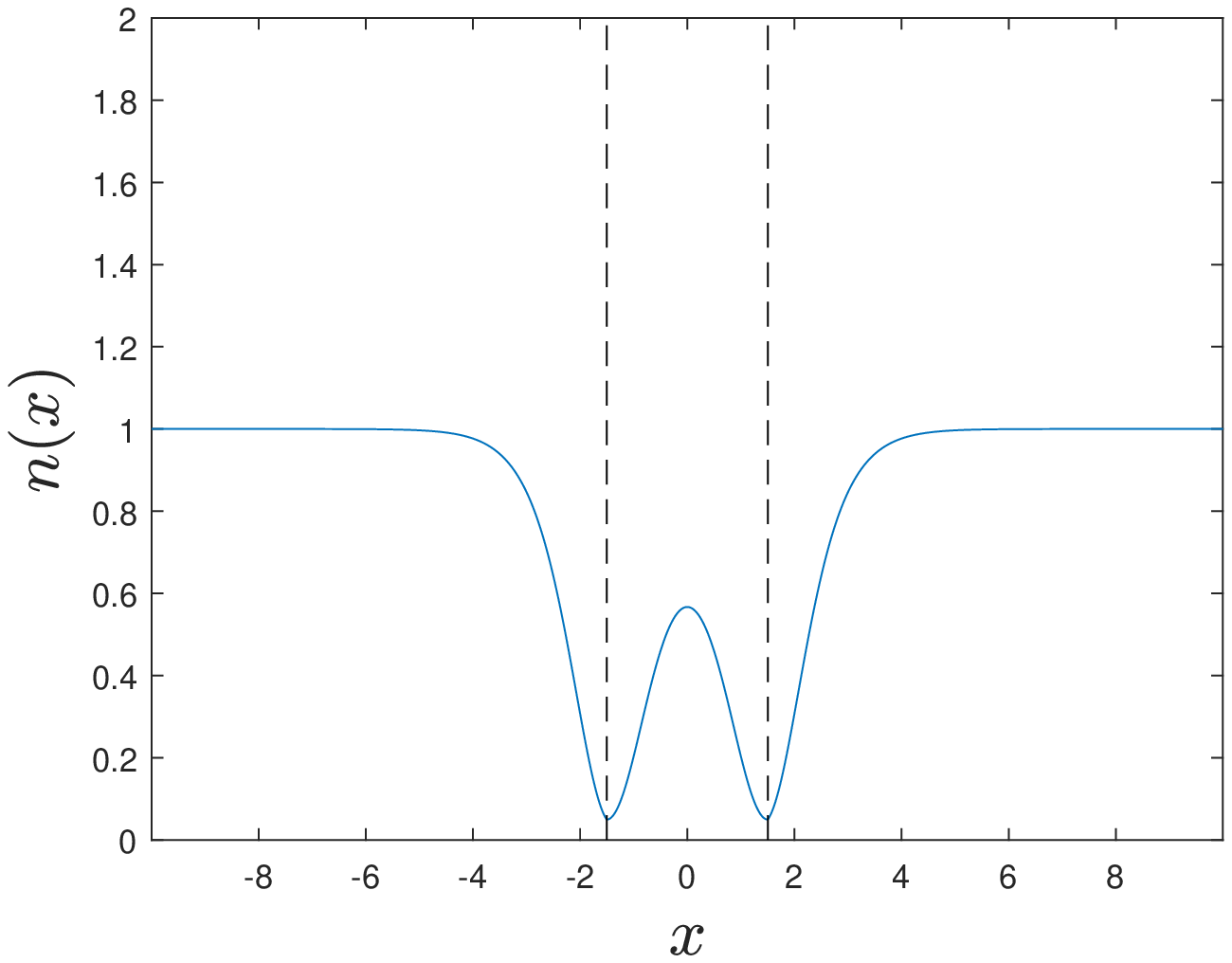} & \includegraphics[width=0.2\columnwidth]{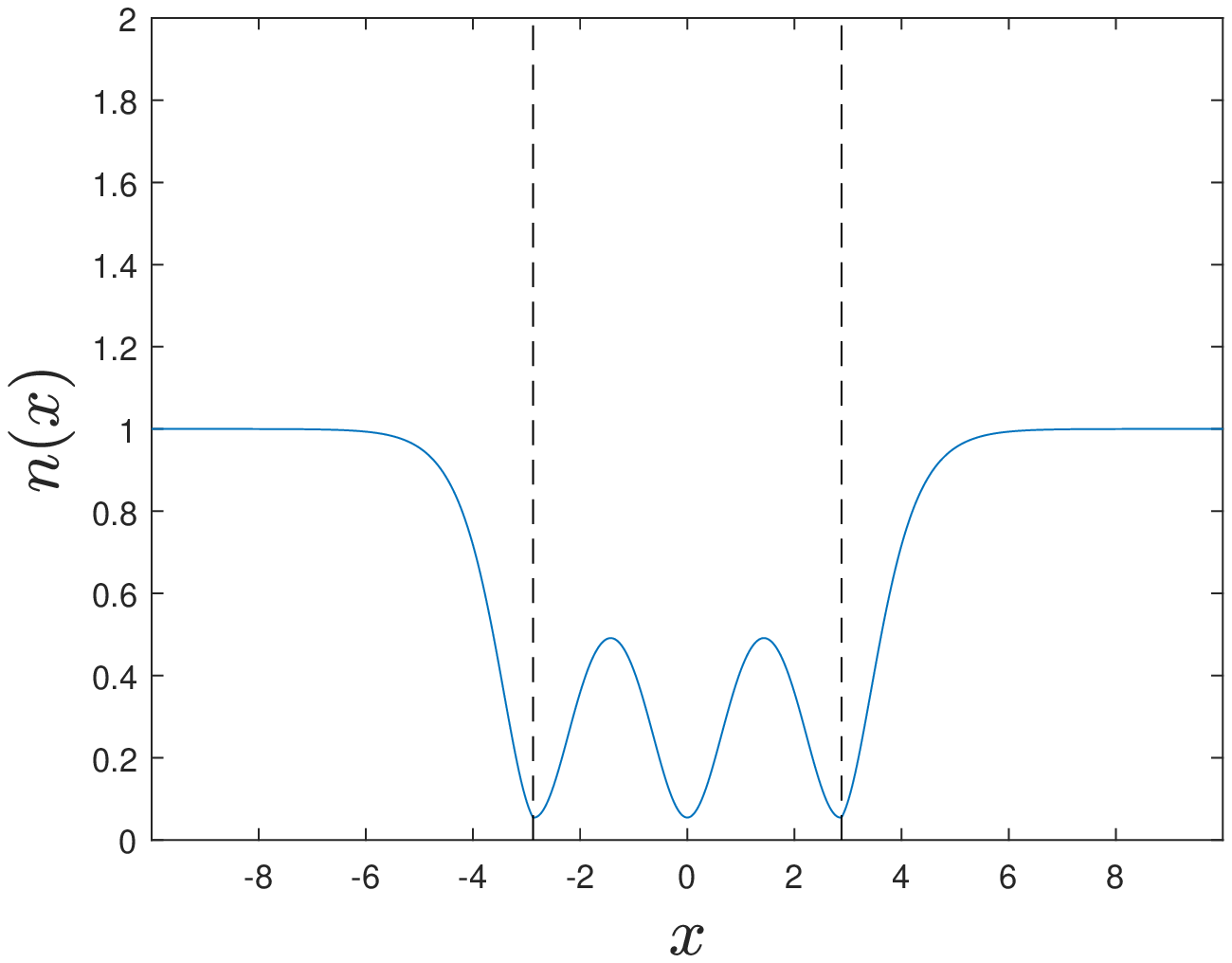} \includegraphics[width=0.2\columnwidth]{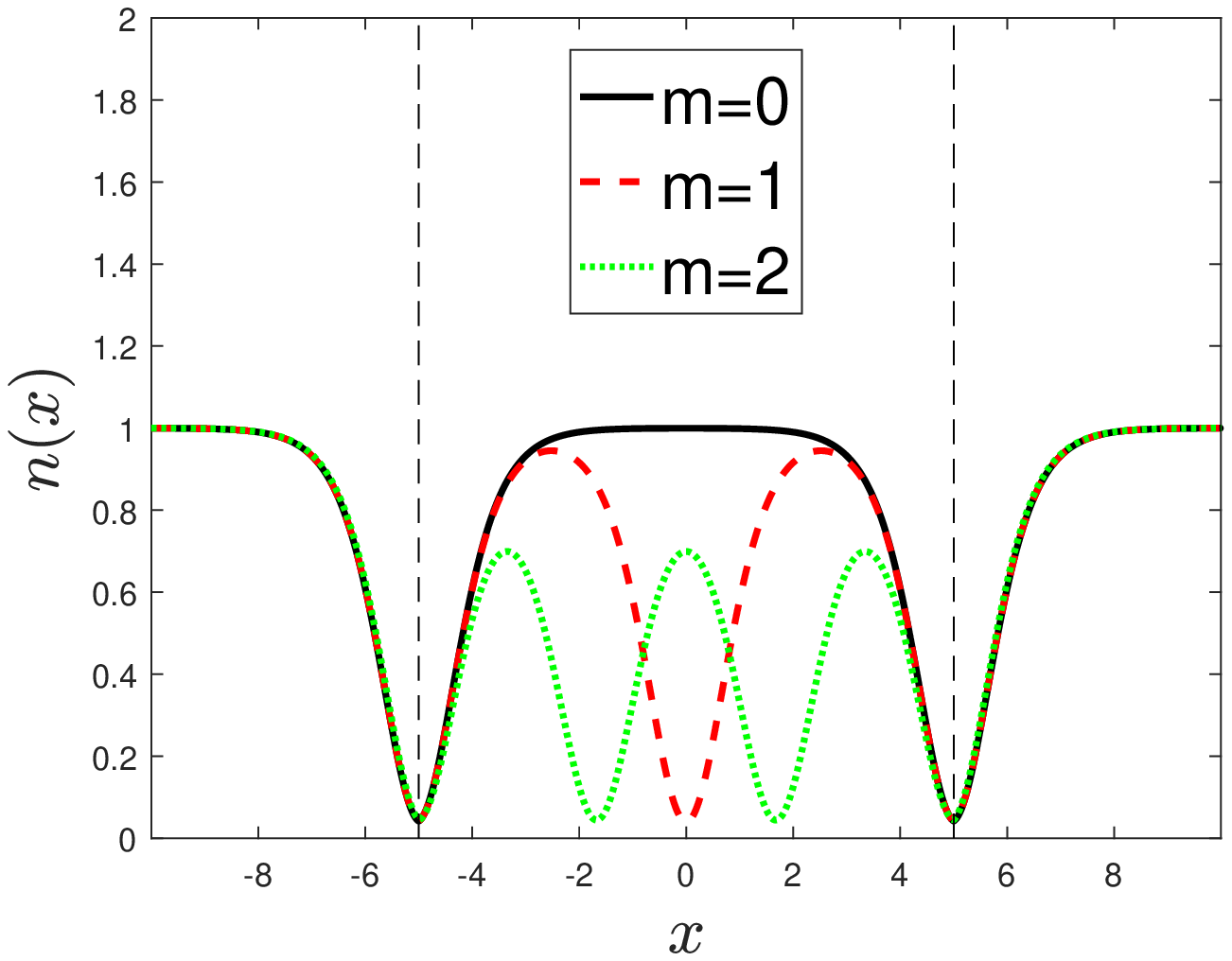}
\end{tabular}
\caption{Density profile of the symmetric solutions for the double-delta configuration. The vertical dashed lines mark the limits of the cavity. Left panel: $m=0,1,2$ $S+$ solutions for $v=0.2$ and $X=7$. Central panels: $m=0,1,2$ $S-$ solutions for $v=0.2$ and $X=0.25,3,5.76$, respectively. Right panel: $m=0,1,2$ $SD$ solutions for $v=0.2$ and $X=10$. }
\label{fig:DeltaI}
\end{figure}

In this case:
\begin{equation}
A'\left(-\frac{X^{+}}{2}\right)>0,~A_{p}<A_W<A_{\sup},~E_p<E_2<E_1
\end{equation}
The corresponding wave function for this family of solutions is:
\begin{equation}\label{eq:SymmetricWaveFunctionDelta}
\psi_0(x)=\psi^{S+}_m(x)=\psi^{S}_m(x,x_0,\phi_0,n^{m}_1,n^{m}_2,n^{m}_3,\alpha_m)\equiv\left\{\begin{array}{cc}
e^{ivx}e^{-i\phi_0}\left(v+i\sqrt{1-v^2}\text{tanh}\left[\sqrt{1-v^2}(x+x_0)\right]\right)& x< -\frac{X}{2}\\
\Lambda(x,n^{m}_1,n^{m}_2,n^{m}_3,\alpha_m) & -\frac{X}{2}< x < \frac{X}{2}\\
e^{ivx}e^{i\phi_0}\left(v+i\sqrt{1-v^2}\text{tanh}\left[\sqrt{1-v^2}(x-x_0)\right]\right) & x>\frac{X}{2}
\end{array}\right.
\end{equation}

We proceed to discuss the two limit values for the energy $E^m_2$. In analogy to the square well, $E^m_2=E_p$ gives $\psi^{S+}_m(x)=\psi^{BHL}(x)$, which exists for an arbitrary length. Following the results of Sec. \ref{subsec:subsonicsolitons}, we analyze the limit $E^m_2\rightarrow E_p$ (see Appendix \ref{appsub:S+} for the details), in which we find that these solutions appear at the critical lengths $X=X^{C}_m$, given by Eq. (\ref{eq:CriticallengthsIDelta}). As for the square well, they can be understood as small perturbations on top of the GP wave function $\psi^{BHL}(x)$, described by the zero-frequency BdG plane waves with wavevector $k_0$ in the supersonic region.

Reasoning in the same way, the critical lengths $X^{C}_m$ are expected to also describe the appearance of new dynamical instabilities. Indeed, the family of solutions $\psi^{S+}_m(x)$ has lower grand-canonical energy than $\psi^{BHL}(x)$; specifically, $\psi^{S+}_0(x)$ is the ground state of the system. Note that, in contrast to the square well, here the ground state is continuously connected to $\psi^{BHL}(x)$ and $\psi^{BHL}(x)$ should only be dynamically unstable for finite lengths $X>X^C_0>0$. Thus, we expect to find in this case a perfect correspondence between dynamical instabilities and stationary solutions with lower grand-canonical energy than the BHL solution (\ref{eq:CompactBHLDelta}), in the same lines of Ref. \cite{Michel2013}.

The upper limit, $E^m_2=E_1$, corresponds to the soliton solution, that gives an infinite value for the cavity length $X$. Hence, the $m$-th $S+$ solution only exists for
\begin{equation}\label{eq:lengthlimitIDelta}
X^{C}_m\leq X \leq \infty
\end{equation}

The density profile of $\psi^{S+}_m(x)$ for $m=0,1,2$ is represented in left Fig. \ref{fig:DeltaI}.

\subsubsection{$S-$ solutions}\label{subsubsec:S-}

In this case:
\begin{equation}
A'\left(-\frac{X^{+}}{2}\right)<0,~A_{q}< A_W<A_{p},~E_p<E_2<E_q
\end{equation}
The wave function is given by the same formal expression of Eq. (\ref{eq:SymmetricWaveFunctionDelta}). This family of solutions is also continuously connected to $\psi^{BHL}(x)$ and the small amplitude limit near $E_p$ gives the same critical lengths $X=X^{C}_m$ as the $S+$ solutions.

The upper limit, $E^m_2=E_q$, only appears for discrete values of the length, $X=X^q_m$, given by Eq. (\ref{eq:lengthsIILimitDelta}); note that $X=X^q_0=0$ corresponds to a trivial single delta-barrier configuration with amplitude $2Z$.

The conditions for the existence of these solutions satisfy similar properties as those of complete-soliton solutions for the square well: for $m=0$, $X^C_0>X^q_0=0$ while for $m$ sufficiently large, $X^C_m<X^q_m$. Hence, reasoning in the same way, the $m=0$ $S-$ solution is limited to the range
\begin{equation}\label{eq:completelengths0}
X\in[0,X^C_0]
\end{equation}
while for $m\geq 1$ we can only generally say that
\begin{equation}\label{eq:completelengths}
X\in[\tilde{X}_m,\max (X^C_m,X^q_m)]
\end{equation}
with $\tilde{X}_m\leq\min (X^C_m,X^H_m)$; check Appendix \ref{appsub:S-} for the details.

The density profile of the $S-$ solutions for $m=0,1,2$ is represented in the central panels of Fig. \ref{fig:DeltaI}.

\subsubsection{$SD$ solutions}\label{subsubsec:SD}

In this case:
\begin{equation}
A'\left(-\frac{X^{+}}{2}\right)>0,~A_{\inf}< A_W<A_{q},~E_q<E_2<E_1
\end{equation}
and this family of solutions is disconnected from the supersonic homogeneous solution with $E_2=E_p$. The wave function is also formally given by Eq. (\ref{eq:SymmetricWaveFunctionDelta}).

The limit solution $E^m_2=E_q$ corresponds to the upper limit of the $m+1$ $S-$ solutions, while $E^m_2=E_1$ is the soliton solution giving infinite cavity length so the condition for the existence of the $m$-th $SD$ solution is
\begin{equation}\label{eq:lengthlimitIIIDelta}
X^{q}_{m+1}\leq X \leq \infty
\end{equation}

The density profile of the $SD$ solutions for $m=0,1,2$ is represented in right Fig. \ref{fig:DeltaI}.

\subsection{Asymmetric solutions}

\begin{figure}[!htb]
\begin{tabular}{@{}cccc@{}}
    \includegraphics[width=0.25\columnwidth]{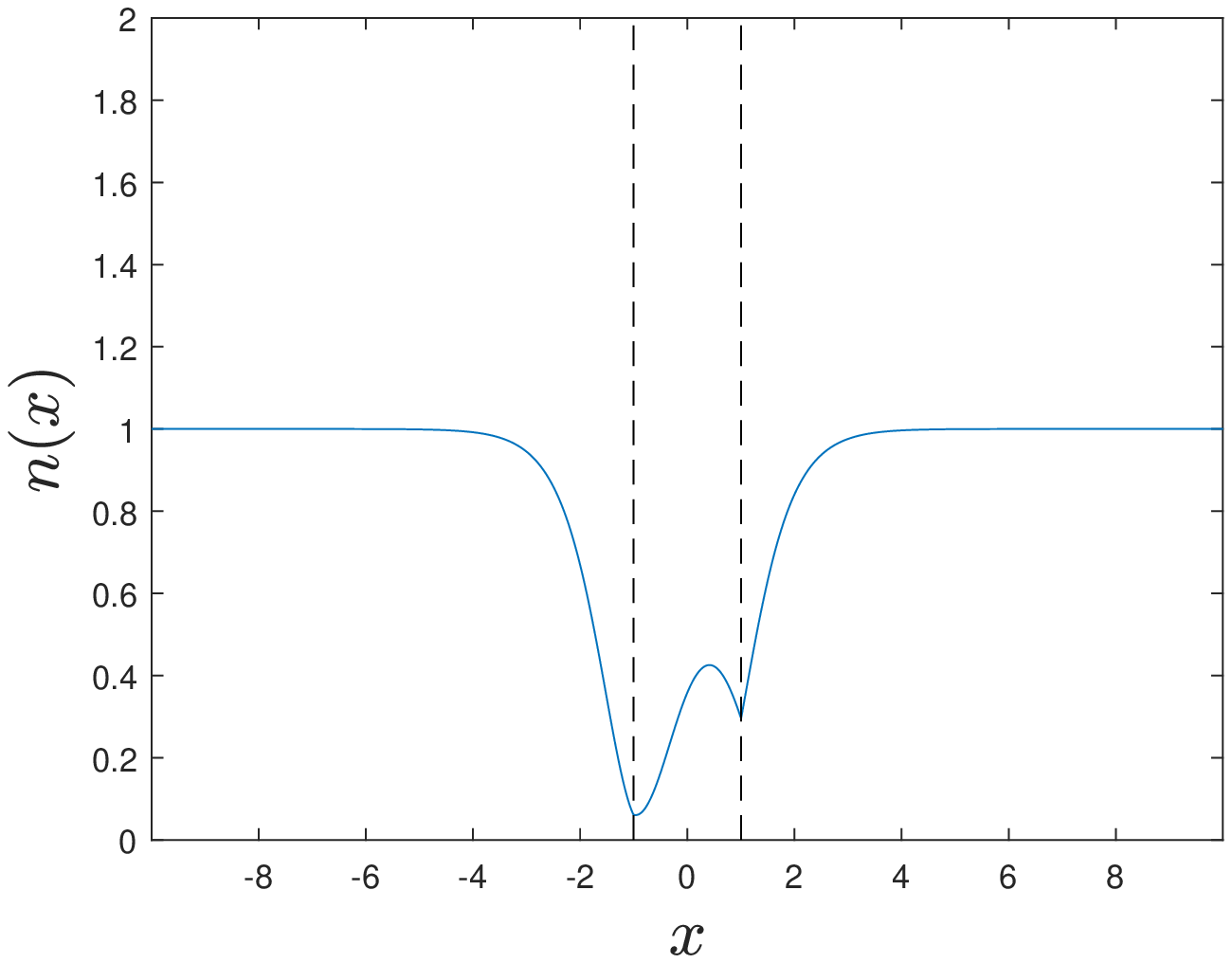} & \includegraphics[width=0.25\columnwidth]{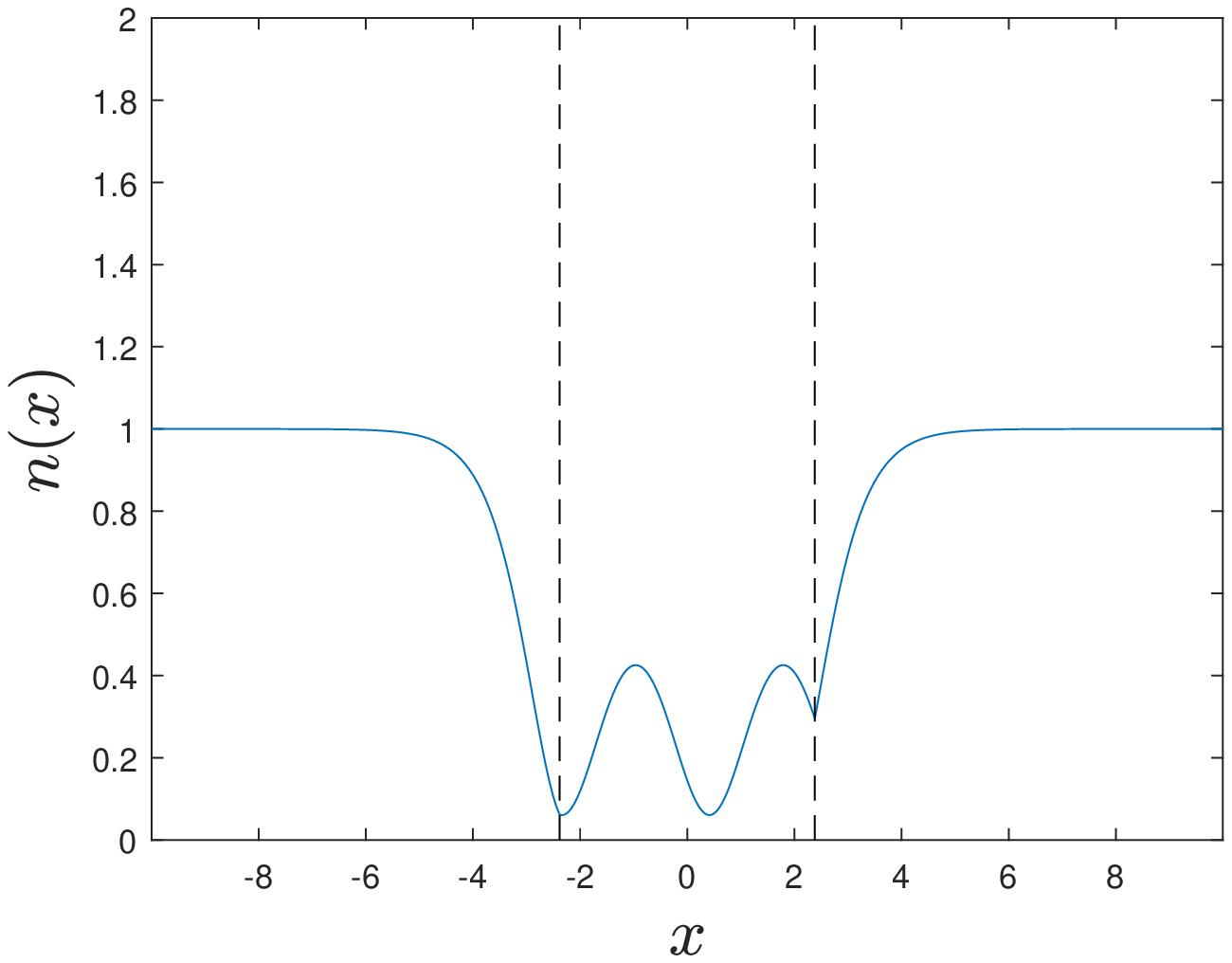} & \includegraphics[width=0.25\columnwidth]{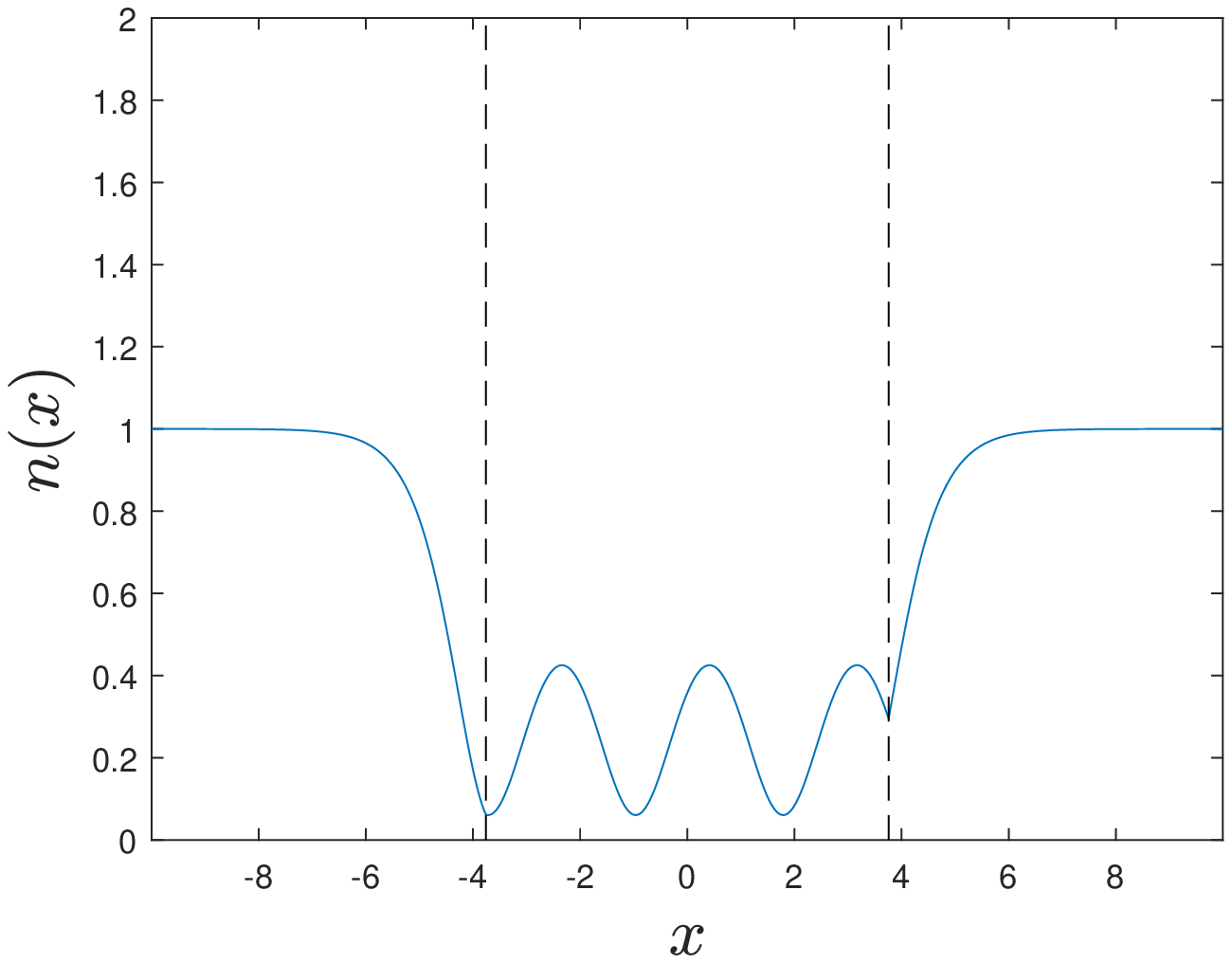} & \includegraphics[width=0.25\columnwidth]{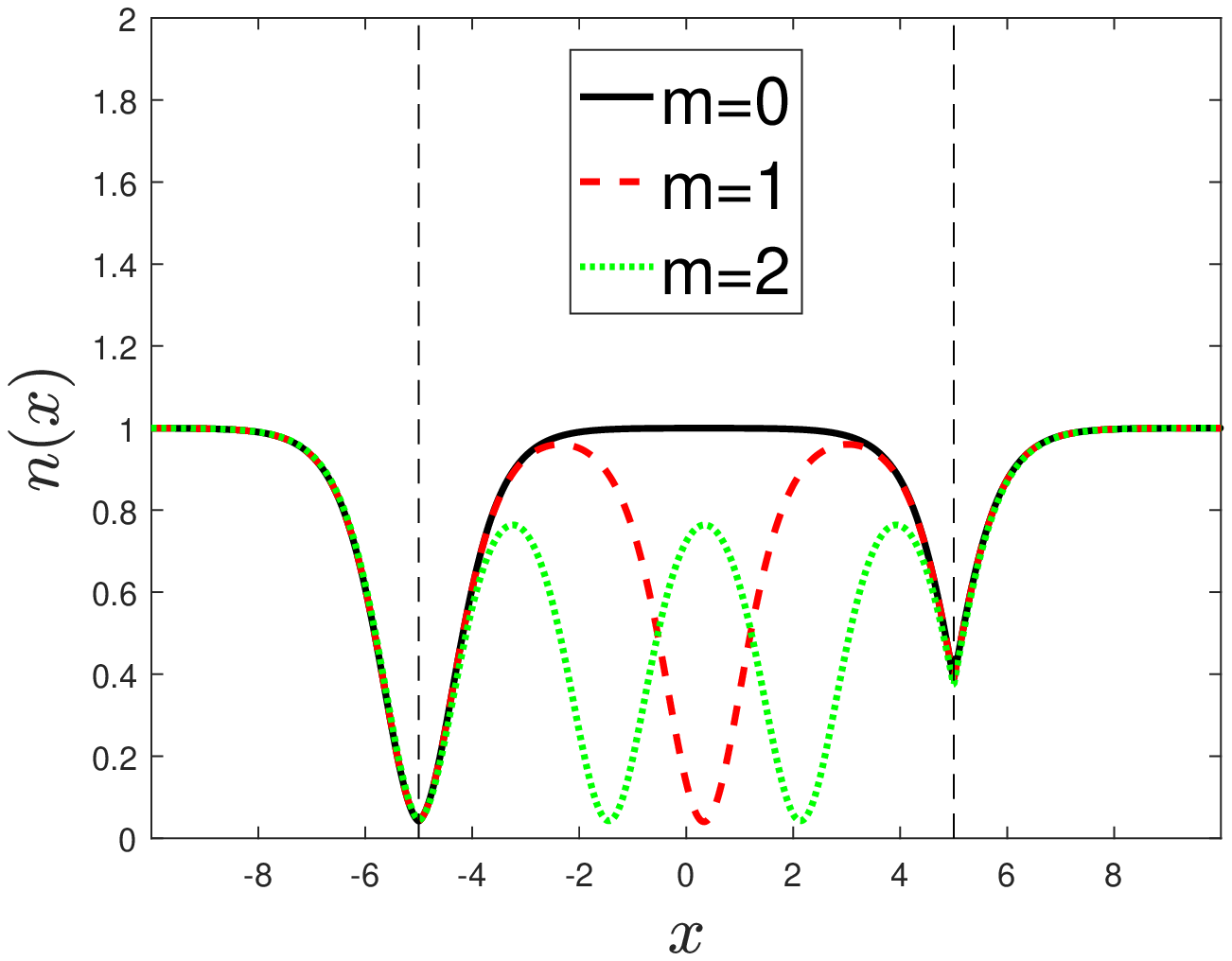}
\end{tabular}
\caption{Density profile of the asymmetric solutions for the double-delta configuration. The vertical dashed lines mark the limits of the cavity. Leftmost panels: $m=0,1,2$ $AC$ solutions for $v=0.2$ and $X=2,4.76,7.51$, respectively. Rightmost panel: $m=0,1,2$ $AD$ solutions for $v=0.2$ and $X=10$.}
\label{fig:DeltaAC}
\end{figure}

For asymmetric solutions:
\begin{equation}
A_{+}\neq A_{-}
\end{equation}
and the matching equations at the edges of the cavity are Eq. (\ref{eq:energyDelta}) and
\begin{equation}\label{eq:deltaasymmetricmatching}
n_\pm=n\left(\pm\frac{X}{2},n_1,n_2,n_3,\alpha\right),~n_\pm=A^2_\pm
\end{equation}
Following the discussion after Eq. (\ref{eq:deltasymmetricmatching}), there are two possible values for $A_{\pm}$ for $E_p<E_2<E_1$. Since $A_{+}\neq A_{-}$, either $A_{-}<A_{+}$ or $A_{+}<A_{-}$; we fix them by imposing $A_{-}<A_{+}$ [the contrary case would just give the wave function resulting from the spatial inversion of $n(x)$]. This choice implies that $A_p<A_{+}<A_{\sup}$ and we distinguish two families of solutions according to the value of $A_{-}$: $AC$, for $A_{q}<A_{-}<A_{p}$; and $AD$, for $A_{\inf}< A_{-}<A_{q}$.

\subsubsection{$AC$ solutions}

The wave function takes the form:
\begin{equation}\label{eq:AsymmetricWaveFunctionDelta}
\psi_0(x)=\psi^{AC}_m(x)\equiv\left\{\begin{array}{cc}
e^{ivx}e^{-i\phi_{-}}\left(v+i\sqrt{1-v^2}\text{tanh}\left[\sqrt{1-v^2}(x+x_{-})\right]\right)& x< -\frac{X}{2}\\
\Lambda(x,n^{m}_1,n^{m}_2,n^{m}_3,\alpha_m) & -\frac{X}{2}<x < \frac{X}{2}\\
e^{ivx}e^{i\phi_{+}}\left(v+i\sqrt{1-v^2}\text{tanh}\left[\sqrt{1-v^2}(x-x_{+})\right]\right) & x>\frac{X}{2}
\end{array}\right.
\end{equation}
with $x_{\pm},\phi_{\pm}$ chosen such that the wave function and its derivative are continuous.

As the energy satisfies $E_{p}<E^m_2\leq E_{q}$, this family of solutions is also continuously connected to the BHL solution (\ref{eq:CompactBHLDelta}). In this limit, the critical lengths of the cavity are $X=X^{A,p}_m$, with $X^{A,p}_m$ given by Eq. (\ref{eq:CriticallengthsAsDeltaLimitTri}), while the opposite limit, $E^m_2=E_q$, gives the critical lengths $X=X^{A,q}_m$, with $X^{A,q}_m$ given by Eq. (\ref{eq:CriticallengthsAsDeltaLimitSup}).

Therefore, the $m$-th $AC$ solution only exists for
\begin{equation}\label{eq:lengthlimitAIDelta}
X\in[X^{A,p}_m, X^{A,q}_{m}]
\end{equation}

The density profile of $AC$ solutions for $m=0,1,2$ is represented in Fig. \ref{fig:DeltaAC}.

\subsubsection{$AD$ solutions}

The wave function for this family of solutions is given by the same formal expression of Eq. (\ref{eq:AsymmetricWaveFunctionDelta}) and the energy satisfies $E_{q}<E^m_2<E_{1}$. Reasoning as for the $SD$ solutions, the $m$-th solution only exists for:
\begin{equation}\label{eq:lengthlimitAs2Delta}
X^{A,q}_{m}\leq X \leq \infty
\end{equation}

The density profile of $AD$ solutions for $m=0,1,2$ is represented in the rightmost panel Fig. \ref{fig:DeltaAC}.

\section{Conclusions and outlook}\label{sec:conclu}

In this work we have analyzed the use of more realistic models for black-hole lasers in Bose-Einstein condensates. First, we have proven a general result that associates a black-hole laser configuration to every compact black-hole solution. As an application, we have proposed two new black-hole laser configurations based on the waterfall and the delta-barrier configurations usually considered for studying analog black holes. In order to characterize them, we have provided a complete classification of the different families of non-linear stationary solutions as they are key to understand the stability of the system as well as its non-linear behavior.

Future works should explore in greater detail these configurations. For instance, a computation of the linear BdG spectrum would provide a further insight on the stability of the system and on the dynamics of the system at short times. Once obtained, a natural task would be to relate the appearance of dynamical instabilities with some of the families of stationary solutions presented in this work, following the ideas outlined in the main text.

On the other hand, in a similar way to Refs. \cite{Michel2015,deNova2016}, the non-linear black-hole lasing regime could be explored by a numerical simulation of the time-dependent Gross-Pitaevskii equation describing the evolution of the instability of the initial black-hole laser solutions $\psi^{BHL}(x)$. According to the results of Ref. \cite{deNova2016}, only two scenarios are expected for late times: either the system converges to the ground state of the system or it enters in a regime of continuous emission of solitons (CES). The characterization of the resulting phase diagram would provide more numerical data that could be useful for the elaboration of a more quantitative theory of the CES regime, which is currently lacking. We note that the production of such a soliton laser is of potential interest in quantum transport scenarios or the emergent field of atomtronics \cite{Seaman2007,Labouvie2015}.

From an experimental point of view, the two black-hole laser configurations here presented describe more realistic scenarios than the typically used flat-profile configuration, as they provide simple models of external potentials easy to implement with standard experimental tools. Consistently, a more realistic numerical simulation should also take into account the complete time evolution of the configuration from the beginning, not just starting from the black-hole laser solution $\psi^{BHL}(x)$: as discussed in Refs. \cite{Tettamanti2016,Wang2016,Steinhauer2017}, the time dependence of the problem is essential in the determination of the mechanism triggering the instability. However, regardless the specific transient of the system, the obtained stationary solutions should still be of great relevance in the non-linear dynamics occurring at long times after the onset of the instability. Among all of them, the family of $SH$ solutions for the attractive well and $S+$ solutions for the double delta-barrier are of special importance as they represent the true ground state of the system.

In addition, as a direct application of the results of the work, the black-hole laser model using an attractive square well is particularly interesting as it is expected to also provide a good description of the actual experimental configuration of Ref. \cite{Steinhauer2014}, much more accurate than the flat-profile configuration. In particular, following the reasoning of the above paragraph, the corresponding stationary states are expected play a key role in the description of future extensions of the experiment \cite{Steinhauer2014} exploring the non-linear dynamics.

\acknowledgments

I thank I. Carusotto, R. Parentani, F. Sols, J. Steinhauer and I. Zapata for useful discussions on the topic. This work has been supported by the Israel Science Foundation.

\appendix

\section{Elliptic functions} \label{app:elliptic}

We briefly present in this Appendix the definition of the different elliptic functions appearing throughout this work; see Refs. \cite{Abramowitz1988,Byrd1971} for more details. The {\em incomplete} elliptic integral of the first kind is defined as
\begin{equation}\label{eq:incompleteelliptic}
F(\phi,\nu)\equiv\int_0^\phi\frac{\mathrm{d}\varphi}{\sqrt{1-\nu\sin^2\varphi}},~0\leq \nu \leq 1
\end{equation}
and the Jacobi amplitude $\text{am}(u,\nu)$ is defined as its inverse function in the argument $\phi$ for fixed $\nu$ so $u=F\left[\text{am}(u,\nu),\nu\right]$.

The Jacobi elliptic functions $\text{sn}(u,\nu), \text{cn}(u,\nu)$ are defined as:
\begin{eqnarray}\label{eq:Jacobielliptic}
\text{sn}(u,\nu)&\equiv&\sin\left[\text{am}(u,\nu)\right]\\
\nonumber \text{cn}(u,\nu)&\equiv&\cos\left[\text{am}(u,\nu)\right]
\end{eqnarray}
As a consequence of the previous relations, $\text{sn}(u,\nu)$ and $\text{cn}(u,\nu)$ are periodic functions with period $4K(\nu)$, $K(\nu)\equiv F(\frac{\pi}{2},\nu)$ being the {\em complete} elliptic integral of the first kind, and satisfy $\text{sn}^2(u,\nu)+\text{cn}^2(u,\nu)=1$. In particular, in the limit of $\nu=0$, they reduce to the usual trigonometric functions,  $\text{sn}(u,0)=\sin~u$, $\text{cn}(u,0)=\cos~u$ and $K(0)=\frac{\pi}{2}$ while in the upper limit, $\nu=1$, $\text{sn}(u,1)=\tanh~u$ and $K(1)=\infty$.

Another interesting function that appears when studying stationary solutions of the GP equation in an infinite well \cite{deNova2014a} is the {\em incomplete} elliptic integral of the second kind
\begin{equation}\label{eq:incompleteelliptic2kind}
E(\phi,\nu)\equiv\int_0^\phi\mathrm{d}\varphi\sqrt{1-\nu\sin^2\varphi}
\end{equation}
with $E(\nu)\equiv E(\frac{\pi}{2},\nu)$ being the {\em complete} elliptic integral of the second kind.

Finally, the function $\Pi(\phi,m,\nu)$ is the {\em incomplete} elliptic integral of the third kind:
\begin{equation}\label{eq:incompleteellipticthird}
\Pi(\phi,m,\nu)\equiv\int_0^\phi\frac{\mathrm{d}\varphi}{(1-m\sin^2\varphi)\sqrt{1-\nu\sin^2\varphi}}
\end{equation}
and it appears when computing the phase of a cnoidal wave, see Eq. (\ref{eq:cnoidalwave}).

\section{Computation of the non-linear solutions of the attractive square well} \label{app:technicalwell}

We provide in this Appendix the technical details of the computation of the different families of stationary states for the attractive square well.

\subsection{Homogeneous plane wave}\label{appsub:homogeneous}

As the wave function outside the well is the homogeneous subsonic plane wave, the amplitude at the edges of the well is fixed to $n_W=1$, so the amplitude energy inside the well is also fixed to the value:
\begin{equation}
E^{H}_2=E_1+V_0=\frac{3}{2}v^2+\frac{1}{2v^2}-\frac{1}{2}
\end{equation}
The roots of Eq. (\ref{eq:rootsWF}) are now $n^H_2=n_W=1$ and
\begin{equation}
n^H_{1,3}=\frac{2E^{H}_2-v^2\mp \sqrt{(2E^{H}_2-v^2)^2-4v^2}}{2}
\end{equation}
As $E_2$ is fixed, this solution only exists for certain critical lengths $X=X^H_m$, computed from Eq. (\ref{eq:matchingWFelliptic}):
\begin{equation}\label{eq:limitlengthsWF}
X^H_m=\frac{2mK\left(\nu^H\right)}{\sqrt{n^H_{3}-n^H_{1}}},~\nu^H=\frac{n^H_{2}-n^H_{1}}{n^H_{3}-n^H_{1}}
\end{equation}
while $\alpha_m=(m+1)K(\nu_m)$.

\subsection{Subsonic shadow solitons}\label{appsub:subsonicsh}

In the case where the wave function outside the well corresponds to shadow solitons, the matching condition (\ref{eq:matchingWFelliptic}) takes the form:
\begin{eqnarray}\label{eq:matchingshWF}
\sqrt{n_3-n_1}\frac{X}{2}+\alpha&=&2(m+1)K(\nu)-\text{sn}^{-1}\left(\sqrt{\frac{n_W-n_1}{n_2-n_1}},\nu\right)\\
\nonumber -\sqrt{n_3-n_1}\frac{X}{2}+\alpha&=&\text{sn}^{-1}\left(\sqrt{\frac{n_W-n_1}{n_2-n_1}},\nu\right)
\end{eqnarray}
from which we find that the amplitude energy $E^m_2$ is computed through the implicit equation
\begin{equation}\label{eq:lengthshWF}
X=X^{SH}_m(E^m_2),~X^{SH}_m(E_2)\equiv2\frac{(m+1)K(\nu)-\text{sn}^{-1}\left(\sqrt{\frac{n_W-n_1}{n_2-n_1}},\nu \right)}{\sqrt{n_3-n_1}},~
\end{equation}
and $\alpha_m=(m+1)K(\nu_m)$.

\subsection{Subsonic solitons}\label{appsub:subsonicsolitons}

\subsubsection{Symmetric solutions}

\paragraph{Incomplete-soliton solutions.}

In this case, the matching condition (\ref{eq:matchingWFelliptic}) gives:
\begin{eqnarray}\label{eq:matchingincompletesolitonWF}
\sqrt{n_3-n_1}\frac{X}{2}+\alpha&=&2mK(\nu)+\text{sn}^{-1}\left(\sqrt{\frac{n_W-n_1}{n_2-n_1}},\nu\right)\\
\nonumber -\sqrt{n_3-n_1}\frac{X}{2}+\alpha&=&-\text{sn}^{-1}\left(\sqrt{\frac{n_W-n_1}{n_2-n_1}},\nu\right)
\end{eqnarray}
from which we obtain that
\begin{equation}\label{eq:constantequationincompletesolitonWF}
X=X^{SOL}_m(E^m_2),~X^{SOL}_m(E_2)\equiv2\frac{mK(\nu)+\text{sn}^{-1}\left(\sqrt{\frac{n_W-n_1}{n_2-n_1}},\nu \right)}{\sqrt{n_3-n_1}}
\end{equation}
and $\alpha_m=mK(\nu_m)$.

The limit $n^m_W\rightarrow v^2$ gives $\nu_m\rightarrow0$ and then all the elliptic functions are reduced to trigonometric functions. Specifically, for small values of the parameter
\begin{equation}\label{eq:smallamplitudeexpansion}
\delta^m_2\equiv E^m_2-E^C_2
\end{equation}
we find from Eqs. (\ref{eq:matchingWF}), (\ref{eq:rootsWF}) that
\begin{eqnarray}\label{eq:smallamplitudeperturbations}
n^m_W&=&v^2+\frac{\delta^m_2}{V_0}\\
\nonumber n^m_{1,2}&\simeq& v^2\pm\delta n^m,~\delta n^m=\sqrt{\frac{2\delta^m_2}{M^2_{p}-1}},~M_p=\frac{1}{v^2}\\
\nonumber n^m_{3}&\simeq& \frac{1}{v^2},\\
\end{eqnarray}
$M_p$ being the supersonic Mach number. Then, as $K(0)=\frac{\pi}{2}$, we find from Eq. (\ref{eq:constantequationincompletesolitonWF}) that the lengths in this limit are:
\begin{equation}\label{eq:criticallengths}
X=X^C_m\equiv X^{SOL}_m(E^m_2\rightarrow E^C_2)=\left(m+\frac{1}{2}\right)\frac{\pi v}{\sqrt{1-v^4}}=\frac{(2m+1)\pi}{k_0},~k_0=2\sqrt{\frac{1}{v^2}-v^2}
\end{equation}

\paragraph{Complete-soliton solutions.}

\begin{figure}[tb!]
\includegraphics[width=1\columnwidth]{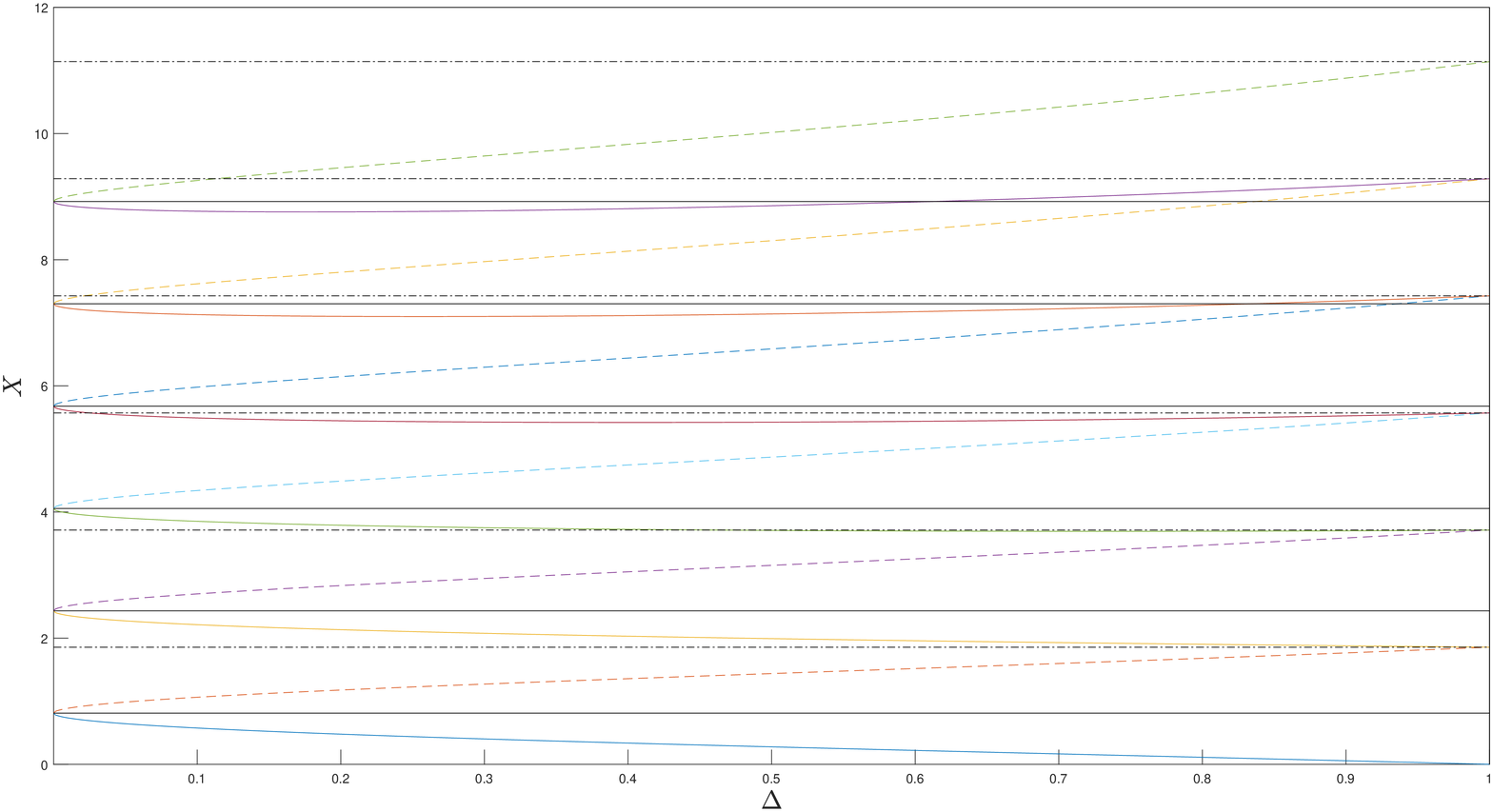}
\caption{Plot of $X^{SH}_m(E_2)$ (solid lines) and $X^{SOL}_m(E_2)$ (dashed lines) as a function of $\Delta\equiv(E_2-E^C_2)/(E^H_2-E^C_2)$ in the range $\delta\in[0,1]$. The horizontal solid and dashed-dotted lines correspond to the limit values $X=X^C_m$ and $X=X^H_m$, respectively. }
\label{fig:LengthWF}
\end{figure}

The matching condition here is formally analog to that of Eq. (\ref{eq:matchingshWF}) so $E^m_2$ is computed from Eq. (\ref{eq:lengthshWF}) and $\alpha_m$ takes the same value; the difference is that now one has to take into account that the solutions outside the well are solitons and then $n^m_W$ is in the range $v^2\leq n^m_W \leq 1$. From Eqs. (\ref{eq:completelengths0}), (\ref{eq:completelengths}), we can expect the behavior of $X^{SH}_m(E_2)$ in Eq. (\ref{eq:matchingshWF}) to be highly non-monotonic in the range $E^C_2\leq E_2\leq E^H_2$ for $m\geq1$. This trend can be observed in Fig. \ref{fig:LengthWF}, where $X^{SH}_m(E_2)$ and $X^{SOL}_m(E_2)$ are represented.

\subsubsection{Asymmetric solutions}

The matching condition (\ref{eq:matchingWFelliptic}) reads:
\begin{eqnarray}\label{eq:matchingasWF}
\sqrt{n_3-n_1}\frac{X}{2}+\alpha&=&2mK(\nu)\pm\text{sn}^{-1}\left(\sqrt{\frac{n_W-n_1}{n_2-n_1}},\nu\right)\\
\nonumber -\sqrt{n_3-n_1}\frac{X}{2}+\alpha&=&\pm\text{sn}^{-1}\left(\sqrt{\frac{n_W-n_1}{n_2-n_1}},\nu\right)
\end{eqnarray}
with $m=1,2,3\ldots$ the number of complete periods inside the well. The value of $E^m_2$ is computed from
\begin{equation}\label{eq:constantequationasWF}
X=X^{A}_m(E^m_2),~X^{A}_m(E_2)\equiv\frac{2mK(\nu)}{\sqrt{n_3-n_1}}
\end{equation}
and then
\begin{equation}\label{eq:constantequationasWF}
\alpha_m=mK(\nu_m)\pm\text{sn}^{-1}\left(\sqrt{\frac{n^m_W-n^m_1}{n^m_2-n^m_1}},\nu\right)
\end{equation}
Following a similar calculation as for the symmetric families, the small amplitude limit $n^m_W\rightarrow v^2$ gives the critical lengths
\begin{equation}\label{eq:criticallengthsAs}
X^A_m\equiv X^{A}_m(E^m_2\rightarrow E^C_2)=\frac{m\pi v}{\sqrt{1-v^4}}
\end{equation}

\section{Computation of the non-linear solutions of the double delta-barrier} \label{app:technicaldelta}

\subsection{Symmetric solutions}

We provide in this Appendix the technical details of the computation of the different families of stationary states for the double delta-barrier.

\subsubsection{$S+$ solutions}\label{appsub:S+}

The matching equations (\ref{eq:deltasymmetricmatching}) give the same formal relations of Eq. (\ref{eq:matchingshWF}) so $\alpha_m=(m+1)K(\nu_m)$ and the possible energies inside the cavity are obtained from:
\begin{equation}\label{eq:lengthsIDelta}
X=X^{S+}_m(E^m_2),~X^{S+}_m(E_2)\equiv2\frac{(m+1)K(\nu)-\text{sn}^{-1}\left(\sqrt{\frac{n_W-n_1}{n_2-n_1}},\nu \right)}{\sqrt{n_3-n_1}},~
\end{equation}
We recall that $n_W(E_2)$ is now a much more complicated function of $E_2$.

Following the discussion associated to Eq. (\ref{eq:smallamplitudeexpansion}), we compute the solution in the small amplitude limit near $\Psi^{BHL}(x)$, expanding all the magnitudes in terms of the parameter $\delta^m_2\equiv E^m_2-E_p$. The roots $n^m_{1,2}$ are computed from Eq. (\ref{eq:smallamplitudeperturbations}), where the supersonic Mach number is now
\begin{eqnarray}\label{eq:smallamplitudeperturbationsDelta}
M_p=\frac{v}{A_p^3}
\end{eqnarray}
The remaining root is $n^m_3\simeq v^2/n^2_p$. On the other hand, as
\begin{equation}
A'\left(-\frac{X^{+}}{2}\right)\simeq 2Z(A-A_p),~\left.\frac{d^2W}{dA^2}\right|_{A=A_p}=2\sqrt{v^4+8v^2}\frac{1-n_p}{n_p}
\end{equation}
from Eq. (\ref{eq:MechanicalEnergyConservationDelta}) we get:
\begin{equation}
n^m_W-n_p\simeq 2A_p(A_W-A_p)\simeq2n_p\sqrt{\frac{\delta^m_2}{2Z^2n_p+\sqrt{v^4+8v^2}(1-n_p)}}
\end{equation}
Using these results, we find that the critical lengths are
\begin{eqnarray}\label{eq:CriticallengthsIDelta}
X=X^{C}_m&\equiv&X^{S+}_m(E^m_2\rightarrow E_p)=\frac{m\pi+\varphi_0}{\sqrt{\frac{v^2}{n^2_p}-n_p}}=\frac{2m\pi+2\varphi_0}{k_0},~k_0=2\sqrt{\frac{v^2}{n^2_p}-n_p}\\
\nonumber \varphi_0&=&2\arcsin\left(\sqrt{\frac{1-r}{2}}\right),~r=n_p\sqrt{\frac{2(M^2_{p}-1)}{2Z^2n_p+\sqrt{v^4+8v^2}(1-n_p)}}
\end{eqnarray}
where we have used the identity
\begin{equation}\label{eq:trigonometricidentity}
\arcsin\left(\sqrt{\frac{1+x}{2}}\right)+\arcsin\left(\sqrt{\frac{1-x}{2}}\right)=\frac{\pi}{2}
\end{equation}

\subsubsection{$S-$ solutions}\label{appsub:S-}

The matching equations (\ref{eq:deltasymmetricmatching}) give the same formal relations of Eq. (\ref{eq:matchingincompletesolitonWF}) so $\alpha_m=mK(\nu_m)$ and the possible energies inside the cavity are obtained from:
\begin{equation}\label{eq:lengthsIIDelta}
X=X^{S-}_m(E^m_2),~X^{S-}_m(E_2)\equiv2\frac{mK(\nu)+\text{sn}^{-1}\left(\sqrt{\frac{n_W-n_1}{n_2-n_1}},\nu \right)}{\sqrt{n_3-n_1}},~
\end{equation}

The small amplitude limit near $E_p$ for this family of solutions gives the same critical lengths $X=X^{S-}_m(E^m_2\rightarrow E_p)=X^{C}_m$ as the $S+$ solutions. The upper limit, $E^m_2=E_q$, only appears for discrete values of the length,
\begin{equation}\label{eq:lengthsIILimitDelta}
X=X^q_m\equiv X^{S-}_m(E_q)=\frac{2mK(\nu^q)}{\sqrt{n^q_3-n^q_1}},~\nu^q=\frac{n^q_{2}-n^q_{1}}{n^q_{3}-n^q_{1}}
\end{equation}
$n^q_i$ being the corresponding roots for $E_2=E_q$, with $n^q_1=n_q$.

The behavior of $X^{S-}_m(E_2)$ in Eq. (\ref{eq:lengthsIIDelta}) in the range $E_p\leq E_2\leq E_q$ is also highly non-monotonic for $m\geq1$, similar to that of $X^{SH}_m(E_2)$ in Fig. \ref{fig:LengthWF}.

\subsubsection{$SD$ solutions}\label{subsubsec:SD}

The matching equations (\ref{eq:deltasymmetricmatching}) are the same as for $S+$ solutions so $\alpha_m=(m+1)K(\nu_m)$ and the possible energies $E^m_2$ inside the cavity are computed from Eq. (\ref{eq:lengthsIDelta}).

\subsection{Asymmetric solutions}

\subsubsection{$AC$ solutions}

In this case, the asymmetric matching condition of Eq. (\ref{eq:deltaasymmetricmatching}) gives:
\begin{eqnarray}\label{eq:matchingasy1Delta}
\sqrt{n_3-n_1}\frac{X}{2}+\alpha&=&2(m+1)K(\nu)-\text{sn}^{-1}\left(\sqrt{\frac{n_{+}-n_1}{n_2-n_1}},\nu\right)\\
\nonumber -\sqrt{n_3-n_1}\frac{X}{2}+\alpha&=&-\text{sn}^{-1}\left(\sqrt{\frac{n_{-}-n_1}{n_2-n_1}},\nu\right)
\end{eqnarray}
from which we obtain that:
\begin{equation}\label{eq:lengthsA1Delta}
X=X^{AC}_m(E^m_2),~X^{AC}_m(E_2)\equiv\frac{2(m+1)K(\nu)+\text{sn}^{-1}\left(\sqrt{\frac{n_{-}-n_1}{n_2-n_1}},\nu\right)-\text{sn}^{-1}\left(\sqrt{\frac{n_{+}-n_1}{n_2-n_1}},\nu \right)}{\sqrt{n_3-n_1}},~
\end{equation}
and
\begin{equation}\label{eq:constantequationasDelta1}
\alpha_m=(m+1)K(\nu_m)-\frac{1}{2}\left[\text{sn}^{-1}\left(\sqrt{\frac{n^m_{+}-n^m_1}{n^m_2-n^m_1}},\nu\right)+\text{sn}^{-1}\left(\sqrt{\frac{n^m_{-}-n^m_1}{n^m_2-n^m_1}},\nu\right)\right]
\end{equation}

The limit $E^m_2\rightarrow E_p$ gives the critical lengths
\begin{equation}\label{eq:CriticallengthsAsDeltaLimitTri}
X=X^{A,p}_m\equiv X^{AC}_m(E^m_2\rightarrow E_p)=\frac{\left(m+\frac{1}{2}\right)\pi+\varphi_0}{\sqrt{\frac{v^2}{n^2_p}-n_p}}
\end{equation}
while, in the opposite limit, $E^m_2=E_q$,
\begin{equation}\label{eq:CriticallengthsAsDeltaLimitSup}
X=X^{AC}_m(E_q)=X^{A,q}_m,~X^{A,q}_m\equiv\frac{2(m+1)K(\nu^q)-\text{sn}^{-1}\left(\sqrt{\frac{n^q_{+}-n^q_1}{n^q_2-n^q_1}},\nu^q \right)}{\sqrt{n^q_3-n^q_1}},~
\end{equation}
with $n^q_{+}$ the density of the $A^q_{+}>A_p$ solution to Eq. (\ref{eq:energyDelta}) for $E_2=E_q$.

\subsubsection{$AD$ solutions}

The asymmetric matching condition of Eq. (\ref{eq:deltaasymmetricmatching}) gives now:
\begin{eqnarray}\label{eq:matchingasy2Delta}
\sqrt{n_3-n_1}\frac{X}{2}+\alpha&=&2(m+1)K(\nu)-\text{sn}^{-1}\left(\sqrt{\frac{n_{+}-n_1}{n_2-n_1}},\nu\right)\\
\nonumber -\sqrt{n_3-n_1}\frac{X}{2}+\alpha&=&\text{sn}^{-1}\left(\sqrt{\frac{n_{-}-n_1}{n_2-n_1}},\nu\right)
\end{eqnarray}
which yields
\begin{equation}\label{eq:lengthsA2Delta}
X=X^{AD}_m(E^m_2),~X^{AD}_m(E_2)\equiv\frac{2(m+1)K(\nu)-\text{sn}^{-1}\left(\sqrt{\frac{n_{+}-n_1}{n_2-n_1}},\nu \right)-\text{sn}^{-1}\left(\sqrt{\frac{n_{-}-n_1}{n_2-n_1}},\nu \right)}{\sqrt{n_3-n_1}},~
\end{equation}
and
\begin{equation}\label{eq:constantequationasDelta2}
\alpha_m=(m+1)K(\nu_m)+\frac{1}{2}\left[\text{sn}^{-1}\left(\sqrt{\frac{n^m_{-}-n^m_1}{n^m_2-n^m_1}},\nu\right)-\text{sn}^{-1}\left(\sqrt{\frac{n^m_{+}-n^m_1}{n^m_2-n^m_1}},\nu\right)\right]
\end{equation}
where $E_{q}<E^m_2<E_{1}$.

\bibliographystyle{apsrev}
\bibliography{Hawking}

\end{document}